\documentclass{elsartNoFoot}

\usepackage{graphicx}
\usepackage{jb}
\usepackage{epsf}
\usepackage{amssymb}
\usepackage{url}

\newcommand{\LABEL}[1]{\label{#1}}

\newcommand{\la}{\leftarrow}
\newcommand{\laa}{\longleftarrow}
\newcommand{\La}{\Leftarrow}

\newcommand{\Lra}{\Leftrightarrow}
\newcommand{\lra}{\leftrightarrow}
\newcommand{\ra}{\rightarrow}
\newcommand{\raa}{\longrightarrow}
\newcommand{\Ra}{\Rightarrow}
\renewcommand{\leq}{\leqslant}
\renewcommand{\geq}{\geqslant}

\newcommand{\tpl}[1]{\langle #1 \rangle}	
\newcommand{\set}[1]{\{ #1 \}}			
\newcommand{\subst}[1]{\{ #1 \}}		
\newcommand{\abs}[1]{\raisebox{0.04cm}{\scriptsize \#}#1}
\newcommand{\sortdef}{::=}

\newcommand{\dlr}[1]{{\it dlr}(#1)}

\newcommand{\eqc}[3]{[#1]_{#2}^{#3}}

\renewcommand{\emph}[1]{{\em #1\/}}
\renewcommand{\leadsto}{\rightsquigarrow}

\newcommand{\V}{{\cal V}}		
\renewcommand{\L}{{\cal L}}		
\newcommand{\NT}{{\cal N}}		
\renewcommand{\O}{{\cal O}}		
\newcommand{\G}{{\cal G}}		
\newcommand{\R}{{\cal R}}		
\newcommand{\N}{I\!\!N}			
\newcommand{\suc}{s}			

\newcommand{\bigmid}{\rule[-0.07cm]{0.04cm}{0.4cm}\hspace*{0.1cm}}

\newcommand{\vN}[2]{{\sf v}(#1,#2)}	

\newcommand{\NN}{{\sf N}}		
\newcommand{\NM}{\NT_{\max}}		
\newcommand{\NNt}[1]{{\sf N}(#1)}	
\newcommand{\vNN}[1]{{\sf v}(#1)}	
\newcommand{\tNN}[1]{{\sf t}(#1)}	

\renewcommand{\mod}{\mathop{\scriptstyle \%}}
\newcommand{\idiv}{\mathop{/\!/}}

\newcommand{\I}{{\cal I}}

\newcommand{\eq}{\mathrel{=_E}}

\newcommand{\hsigma}{\chi}

\newcommand{\T}[3]{%
	\Ite{#1}%
		{\Ite{#2}%
			{\Ite{#3}%
				{{\cal T}_{#3,#2}(#1)}
				{{\cal T}_{#2}(#1)}
			}%
			{\Ite{#3}%
				{{\cal T}_{#3}(#1)}
				{{\cal T}(#1)}
			}%
		}%
		{\Ite{#2}%
			{\Ite{#3}%
				{{\cal T}_{#3,#2}}
				{{\cal T}_{#2}}
			}%
			{\Ite{#3}%
				{{\cal T}_{#3}}
				{{\cal T}}
			}%
		}%
	}

\renewcommand{\:}[4]{%
        {%
        \renewcommand{\:}[4]{%
                {%
                \renewcommand{\:}[4]{error\error}%
                \renewcommand{\j}{{##2}}%
                {##4}%
                ##1...##1%
                \renewcommand{\j}{{##3}}%
                {##4}%
                }%
        }%
        \renewcommand{\i}{{#2}}%
        {#4}%
        #1\ldots#1%
        \renewcommand{\i}{{#3}}%
        {#4}%
        }%
}

\newcommand{\bigOp}[5]{%
	{%
	\renewcommand{\i}{#1}%
	\renewcommand{\bigOp}[5]{%
		{%
		\renewcommand{\j}{#1}%
		\renewcommand{\bigOp}[5]{error\error}%
		#2_{#1=#3}^{#4} #5%
		}%
	}%
	#2_{#1=#3}^{#4} #5%
	}%
}

\newcommand{\THM}[3]{%
	\begin{#1}%
	\Ite{#2}{ (#2) }{ }%
	{#3}%
	\end{#1}%
	}
\newcommand{\PRF}[2]{%
	\begin{pf}%
	{#2}%
	\end{pf}%
	}
\newcommand{\EXAMPLE}[2]   {\THM{example}   {#1}{#2}}
\newcommand{\LEMMA}[2]     {\THM{lemma}     {#1}{#2}}
\newcommand{\THEOREM}[2]   {\THM{theorem}   {#1}{#2}}
\newcommand{\DEFINITION}[2]{\THM{definition}{#1}{#2}}

\newcommand{\PROOF}[1]     {\PRF{\em Proof. }       {#1}}

\newcommand{\cal}{\mathcal}

\newcommand{\tNNs}[1]{\overline{#1}}
\newcommand{\tNNt}[1]{\overline{#1}}

\newcommand{\impl}{\models}
\newcommand{\hypp}[2]{{\sf h}^+\Ite{#1}{\Ite{#2}{(#1,#2)}{(#1)}}{}}
\newcommand{\hypn}[2]{{\sf h}^-\Ite{#1}{\Ite{#2}{(#1,#2)}{(#1)}}{}}
\newcommand{\Hyp}[3]{{\sf H}\Ite{#1}{_{#1}}{}%
		\Ite{#2}%
		{%
			\Ite{#3}%
			{(#2,#3)}%
			{(#2)}%
		}%
		{}%
}

\newcommand{\eps}{\varepsilon}

\newcommand{\function}[1]{{\sf #1}}

\renewcommand{\.}[1]{\!#1\!}

\begin{document}

\begin{frontmatter}
{\scriptsize\sl FIRST Technical Report 002 \hfill \today}

\title{$E$-Generalization Using Grammars}
\author{Jochen Burghardt}
\address{jochen@first.fhg.de}
\begin{abstract}
We extend the notion of anti-unification to cover equational
theories and present a method based on regular tree grammars to
compute a finite representation of $E$-generalization sets.
We present 
a framework to combine 
Inductive Logic Programming and $E$-generalization
that includes an extension of Plotkin's ${\it lgg}$ theorem to the
equational case.
We demonstrate the potential power of $E$-generalization
by three example
applications: computation of suggestions for auxiliary lemmas in
equational inductive proofs, computation of construction laws for given
term sequences, and learning of screen editor command sequences.
\end{abstract}
\end{frontmatter}


\renewcommand{\,}[3]{\:{,}{#1}{#2}{#3}}

\section{Introduction}
\LABEL{Introduction}

Many learning techniques in the field of symbolic Artificial
Intelligence are based on
adopting the features common to the given examples, called
\emph{selective induction} in the classification of
\cite{Dietterich.Michalski.1984}, for example.
Syntactical \emph{anti-unification}
reflects these abstraction
techniques in the theoretically elegant domain of term algebras.

In this article, we propose an extension, called
$E$-anti-unification or $E$-generalization, which also
provides a way of coping with the well-known problem of
representation change \cite{OHara.1992,Dastani.Indurkhya.Scha.1997}.
It allows us to perform abstraction while
modeling equivalent representations
using appropriate equations between terms.
This means that
all equivalent representations are considered simultaneously in
the abstraction process.
Abstraction becomes insensitive to representation changes.

In 1970, Plotkin and Reynolds
\cite{Plotkin.1970,Plotkin.1971,Reynolds.1970}
introduced the notion of (syntactical)
anti-unification of terms as the dual
operation to unification: while the latter computes the most general
common specialization of the given terms, if it exists,
the former computes the most
special generalization of them, which always exists and is unique up to
renaming.
For example, using the usual $0$-$\suc$ representation of natural
numbers
and abbreviating $\suc(\suc(0))$ to $\suc^2(0)$,
the terms $0*0$ and $\suc^2(0)*\suc^2(0)$ anti-unify to $x*x$,
retaining the common function symbol $*$ as well as the equality of its
arguments.

While extensions of unification to equational theories and classes
of them have been investigated
\cite{Fay.1979,Siekmann.1985,Gallier.Snyder.1989},
anti-unification has long been neglected in this respect,
except for the theory of associativity and commutativity
\cite{Pottier.1989b}
and so-called \emph{commutative theories} \cite{Baader.1991}.
For an arbitrary equational theory $E$,
the set of all $E$-generalizations
of given terms is usually infinite.
Heinz \cite{Heinz.1995,Burghardt.Heinz.1996} presented a
specially tailored algorithm that uses
regular tree grammars to
compute a finite representation of this set,
provided $E$ leads to regular congruence classes.
However, this work has never been internationally published.
In this paper, we try to make up for this neglect,
giving an improved presentation using standard grammar
algorithms only, and adding some new theoretical results and
applications
(Sect.~\ref{More expressive grammar formalisms},
\ref{Learning predicate definitions},
\ref{Generalizing screen editor commands} below).

\begin{figure}
\begin{center}
\begin{tabular}{@{}r@{\hspace*{0.5cm}}r@{$\;$}l@{}}
1. & $x+0$ & $= x$	\\
2. & $x+\suc(y)$ & $= \suc(x+y)$	\\
3. & $x*0$ & $= 0$	\\
4. & $x*\suc(y)$ & $= x*y+x$	\\
\end{tabular}
\hspace*{2.5cm}
\begin{tabular}{@{}c@{}c@{}c@{}l@{}}
$0$ & $\eq$ & $0*0$
	\\
$\suc^4(0)$ & ~$\eq$~
	& $\suc^2(0)*\suc^2(0)$
	\\
\cline{1-1}
\cline{3-3}
{\footnotesize syn.\ anti-un.}
	&& {\footnotesize syn.\ anti-un.} \\
\cline{1-1}
\cline{3-3}
$y$ && $x*x$\rule[0.4cm]{0cm}{0cm}	\\
\end{tabular}
\caption{Equations Defining $(+)$ and $(*)$ \hspace*{1.0cm}
	$E$-Generalization of $0$ and $\suc^4(0)$}
\LABEL{$E$-generalization of $0$ and $\suc^4(0)$}
\end{center}
\end{figure}

In general, $E$-anti-unification
provides a means to find correspondences that are only detectable
using an equational theory as background knowledge.
By way of a simple example, consider the terms $0$ and $\suc^4(0)$.
Anti-unifying them purely syntactically, without considering an
equational
theory, we obtain the term $y$, which indicates 
that there is no common structure.
If, however, we consider the usual defining equations 
for $(+)$ and $(*)$,
see Fig.~\ref{$E$-generalization of $0$ and $\suc^4(0)$} (left),
the terms may be rewritten nondeterministically as shown in
Fig.~\ref{$E$-generalization of $0$ and $\suc^4(0)$} (right), and then
syntactically anti-unified to $x*x$ as one possible result.
In other words, it is recognized that both terms are quadratic numbers.

Expressed in
predicate logic,
this means we can learn a definition $p(x*x)$ from the examples
$p(0)$ and $p(s^4(0))$.
Other possible results are $p(s^4(0)*x)$,
and, less meaningfully, $p(x+y*z)$.
The closed representation of generalization sets by grammars allows us
to filter out generalizations with certain properties that
are undesirable in a given application context.

After some formal definitions in Sect.~\ref{Definitions},
we introduce our method of $E$-generalization based on regular tree
grammars in Sect.~\ref{$E$-generalization} and briefly discuss
extensions to more sophisticated grammar formalisms.
As a first step toward integrating $E$-generalization into Inductive
Logic Programming (ILP),
we provide, in Sect.~\ref{Learning predicate definitions},
theorems for learning determinate or nondeterminate
predicate definitions using atoms or clauses.
In Sect.~\ref{Applications}, we present applications of determinate atom
learning in different areas,
including inductive equational theorem-proving, 
learning of series-construction laws and user support for
learning advanced screen-editor commands.
Section \ref{Conclusions} draws some conclusions.

\section{Definitions}
\LABEL{Definitions}

We assume familiarity with the classical definitions of terms,
substitutions \cite{Dershowitz.Jouannaud.1990},
and Horn clauses \cite{Kowalski.1973}.
The cardinality of a finite set $S$ is denoted by $\abs{S}$.

A signature $\Sigma$ is a set of function symbols $f$,
each of which has a fixed arity;
if some $f$ is nullary, we call it a constant.
Let $\V$ be an infinite set of variables.
$\T{}{}{V}$ denotes the set of all terms over
$\Sigma$ and a given $V \subseteq \V$.
%
For a term $t$,
$\var(t)$ denotes the set of variables occurring in $t$;
if it is empty, we call $t$ a ground term.
We call a term linear if each variable occurs at most once in it.

By $\subst{\,1n{x_\i \mapsto t_\i}}$,
or $\subst{x_i \mapsto t_i \allowbreak \mid 1 \leq i \leq n}$,
we denote a substitution that
maps each variable $x_i$ to the term $t_i$.
We call it ground if all $t_i$ are ground.
We use the
postfix notation $t \sigma$ for application of $\sigma$ to $t$,
and $\sigma \tau$ for the composition of $\sigma$ (to be applied first)
and $\tau$ (second).
The domain of $\sigma$ is denoted by $\dom \sigma$.

A term $t$ is called an instance of a term $t'$ if
$t = t' \sigma$ for some substitution $\sigma$.
In this case, we call $t$ more special than $t'$,
and $t'$ more general than $t$.
We call $t$ a renaming of $t'$ if $\sigma$ is a bijection that maps
variables to variables.

A term $t$ is called a syntactical generalization
of terms
$t_1$ and $t_2$, if there exist substitutions $\sigma_1$ and $\sigma_2$
such that $t \sigma_1 = t_1$ and $t \sigma_2 = t_2$.
In this case, $t$ is called the most specific syntactical
generalization of
$t_1$ and $t_2$, if for each syntactical
generalization $t'$ of $t_1$ and
$t_2$ there exists a substitution $\sigma'$ such that $t = t' \sigma'$.
The most specific syntactical generalization of two terms
is unique up to renaming; we also call it their syntactical anti-unifier
\cite{Plotkin.1970}.

An equational theory $E$ is a finite set of equations between terms.
$(\eq)$ denotes the smallest congruence relation that contains all
equations of $E$.
Define
$$\eqc{t}{E}{} = \set{t' \in \T{}{}{\set{}}\mid t' \eq t}$$
to be the congruence class of $t$ in the algebra of ground terms.
The congruence class of a term is usually infinite;
for example,
using the equational theory from Fig.~\ref{$E$-generalization of
$0$ and $\suc^4(0)$} (left),
we have
$\eqc{0}{E}{} = \set{0, 0*\suc(0), 
0+0*0,
\ldots}$.
Let
$$\eqc{t}{E}{\sigma}
= \set{t' \in \T{}{}{\dom \sigma} \mid t' \sigma \eq t}$$
denote the set of all terms congruent to~$t$ under $\sigma$.

A congruence relation $(=_1)$ is said to be a refinement of another
congruence relation $(=_2)$,
if
$\forall t,t' \in \T{}{}{\V} \;\; t =_1 t' \Ra t =_2 t'$.
In Sect.~\ref{Candidate lemmas in inductive proofs},
we need the definition
$t_1 \equiv_E t_2$ if $t_1 \sigma \eq t_2 \sigma$
for all ground
substitutions $\sigma$ with $\var(t_1) \cup \var(t_2) \subseteq
\dom \sigma$;
this is equivalent to the equality of $t_1$ and $t_2$ being inductively
provable \cite[Sect.~3.2]{Dershowitz.Jouannaud.1990}.

We call an $n$-ary function symbol $f$ a constructor
if $n$ functions $\,1n{\pi^f_\i}$ exist such that
$$\forall t, \,1n{t_\i}:
(\bigwedge_{i=1}^n \pi^f_i(t) \eq t_i) \lra t \eq f(\,1n{t_\i}) .$$
The $\pi^f_i$ are called selectors associated to $f$.
As usual, we assume additionally that
$f( \,1n{s_\i} ) \not\eq g( \,1m{t_\i} ) \not\eq x$
for any two constructors $f \neq g$,
any variable $x$ and arbitrary terms $s_i,t_j$.
On this assumption, some
constants can also be called
constructors.
No selector can be a constructor.
If $f$ is a constructor, then
$\eqc{f(\,1n{t_\i})}{E}{} = f(\,1n{\eqc{t_\i}{E}{}})$.

A term $t$ is called a constructor term if it is built from constructors
and variables only.
Let $t$ and $t'$ be constructor terms.


\LEMMA{Constructor terms}{ \LABEL{cr terms}
Let $t$ and $t'$ be constructor terms.
\begin{enumerate}
\item If $t \sigma \eq t'$,
        then $t \sigma' = t'$ for some $\sigma'$
        such that 
        $x \sigma'$ is a constructor term 
        and $x \sigma' \eq x \sigma$ 
        for each $x \in \V$.
\item If $t \eq t'$, then $t = t'$.
\item If $t \sigma_1 \eq t \sigma_2$,
        then $\forall x \in \var(t): \; x \sigma_1 \eq x \sigma_2$.
\end{enumerate}
}
\PROOF{
\begin{enumerate}
\item Induction on $t$: 
        \begin{itemize}
        \item If $t = x \in \V$,
                then
                choose $\sigma' = \subst{x \mapsto t'}$.
        \item If $t = f( \,1n{t_\i} )$,
                ~
                then
                $f( \,1n{t_\i \sigma} ) \eq t'$.
                Hence,
                $t' = f( \,1n{t'_\i} )$ for some $t'_i$, 
                and
                $t_i \sigma
                \eq \pi^f_i(t \sigma) 
                \eq \pi^f_i(t')
                \eq t'_i$.
                By I.H.,
                $t_i \sigma'_i = t'_i$ for some $\sigma'_i$.
                For any $i$, $j$, and
                $x \in \dom \sigma'_i \cap \dom \sigma'_j$,
                we have $x \sigma'_i \eq x \sigma \eq x \sigma'_j$,
                and therefore $x \sigma'_i = x \sigma'_j$.
                Hence, all $\sigma'_i$ are compatible, 
                and we can unite them into a single $\sigma'$.
        \end{itemize}
\item Follows from~1 with $\sigma = \subst{\;}$.
\item Induction on $t$: 
        \begin{itemize}
        \item For $t = x \in \V$, we have nothing to show.
        \item If $t = f( \,1n{t_\i} )$,
                we have
                $t_i \sigma_1
                \eq \pi^f_i(t \sigma_1)
                \eq \pi^f_i(t \sigma_2)
                \eq t_i \sigma_2$,
                and we are done by I.H.
\qed
        \end{itemize}
\end{enumerate}
}

A (nondeterministic) regular tree grammar
\cite{Thatcher.Wright.1968,Comon.Dauchet.Gilleron.1999}
is a triple $\G = \tpl{\Sigma,\NT,\R}$.
$\Sigma$ is a signature,
$\NT$ is a finite set of nonterminal symbols
and
$\R$ is a finite set of rules of the form
$$N \sortdef \:{\mid}1m{ f_\i( \,{1}{n_\i}{ N_{\i\j} } ) }$$
or, abbreviated,
$$N \sortdef \bigmid_{i=1}^m \; f_i(N_{i1},...,N_{in_i}) .$$
Each $f_i( \,{1}{n_i}{ N_{i\i} } )$
is called an alternative of the rule.
We assume that for each nonterminal $N \in \NT$, there is exactly one
defining rule in $\R$ with $N$ as its left-hand side.
As usual, the rules may be mutually recursive.

Given a grammar $\G$ and a nonterminal $N \in \NT$,
the language $\L_\G(N)$ produced by $N$
is defined in the usual way
as the set of all
ground terms derivable from $N$ as the start symbol.
We omit the index $\G$ if it is clear from the context.
We denote the total number of alternatives in $\G$ by $\abs{\G}$.

In Sect.~\ref{Learning predicate definitions},
we will use the following predicate logic definitions.
To simplify notation,
we sometimes assume all predicate symbols to be unary.
An $n$-ary predicate $p'$ can be simulated by a unary $p$
using an $n$-ary tupling constructor symbol and defining
$p(\tpl{\,1n{t_\i}}) \Lra p'(\,1n{t_\i})$.

An $n$-ary predicate $p$ is called determinate wrt.\ some background
theory $B$ if
there is some $k$ such that w.l.o.g.\
each of the arguments $\,{k+1}n\i$ has only one possible binding,
given the bindings of the arguments $\,1k\i$
\cite[Sect.~5.6.1]{Dzeroski.Lavrac.1994}.
The background theory $B$ may be used to define $p$,
hence $p$'s determinacy depends on $B$.
Similar to the above, we sometimes
write $p(\,1n{t_\i})$ as a binary predicate
$p(\tpl{\,1k{t_\i}} , \tpl{\,{k+1}n{t_\i}})$
to reflect the two classes of arguments.
For a binary determinate predicate $p$, the relation
$\set{ \tpl{s,t} \mid s,t \in \T{}{}{\set{}} \land B \impl p(s,t) }$
corresponds to a function $g$.
We sometimes assume that $g$ is defined by equations from a given
$E$,
i.e.\ that $B \impl p(s,t) \lra g(s) \eq t$.

A literal has the form $p(t)$ or $\lnot p(t)$,
where $p$ is a predicate symbol and $t$ is a term.
We consider a negation to be part of the
predicate symbol.
We say that the literals $L_1$ and $L_2$ fit if both
have the same predicate symbol, including negation.
We extend $(\eq)$ to literals
by defining $p(t_1) \eq p(t_2)$ if $t_1 \eq t_2$.
For example,
$(\lnot {\it divides}(\tpl{1+1,5}))
\eq (\lnot {\it divides}(\tpl{2,5}))$
if $1+1 \eq 2$.

A clause is a finite set $C = \set{\,1n{L_\i}}$ of literals,
with the meaning $C \Lra \:{\lor}1n{L_\i}$.
We consider only nonredundant clauses, i.e.\ clauses that
do not contain congruent literals.
For example, $\set{ p(x+0), p(x) }$ is redundant if $x+0 \eq x$.
We write
$C_1 \subseteq_E C_2$ if
$$\forall L_1 \in C_1 \; \exists L_2 \in C_2: \; L_1 \eq L_2 ;$$
if $C_2$ is nonredundant, $L_2$ is uniquely determined by $L_1$.

We say that
$C_1$ $E$-subsumes $C_2$ if
$C_1 \sigma \subseteq_E C_2$ for some $\sigma$.
In this case, the conjunction of $E$ and $C_1$ implies $C_2$;
however, there are other cases in which $E \land C_1 \impl C_2$
but $C_1$ does not $E$-subsume $C_2$.
For example, $\set{ \lnot p(x), p(f(x))}$ implies, but does not subsume,
$\set{ \lnot p(x), p(f(f(x))) }$, even for an empty $E$.

A Horn clause is a clause $\set{p_0(t_0), \,1n{\lnot p_\i(t_\i)}}$
with exactly one positive literal.
It is also written as $p_0(t_0) \la \:{\land}1n{p_\i(t_\i)}$.
We call $p_0(t_0)$ the head literal,
and $p_i(t_i)$ a body literal for $i=\,1n\i$.
Like
\cite[Sect.~2.1]{Dzeroski.Lavrac.1994},
we call the Horn clause constrained
if $\var(t_0) \supseteq \var(\,1n{t_\i})$.

We call a Horn clause
$$p_0(s_0,t_0) 
\la \bigOp{i}{\bigwedge}1n{p_\i(s_\i,x_\i)}~%
\land~\bigOp{i}{\bigwedge}1m{q_\i(t_\i)}$$
semi-determinate wrt.\ some background theory $B$
if
all $p_i$ are determinate wrt.\ $B$,
all variables $x_i$ are distinct and
do not occur in~$s_0$,
~
$\var(s_i) 
\subseteq \var(s_0) \cup \set{ \,1{i-1}{x_\i} }$,
and
$\var(\,1m{t_\i}) \subseteq \var(s_0,\,1n{x_\i})$.
Semi-determinacy for clauses is a slight extension of determinacy
defined by \cite[Sect.~5.6.1]{Dzeroski.Lavrac.1994},
as it additionally permits arbitrary predicates $q_i$.
On the other hand, \cite{Dzeroski.Lavrac.1994} 
permits $x_i = x_j$ for $i \neq j$;
however,
$p_i(s_i,x_i) \land 
p_j(s_j,x_i)$
can be equivalently transformed into
$p_i(s_i,x_i) \land 
p_j(s_j,x_j)
\land x_i \eq x_j$.

\section{$E$-Generalization}
\LABEL{$E$-generalization}

We treat the problem of $E$-generalization of ground terms
by standard algorithms on regular tree grammars.
Here, we also give a rational reconstruction of the
original approach from \cite{Heinz.1995}, who provided monolithic
specially tailored algorithms for $E$-anti-unification.
We confine ourselves to $E$-generalization of two terms.
All methods work similarly for the simultaneous $E$-generalization of
$n$ terms.

\subsection{The Core Method}
\LABEL{The core method}

\DEFINITION{$E$-Generalization}{ \LABEL{eau}
For an equational theory $E$,
a term $t$ is called an $E$-generalization, or $E$-anti-unifier,
of terms $t_1$ and $t_2$
if there exist substitutions $\sigma_1$ and $\sigma_2$
such that $t \sigma_1 \eq t_1$
and $t \sigma_2 \eq t_2$.
In Fig.~\ref{$E$-generalization of $0$ and $\suc^4(0)$} (right),
we had $t_1 = 0$, $t_2 = \suc^4(0)$,
$t = x*x$,
$\sigma_1= \subst{x \mapsto 0}$,
and $\sigma_2 = \subst{x \mapsto \suc^2(0)}$.

As in unification, a most special
$E$-generalization of arbitrary terms does not normally exist.
A set $G \subseteq \T{}{}{\V}$
is called a set of $E$-generalizations
of $t_1$ and $t_2$ if each member is an $E$-generalization of $t_1$
and $t_2$.
Such a $G$ is called complete if,
for each $E$-generalization $t$ of $t_1$ and $t_2$,
$G$ contains an instance of $t$.
\qed
}

As a first step towards computing $E$-generalization sets, let us
weaken Def.~\ref{eau} by \emph{fixing} the substitutions $\sigma_1$ and
$\sigma_2$.
We will see below, in Sect.~\ref{Learning predicate definitions}
and~\ref{Applications}, that the weakened definition has important
applications in its own right.

\DEFINITION{Constrained $E$-Generalization}{\LABEL{eau v}
Given two terms $t_1, t_2$,
a variable set $V$,
two substitutions $\sigma_1, \sigma_2$
with $\dom \sigma_1 = \dom \sigma_2 = V$
and an equational theory $E$, define the set of $E$-generalizations of
$t_1$ and $t_2$ wrt.\ $\sigma_1$ and $\sigma_2$
as
$\set{t \in \T{}{}{V} \mid
t \sigma_1 \eq t_1 \land t \sigma_2 \eq t_2}$.
This set equals
$\eqc{t_1}{E}{\sigma_1} \cap \eqc{t_2}{E}{\sigma_2}$.
\qed
}

\begin{figure}
\begin{center}
\begin{picture}(10.8,3.0)
\put(1.250,1.500){\makebox(0.000,0.000){$E$}}
	\put(1.500,1.500){\vector(2,1){0.700}}
	\put(1.500,1.500){\vector(2,-1){0.700}}
\put(0.300,2.000){\makebox(0.000,0.000){$t_1$}}
\put(0.300,1.000){\makebox(0.000,0.000){$t_2$}}
	\put(0.500,2.000){\vector(1,0){1.700}}
	\put(0.500,1.000){\vector(1,0){1.700}}
\put(2.700,2.000){\makebox(0.000,0.000){$\eqc{t_1}{E}{}$}}
\put(2.700,1.000){\makebox(0.000,0.000){$\eqc{t_2}{E}{}$}}
	\put(3.200,2.000){\vector(1,0){2.000}}
	\put(3.200,1.000){\vector(1,0){2.000}}
	\multiput(3.200,2.350)(0.100,0.050){7}{%
				\makebox(0,0){$\scriptstyle\cdot$}}
	\multiput(3.200,0.650)(0.100,-0.050){7}{%
				\makebox(0,0){$\scriptstyle\cdot$}}
	\put(3.850,2.680){\vector(2,1){0.100}}
	\put(3.850,0.320){\vector(2,-1){0.100}}
\put(4.200,2.800){\makebox(0.000,0.000){$\sigma_1$}}
\put(4.200,0.200){\makebox(0.000,0.000){$\sigma_2$}}
	\put(4.500,2.700){\vector(2,-1){0.700}}
	\put(4.500,0.300){\vector(2,1){0.700}}
\put(5.700,2.000){\makebox(0.000,0.000){$\eqc{t_1}{E}{\sigma_1}$}}
\put(5.700,1.000){\makebox(0.000,0.000){$\eqc{t_2}{E}{\sigma_2}$}}
	\put(6.200,1.900){\vector(2,-1){0.700}}
	\put(6.200,1.100){\vector(2,1){0.700}}
\put(8.100,1.500){\makebox(0.000,0.000){%
	$\eqc{t_1}{E}{\sigma_1} \cap \eqc{t_2}{E}{\sigma_2}$}}
	\put(9.400,1.500){\vector(1,0){0.700}}
\put(10.500,1.500){\makebox(0.000,0.000){$t$}}
\end{picture}
\\
\renewcommand{\arraystretch}{1.0}
\begin{tabular}[t]{@{}rl@{}}
Constrained $E$-generalization:
	& $\sigma_1$, $\sigma_2$ externally prescribed	\\
Unconstrained $E$-generalization:
	& $\sigma_1$, $\sigma_2$ computed from
	$\eqc{t_1}{E}{}$, $\eqc{t_2}{E}{}$	\\
\end{tabular}
\end{center}
\caption{$E$-Generalization Using Tree Grammars}
\LABEL{Constrained $E$-generalization using tree grammars}
\end{figure}

If we can represent the congruence class $\eqc{t_1}{E}{}$ and
$\eqc{t_2}{E}{}$ as some regular tree language $\L_{\G_1}(N_1)$ and
$\L_{\G_2}(N_2)$, respectively,
we can immediately compute
the set of constrained $E$-generalizations
$\eqc{t_1}{E}{\sigma_1} \cap \eqc{t_2}{E}{\sigma_2}$:
The set of regular tree languages is closed wrt.\ union, intersection
and complement,
as well as under inverse tree homomorphisms,
which cover substitution application as a special case.
Figure~\ref{Constrained $E$-generalization using tree grammars}
gives an overview of our method for computing the \emph{constrained}
set of $E$-generalizations of $t_1$ and $t_2$ wrt.\ $\sigma_1$ and
$\sigma_2$ according to Def.~\ref{eau v}:
\begin{itemize}
\item From $t_i$ and $E$, obtain a grammar for the congruence class
	$\eqc{t_i}{E}{}$, if one exists; the discussion of
	this issue is postponed to
	Sect.~\ref{Setting up grammars for congruence classes}
	below.
\item Apply the inverse substitution $\sigma_i$ to the grammar for
	$\eqc{t_i}{E}{}$ to get a grammar for $\eqc{t_i}{E}{\sigma_i}$,
	using some standard algorithm, e.g.\ that from
	\cite[Thm.7 in Sect.1.4]{Comon.Dauchet.Gilleron.1999}.
	This algorithm takes time
	$\O(\abs{\NT} \cdot {\it size}(\sigma_i))$ for inverse
	substitution application,
	where 
	${\it size}(\sigma_i) 
	= \sum_{x \in \dom \sigma_i} {\it size}(x \sigma_i)$ 
	is the total number of function symbols occurring in $\sigma_i$.
	See Thm.~\ref{include vars} below.
\item Compute a grammar for the intersection
	$\eqc{t_1}{E}{\sigma_1} \cap \eqc{t_2}{E}{\sigma_2}$,
	using the product-automaton construction, e.g., from
	\cite[Sect.~1.3]{Comon.Dauchet.Gilleron.1999},
	which takes time $\O(\abs{\NT_1} \cdot \abs{\NT_2})$.
\item Each member $t$ of the resulting tree language is an actual
	$E$-generalization of $t_1$ and $t_2$.
	The question of enumerating that language is discussed
	later on.
\end{itemize}

\THEOREM{Lifting}{ \LABEL{include vars}
Let a regular tree grammar $\G = \tpl{\Sigma,\NT,\R}$
and a ground substitution $\sigma$ be given.
Define
$\G^\sigma :=
\tpl{\Sigma \cup \dom \sigma,\set{N^\sigma \mid N \in \NT},\R^\sigma}$,
where each $N^\sigma$ is a distinct new nonterminal, and
the rules of $\R^\sigma$ are built as follows:
\\
\begin{tabular}[t]
                {@{}l@{\hspace*{0.5cm}}l@{$\;$}l@{\hspace*{0.5cm}}l@{}}
For each rule
& $N$ & $\sortdef \bigmid_{i=1}^m f_i(\,{1}{n_i}{N_{i\i}})$
& in $\R$       \\      
include the rule
& $N^\sigma$ & $\sortdef \bigmid_{i=1}^m f_i(\,{1}{n_i}{N^\sigma_{i\i}})
        \mid \bigmid_{x \in \dom \sigma, x \sigma \in \L_\G(N)} \; x$
& into $\R'$.   \\
\end{tabular}
\\
Then for all $N \in \NT$ and all $t \in \T{}{}{\V}$,
we have
$t \in \L_{\G^\sigma}(N^\sigma)$
iff $\var(t) \subseteq \dom \sigma$
and $t \sigma \in \L_\G(N)$.
%
\\
The condition $x \sigma \in \L_\G(N)$ in the defining rule of
$N^\sigma$ is decidable.
$\G$ and $\G^\sigma$ have the same number of nonterminals, and of rules.
Each rule of $\G^\sigma$ may have at most
$\abs{\dom \sigma}$ more alternatives.
Note that variables from $\dom \sigma$ occur in the grammar
$\G^\sigma$ like constants.
\qed
}

Based on the above result about constrained $E$-generalization,
we show how to compute the
set of \emph{unconstrained}
$E$-generalizations of $t_1$ and $t_2$ according to Def.~\ref{eau},
where no $\sigma_i$ is given.
It is sufficient to compute two fixed \emph{universal}
substitutions $\tau_1$ and $\tau_2$ from
the grammars for $\eqc{t_1}{E}{}$ and $\eqc{t_2}{E}{}$
and to let them play the role of $\sigma_1$ and $\sigma_2$ in the above
method (cf.\ the dotted vectors in 
Fig.~\ref{Constrained $E$-generalization using tree grammars}).
Intuitively, we introduce one variable for each pair of congruence
classes and map them to a kind of normalform
member of the first and second class by
$\tau_1$ and $\tau_2$, respectively.

We give below a general construction that also accounts for
\emph{auxiliary} nonterminals not representing a congruence class,
and state the universality of $\tau_1,\tau_2$ in a formal way.
For the sake of
technical simplicity, we assume that $\eqc{t_1}{E}{}$ and
$\eqc{t_2}{E}{}$ share the same grammar $\G$; this can easily be
achieved by using the disjoint union of the grammars for
$\eqc{t_1}{E}{}$ and $\eqc{t_2}{E}{}$.

\DEFINITION{Normal Form}{ \LABEL{norm def}
Let an arbitrary tree grammar $\G = \tpl{\Sigma,\NT,\R}$
be given.
A non-empty set of nonterminals
$\NN \subseteq \NT$ is called \emph{maximal} if
$\bigcap_{N \in \NN} \L(N) \neq \set{}$,
but
$\bigcap_{N \in \NN'} \L(N) = \set{}$
for each $\NN' \supset \NN$.
Define
$\NM =
\set{\NN \subseteq \NT
\mid \NN \neq \set{}, \allowbreak \NN \mbox{ maximal}\;}$.
Choose some arbitrary but fixed
\begin{itemize}
\item maximal
        $\NNt{t} \supseteq \set{N \in \NT \mid t \in \L(N)}$
        for each $t \in \bigcup_{N \in \NT} \L(N)$
\item maximal $\NNt{t}$
        for each
        $t \in \T{}{}{\V} \setminus \bigcup_{N \in \NT} \L(N)$
\item ground term
        $\tNN{\NN} \in \bigcap_{N \in \NN} \L(N)$
        for each $\NN \in \NM$
\end{itemize}
The mappings $\NNt{\cdot}$ and $\tNN{\cdot}$
can be effectively computed from $\G$.
We abbreviate $\tNNt{t} = \tNN{\NNt{t}}$;
this is a kind of normalform of $t$.
Each term not in any $\L(N)$, in particular each nonground term,
is mapped to some arbitrary ground term, the choice of which does not
matter.
For a substitution $\sigma = \subst{\,1n{x_\i \mapsto t_\i}}$,
define
$\tNNs{\sigma} = \subst{\,1n{x_\i \mapsto \tNNt{t_\i}}}$.
We always have $x \tNNs{\sigma} = \tNNt{x \sigma}$.
\qed
}

\LEMMA{Substitution Normalization}{ \LABEL{norm}
For all $N \in \NT$, $t \in \T{}{}{\V}$,
and $\sigma$,
\begin{enumerate}
\item $t \in \L(N) \Ra \tNNt{t} \in \L(N)$, ~ and
\item $t \sigma \in \L(N) \Ra t \tNNs{\sigma} \in \L(N)$.
\end{enumerate}
}
\PROOF{
From the definition of $\NNt{\cdot}$ and $\tNN{\cdot}$,
we get $t \in \L(N) \Ra N \in \NNt{t}$
and $N \in \NN \Ra \tNN{\NN} \in \L(N)$,
respectively.
\begin{enumerate}
\item Hence, $t \in \L(N)
	\Ra N \in \NNt{t}
	\Ra \tNNt{t} \in \L(N)$.
\item Induction on the structure of $t$:
	\begin{itemize}
	\item If $t = x \in \V$
		and $x \sigma \in \L(N)$,
		then $x \tNNs{\sigma} = \tNNt{x \sigma} \in \L(N)$ by 1.
	\item Assuming
		$N \sortdef \ldots \:{\mid}1m{ f( \,1n{N_{\i\j}} ) }
		\ldots$,
		we have
		\\
		{
		\renewcommand{\arraystretch}{1.0}
		\begin{tabular}[b]{@{}ll@{\hspace*{1cm}}l@{}}
		& $f(\,1n{t_\i}) \; \sigma \in \L(N)$	\\
		$\Ra$ & $\exists i \leq m \; \forall j \leq n: \;\;
			t_j \sigma \in \L(N_{ij})$
			& by Def.\ $\L(\cdot)$	\\
		$\Ra$ & $\exists i \leq m \; \forall j \leq n: \;\;
			t_j \tNNs{\sigma} \in \L(N_{ij})$
			& by I.H.	\\
		$\Ra$ & $f(\,1n{t_\i}) \; \tNNs{\sigma} \in \L(N)$
			& by Def.\ $\L(\cdot)$	\\
		\end{tabular}
		}
\qed
	\end{itemize}
\end{enumerate}
}

\LEMMA{Universal Substitutions}{ \LABEL{hsg lemma nondet}
For each grammar $\G$,
we can effectively compute two substitutions $\tau_1,\tau_2$ that
are \emph{universal} for $\G$ in the following sense.
For any two
substitutions $\sigma_1,\sigma_2$, a
substitution $\sigma$ exists such that
for $i=1,2$, we have
$\forall t \.\in \T{}{}{\dom \sigma_1 \cap \dom \sigma_2} \;
\forall N \in \NT: \;\;
t \sigma_i \in \L(N) \Ra t \sigma \tau_i \in \L(N)$.
}
\PROOF{
Let $\vNN{\NN_1,\NN_2}$ be a new distinct variable
for each
$\NN_1,\NN_2 \in \NM$.
Define
	$\tau_i =
	\subst{\vNN{\NN_1,\NN_2} \mapsto \tNN{\NN_i}
	\mid \NN_1, \NN_2 \in \NM}$
	for $i=1,2$.
Given $\sigma_1$ and $\sigma_2$,
let
$\sigma
= \subst{x \mapsto \vNN{\NNt{x \sigma_1},\NNt{x \sigma_2}}
\mid x \in \dom \sigma_1 \cap \dom \sigma_2}$.
%
%
Then $\sigma \tau_i$ and $\tNNs{\sigma_i}$ coincide on $\var(t)$,
and hence $t \sigma_i \in \L(N) \Ra t \sigma \tau_i \in \L(N)$
by Lem.~\ref{norm}.2.
\qed
}

\EXAMPLE{}{ \LABEL{exm 0}
We apply Lem.~\ref{hsg lemma nondet}
to the grammar $\G$ consisting of the topmost three rules in
Fig.~\ref{intersection grammar} below.
The result will be used in Ex.~\ref{exm 1} to compute some set of
$E$-generalizations.
We have
$\NM = \set{ \set{N_0,N_t}, \set{N_1,N_t} }$,
since, e.g., $0 \in \L(N_0) \cap \L(N_t)$
and $s(0) \in \L(N_1) \cap \L(N_t)$,
while $\L(N_0) \cap \L(N_1) = \set{}$.
We choose
$$\renewcommand{\arraystretch}{1.0}
\NNt{t} = \left\{	
\begin{array}{ll}
\set{N_0,N_t} & \mbox{ if } t \in \L(N_0)	\\
\set{N_1,N_t} & \mbox{ else }	\\
\end{array}
\right. \mbox{ \hspace*{0.5cm} and \hspace*{0.5cm} }
\begin{array}{lll}
\tNN{\set{N_0,N_t}} & = & 0	\\
\tNN{\set{N_1,N_t}} & = & \suc(0)	\\
\end{array}
\mbox{ \hspace*{0.5cm} . }$$
We abbreviate, e.g., $\vNN{\set{N_0,N_t}}{\set{N_1,N_t}}$ to
$v_{01}$.
This way, we obtain
$$\renewcommand{\arraystretch}{1.0}
\begin{array}{lll *{4}{ccc} ll}
\tau_1 & = & \subst{
	& v_{00} & \mapsto & 0,
	& v_{01} & \mapsto & 0,
	& v_{10} & \mapsto & \suc(0),
	& v_{11} & \mapsto & \suc(0)
	& }
	& \mbox{ and}	\\
\tau_2 & = & \subst{
	& v_{00} & \mapsto & 0,
	& v_{01} & \mapsto & \suc(0),
	& v_{10} & \mapsto & 0,
	& v_{11} & \mapsto & \suc(0)
	& }
	& \mbox{ .}	\\
\end{array}$$
Given $t = x \.* y$,
$\sigma_1 = \subst{x \mapsto 0\.+0,\; y \mapsto 0}$
and $\sigma_2 = \subst{x \mapsto s(0),\; y \mapsto s(0)\.*s(0)}$
for example,
we obtain a proper instance $v_{01} * v_{01}$ of $t$
using $\tau_1$ and $\tau_2$:
$$\renewcommand{\arraystretch}{1.0}
\begin{array}{@{}c@{\;}c@{\;}
	r@{}c@{}l@{} @{\;}c@{\;}
	r@{}c@{}l@{} @{\;}c@{\;}
	r@{}c@{}l@{} @{\;}c@{\;}l@{}}
\L(N_0) & \ni & (0\.+0)&*&0 & \stackrel{\sigma_1}{\laa}
	& x&*&y & \stackrel{\sigma_2}{\raa}
	& s(0)&*&(s(0)\.*s(0)) & \in & \L(N_1)	\\
\L(N_0) & \ni & 0&*&0 & \stackrel{\tau_1}{\laa}
	& v_{01}&*&v_{01} & \stackrel{\tau_2}{\raa}
	& s(0)&*&s(0) & \in & \L(N_1) \mbox{ .\qed}	\\
\end{array}$$
}

The computation of universal substitutions
is very expensive because it involves computing many 
tree-language intersections to determine the mappings $\NNt{\cdot}$ and
$\tNN{\cdot}$.
Assume $\NT = \NT_c \cup \NT_o$,
where $\NT_c$
comprises $n_c$ nonterminals representing congruence classes
and $\NT_o$ comprises $n_o$ other ones.
A maximal set $\NN$ may contain
at most one nonterminal from $\NT_c$ and an arbitrary subset of $\NT_o$;
however, no maximal $\NN$ may be a proper subset of another one.
By some combinatorics, we get
$(n_c+1) \cdot {n_o \choose {n_o/2}}$
as an upper bound on $\abs{\NM}$.
Hence, the cardinality of $\dom \tau_i$
is bounded by the square of that number.
In our experience, $n_o$ is usually small.
In most applications, it does not exceed $1$,
resulting in $\abs{\dom \tau_i} \leq (n_c+1)^2$.
Computing the $\tau_i$ requires
$n_o+1$ grammar intersections in the worst
case, viz.\ when $\NT_o \cup \set{N_c}$ for some $N_c \in \NT_c$
is maximal.
In this case, $\dom \tau_i$ is rather small.
Since the time for testing emptiness is dominated by the intersection
computation time, and
$(n_c+1) \cdot {n_o \choose {n_o/2}} 
\leq (n_c+1) \cdot n_o^{n_o/2} 
\leq \abs{\G}^{n_o+1}$,
we get a time upper bound of $\O(\abs{\G}^{n_o+1})$
for computing the $\tau_i$.

If the grammar is deterministic,
then each nonterminal produces a
distinct congruence class \cite[Sect.2]{McAllester.1992},
and we need compute no intersection at all to
obtain $\tau_1$ and $\tau_2$.
We get $\abs{\dom \tau_i} = \abs{\NT}^2$.
In this case,
$\NNt{\cdot}$, $\tNN{\cdot}$, and $\vN{\cdot}{\cdot}$ can be
computed in linear time from $\G$.
However,
in general a nondeterminstic grammar is
smaller in size than its deterministic counterpart.

\THEOREM{Unconstrained $E$-Generalization}{ \LABEL{hsg nondet}
Let an equational theory $E$ and two ground terms $t_1,t_2$ be given.
Let $\G = \tpl{\Sigma,\NT,\R}$ be a tree grammar
and $N_1,N_2 \in \NT$
such that $\L_\G(N_i) = \eqc{t_i}{E}{}$ for $i = 1,2$.
Let $\tau_1,\tau_2$ be as in Lemma~\ref{hsg lemma nondet}.
Then,
$\eqc{t_1}{E}{\tau_1} \cap \eqc{t_2}{E}{\tau_2}$
is a complete set of $E$-generalizations of $t_1$ and $t_2$.
A regular tree grammar for it can be computed from $\G$
in time $\O(\abs{\G}^2+\abs{\G}^{n_o+1})$.
}
\PROOF{
If $t \in \eqc{t_1}{E}{\tau_1} \cap \eqc{t_2}{E}{\tau_2}$,
then $t \tau_1 \eq t_1$ and $t \tau_2 \eq t_2$,
i.e.\
$t$ is an $E$-generalization of $t_1$ and $t_2$.
~
To show the completeness,
let $t$ be an arbitrary $E$-generalization of $t_1$ and $t_2$,
i.e.\
$t \sigma_i \eq t_i$ for some $\sigma_i$.
Obtain $\sigma$ from Lemma~\ref{hsg lemma nondet} such that
$t \sigma \tau_i \in \eqc{t_i}{E}{}$.
Then, by definition,
$\eqc{t_1}{E}{\tau_1} \cap \eqc{t_2}{E}{\tau_2}$
contains the instance $t \sigma$ of $t$.
\qed
}

Since the set of $E$-generalizations resulting from our method
is given by a regular tree grammar,
it is necessary to enumerate some
terms of the corresponding tree language
in order to actually obtain some results.
Usually, there is a notion of \emph{simplicity} (or \emph{weight}),
depending on the application $E$-generalization is used in,
and it is desirable to enumerate the simplest terms (with least weight)
first.
The minimal weight of a term in the language of each nonterminal can be
computed in time $\O(\abs{\G} \cdot \log \abs{\G})$
by \cite{Burghardt.2002c}.
After that, it is easy to enumerate
a nonterminal's language in order of increasing weight
in time linear to the output size
using a simple {\sc Prolog} program.

\begin{figure}
\begin{center}
\begin{tabular}[t]{@{}c*{23}{@{\hspace*{0.02em}}c}@{}}
$N_0$ & $\sortdef$
	&&&&&&&&& $0$
	&&& $\mid$ & $N_0\.+N_0$
	&&&&& $\mid$ & $N_0\.*N_t$
	& $\mid$ & $N_t\.*N_0$	\\
$N_1$ & $\sortdef$
	&&&&&&&&&&& $\suc(N_0)$
	& $\mid$ & $N_0\.+N_1$
	& $\mid$ & $N_1\.+N_0$
	&&&&& $\mid$ & $N_1\.*N_1$	\\
$N_t$ & $\sortdef$
	&&&&&&&&& $0$
	& $\mid$ & $\suc(N_t)$
	& $\mid$ & $N_t\.+N_t$
	&&&&&&& $\mid$ & $N_t\.*N_t$	\\
\hline
$N_{0*}$ & $\sortdef$
	& $v_{00}$
	& $\mid$ & $v_{01}$
	&&&&& $\mid$ & $0$
	&&& $\mid$ & $N_{0*}\.+N_{0*}$
	&&&&& $\mid$ & $N_{0*}\.*N_{t*}$
	& $\mid$ & $N_{t*}\.*N_{0*}$	\\
$N_{1*}$ & $\sortdef$
	&&&&& $v_{10}$
	& $\mid$ & $v_{11}$
	&&& $\mid$ & $\suc(N_{0*})$
	& $\mid$ & $N_{0*}\.+N_{1*}$
	& $\mid$ & $N_{1*}\.+N_{0*}$
	&&&&& $\mid$ & $N_{1*}\.*N_{1*}$	\\
$N_{t*}$ & $\sortdef$
	& $v_{00}$
	& $\mid$ & $v_{01}$
	& $\mid$ & $v_{10}$
	& $\mid$ & $v_{11}$
	& $\mid$ & $0$
	& $\mid$ & $\suc(N_{t*})$
	& $\mid$ & $N_{t*}\.+N_{t*}$
	&&&&&&& $\mid$ & $N_{t*}\.*N_{t*}$	\\
\hline
$N_{*0}$ & $\sortdef$
	& $v_{00}$
	&&& $\mid$ & $v_{10}$
	&&& $\mid$ & $0$
	&&& $\mid$ & $N_{*0}\.+N_{*0}$
	&&&&& $\mid$ & $N_{*0}\.*N_{*t}$
	& $\mid$ & $N_{*t}\.*N_{*0}$	\\
$N_{*1}$ & $\sortdef$
	&&& $v_{01}$
	&&& $\mid$ & $v_{11}$
	&&& $\mid$ & $\suc(N_{*0})$
	& $\mid$ & $N_{*0}\.+N_{*1}$
	& $\mid$ & $N_{*1}\.+N_{*0}$
	&&&&& $\mid$ & $N_{*1}\.*N_{*1}$	\\
$N_{*t}$ & $\sortdef$
	& $v_{00}$
	& $\mid$ & $v_{01}$
	& $\mid$ & $v_{10}$
	& $\mid$ & $v_{11}$
	& $\mid$ & $0$
	& $\mid$ & $\suc(N_{*t})$
	& $\mid$ & $N_{*t}\.+N_{*t}$
	&&&&&&& $\mid$ & $N_{*t}\.*N_{*t}$	\\
\hline
$N_{00}$ & $\sortdef$
	& $v_{00}$
	&&&&&&& $\mid$ & $0$
	&&& $\mid$ & $N_{00}\.+N_{00}$
	& $\mid$ & $N_{00}\.*N_{tt}$
	& $\mid$ & $N_{0t}\.*N_{t0}$
	& $\mid$ & $N_{t0}\.*N_{0t}$
	& $\mid$ & $N_{tt}\.*N_{00}$	\\
$N_{01}$ & $\sortdef$
	&&& $v_{01}$
	&&&&&&&&
	& $\mid$ & $N_{00}\.+N_{01}$
	& $\mid$ & $N_{01}\.+N_{00}$
	&&& $\mid$ & $N_{01}\.*N_{t1}$
	& $\mid$ & $N_{t1}\.*N_{01}$ \\
$N_{0t}$ & $\sortdef$
	& $v_{00}$
	& $\mid$ & $v_{01}$
	&&&&& $\mid$ & $0$
	&&& $\mid$ & $N_{0t}\.+N_{0t}$
	&&&&& $\mid$ & $N_{0t}\.*N_{tt}$
	& $\mid$ & $N_{tt}\.*N_{0t}$	\\
$N_{t0}$ & $\sortdef$
	& $v_{00}$
	&&& $\mid$ & $v_{10}$
	&&& $\mid$ & $0$
	&&& $\mid$ & $N_{t0}\.+N_{t0}$
	&&&&& $\mid$ & $N_{t0}\.*N_{tt}$
	& $\mid$ & $N_{tt}\.*N_{t0}$	\\
$N_{t1}$ & $\sortdef$
	&&& $v_{01}$
	&&& $\mid$ & $v_{11}$
	&&& $\mid$ & $\suc(N_{t0})$
	& $\mid$ & $N_{t0}\.+N_{t1}$
	& $\mid$ & $N_{t1}\.+N_{t0}$
	&&&&& $\mid$ & $N_{t1}\.*N_{t1}$ \\
$N_{tt}$ & $\sortdef$
	& $v_{00}$
	& $\mid$ & $v_{01}$
	& $\mid$ & $v_{10}$
	& $\mid$ & $v_{11}$
	& $\mid$ & $0$
	& $\mid$ & $\suc(N_{tt})$
	& $\mid$ & $N_{tt}\.+N_{tt}$
	&&&&&&& $\mid$ & $N_{tt}\.*N_{tt}$	\\
\end{tabular}
\caption{Grammars $\G$ (Top), $\G^{\tau_1}$, $\G^{\tau_2}$
	and $\G_{12}$ (Bottom) in Exs.~\ref{exm 1} and~\ref{exm 2}}
\LABEL{intersection grammar}
\end{center}
\end{figure}

\EXAMPLE{}{ \LABEL{exm 1}
To give a simple example, we
generalize $0$ and $\suc(0)$ wrt.\ the equational
theory from Fig.~\ref{$E$-generalization of $0$ and $\suc^4(0)$}
(left).
Figure~\ref{intersection grammar} shows all grammars that will
appear during the computation.
For now, assume that the grammar $\G$ defining
the congruence classes $\eqc{0}{E}{}$
and $\eqc{s(0)}{E}{}$ is already given by the topmost three rules in
Fig.~\ref{intersection grammar}.
In Ex.~\ref{exm 2} below, we
discuss in detail how it can be obtained from $E$.
Nevertheless, the rules of $\G$ are intuitively understandable even now;
e.g.\ the rule for $N_0$ in the topmost line reads:
\emph{A term of value $0$ can be built by the constant $0$,
the sum of two terms of value $0$, the product of a term of value $0$
and any other term, or vice versa}.
Similarly, $\L(N_1) = \eqc{s(0)}{E}{}$ and
$\L(N_t) = \T{}{}{\set{}}$.
In Ex.~\ref{exm 0}, we already computed the universal substitutions
$\tau_1$ and $\tau_2$ from $\G$.

Figure~\ref{intersection grammar} shows the grammars $\G^{\tau_1}$
and $\G^{\tau_2}$
resulting from inverse substitution application,
defining the nonterminals $N_{0*},N_{1*},N_{t*}$ and
$N_{*0},N_{*1},N_{*t}$, respectively.
For example,
$\L_{\G^{\tau_1}}(N_{0*}) = \eqc{0}{E}{\tau_1}$,
where the rule for $N_{0*}$ is obtained from that for $N_0$ by
simply including all variables that are mapped to a member of
$\L_\G(N_0)$ by
$\tau_1$.
They appear as new constants, i.e.\
$\Sigma_{\G^{\tau_1}}
= \Sigma_{\G^{\tau_2}}
= \Sigma_{\G} \cup \set{v_{00},\ldots,v_{11}}$.
For each $t \in \L_{\G^{\tau_1}}(N_{0*})$, we have $t \tau_1 \eq 0$.

The bottommost 6 rules in
Fig.~\ref{intersection grammar} show the intersection
grammar $\G_{12}$ obtained from a kind of product-automaton
construction and defining $N_{00},\ldots,N_{tt}$.
We have
$\L_{\G_{12}}(N_{ij})
= \L_{\G^{\tau_1}}(N_i) \cap \L_{\G^{\tau_2}}(N_j)$
for $i,j \in \set{0,1,t}$.
By Thm.~\ref{hsg nondet},
$\L_{\G_{12}}(N_{01})$
is a complete set of $E$-generalizations of $0$ and
$\suc(0)$ wrt.\ $E$.
We have, e.g.,
$N_{01}
\ra N_{01} * N_{t1}
\ra v_{01} * v_{01}$,
showing that $0$ and $\suc(0)$ are both quadratic numbers,
and
$(v_{01} * v_{01}) \tau_1 = 0 * 0 \eq 0$,
$(v_{01} * v_{01}) \tau_2 = \suc(0) * \suc(0) \eq \suc(0)$.
By repeated application of the rules
$N_{01} \sortdef \ldots N_{01}*N_{t1} \ldots$
and $N_{t1} \sortdef \ldots N_{t1}*N_{t1} \ldots$,
we can obtain any generalization of the form
$\:{*}{}{}{v_{01}}$, with arbitrary paranthesation.
By way of a less intuitive example, we have
$v_{01} * \suc(v_{10}+v_{10}) \in \L_{\G_{12}}(N_{01})$.
\qed
}

\subsection{Setting up Grammars for Congruence Classes}
\LABEL{Setting up grammars for congruence classes}

The question of how to obtain a grammar representation of the initial
congruence classes $\eqc{t_1}{E}{}$ and $\eqc{t_2}{E}{}$ was
deferred in Sect.~\ref{The core method}.
It is discussed below.

Procedures that help
to compute a grammar representing the congruence class of a
ground term are given, e.g., in
\cite[Sect.3]{McAllester.1992}.
%
This paper also provides a criterion for an equational
theory inducing regular tree languages as congruence classes:

\THEOREM{Ground Equations}{ \LABEL{eqc 1}
An equational theory induces regular congruence classes iff it is the
deductive closure of finitely many ground equations $E$.
}
\PROOF{
To prove the \emph{if} direction,
start with a grammar consisting of
a rule $N_{f(\,1n{t_\i})} \sortdef f(\,1n{N_{t_\i}})$
for each subterm $f(\,1n{t_\i})$ occurring in $E$.
For each equation $(t_1 \eq t_2) \in E$,
fuse 
the nonterminals $N_{t_1}$ and $N_{t_2}$ everywhere in the
grammar.
Then successively fuse 
every pair of nonterminals whose rules
have the same right-hand sides.
The result is a deterministic grammar $\G$ such that
$t_1 \eq t_2$ iff $t_1,t_2 \in \L_\G(N)$ for some $N$.

In addition to this nonoptimal but intuitive algorithm,
McAllester gives an $\O(n \cdot \log n)$ algorithm based on congruence
closure \cite{Downey.Sethi.Tarjan.1980},
where $n$ is the written length of $E$.
The proof of the \emph{only if} direction need not be sketched here.
\qed
}

In order to compute a complete set of $E$-generalizations of two given
ground terms $t_1$ and $t_2$, we do not need \emph{all} congruence
classes to be regular;
it is sufficient that those of $t_1$ and $t_2$ are.
More precisely, it is sufficient that
$(\eq)$ is a refinement of some congruence relation
$(=_\G)$ with finitely many classes only, such that
$\eqc{t_i}{E}{} = \eqc{t_i}{\G}{}$ for $i=1,2$.

\EXAMPLE{}{ \LABEL{exm 2}
Let $\Sigma = \set{0,s,(+),(*)}$, and let $E$ be the equational theory
from Fig.~\ref{$E$-generalization of $0$ and $\suc^4(0)$} (left).
To illustrate the application of Thm.~\ref{eqc 1},
we show how to obtain
a grammar defining $\eqc{0}{E}{}, \ldots, \eqc{s^n(0)}{E}{}$
for arbitrary $n \in \N$.
Obviously, $(\eq)$ itself has infinitely many congruence classes.
However, in order to consider the terms $0,\ldots,\suc^n(0)$ only,
the relation
$(=_\G)$, defined by the classes
$\eqc{0}{E}{}, \ldots, \eqc{\suc^n(0)}{E}{}$ and
$C = \bigcup_{i > n} \eqc{\suc^i(0)}{E}{}$,
is sufficient.
This relation is, in fact, a congruence wrt.\
$0$, $\suc$, $(+)$, and $(*)$:
	in a term $t \in \bigcup_{i=0}^n \eqc{\suc^i(0)}{E}{}$,
	a member $t_c$ of $C$ can occur only
	in some subterm of the form $0*t_c$ or similar,
	which always equals $0$, regardless of the choice of $t_c$.

For $n=1$, we may choose a representative term $0$, $s(0)$, and $c$
for the classes $\eqc{0}{E}{}$, $\eqc{s(0)}{E}{}$, and $C$,
respectively.
We may instantiate each equation from
Fig.~\ref{$E$-generalization of $0$ and $\suc^4(0)$} (left)
with all possible combinations of representative terms,
resulting in $3+3^2+3+3^2$ ground equations.
After adding the equation $c = s(s(0))$,
we may apply Thm.\ref{eqc 1} to obtain a deterministic grammar
$\G_d$ with $\L_{\G_d}(N_0) = \eqc{0}{E}{}$,
$\L_{\G_d}(N_{s(0)}) = \eqc{s(0)}{E}{}$,
and $\L_{\G_d}(N_c) = C$.
The equations' ground instances, and thus
$\G_d$, can be built automatically.
The grammar $\G_d$ is equivalent to $\G$
from Fig.~\ref{intersection grammar} (top),
which we used in Ex.~\ref{exm 1}.
The latter is nondeterministic
for the sake of brevity.
It describes $\eqc{0}{E}{}$ and $\eqc{\suc(0)}{E}{}$ by
$N_0$ and $N_1$, respectively,
while
$\L(N_t)
= \eqc{0}{E}{} \cup \eqc{\suc(0)}{E}{} \cup C$.
\qed
}

In \cite[Cor.~16]{Burghardt.Heinz.1996},
another sufficient criterion is given.
Intuitively, it allows us to construct a grammar from each equational
theory that describes only
operators \emph{building up} larger values (normal forms)
from smaller ones:

\THEOREM{Constructive Operators}{ \LABEL{eqc 2}
Let $E$ be given by a ground
confluent and Noetherian term-rewriting system.
For each term $t \in \T{}{}{\set{}}$,
let ${\it nf}(t)$ denote its unique normal form;
let ${\it NF}$ be the set of all normal forms.
Let $(\prec)$ be a well-founded partial ordering on ${\it NF}$
with a finite branching degree,
and let $(\preceq)$ be its reflexive closure.
If $t_i \preceq {\it nf}(f( \,1n{t_\i} ))$
for all $f \in \Sigma$, $\,1n{t_\i} \in {\it NF}$
and $i = \,1n\i$,
then for each $t \in \T{}{}{\set{}}$,
the congruence class $\eqc{t}{E}{}$ is a regular tree language.
}
\PROOF{
Define one nonterminal $N_t$ 
for each normal-form term $t \in {\it NF}$.
Whenever $t \eq f(\,1n{t_\i})$ for some $\,1n{t_\i} \in {\it NF}$,
include an alternative $f(\,1n{N_{t_\i}})$ into the right-hand side of
the rule for $N_t$.
Since $t_i \preceq t$ for all $i$,
there are only finitely many such alternatives.
This results in a finite grammar $\G$, and we have
$\L_\G(N_{{\it nf}(t)}) = \eqc{t}{E}{}$ for all terms $t$.
Let $a_i$ denote the number of $i$-ary function symbols in $\Sigma$
and $m$ denote the maximal arity in $\Sigma$.
Then we need $\sum_{i=0}^m a_i \cdot \abs{\it NF}^i$ 
normal-form computations
to build $\G$,
which has a total of these many alternatives and $\abs{\it NF}$ rules.
Its computation takes $\O(\abs{\G})$ time.
\qed
}

By way of an example,
consider the theory $E$ consisting of only equations 1.\ and 2.\ in
Fig.~\ref{$E$-generalization of $0$ and $\suc^4(0)$} (left).
$E$ is known to be ground-confluent and Noetherian
and to lead to ${\it NF} = \set{s^n(0) \mid n \in \N}$.
Defining $s^i(0) \prec s^j(0) \Lra i < j$,
we observe that $s^i(0) \preceq s^{i+j}(0) = {\it nf}(s^i(0)+s^j(0))$;
similarly, $s^j(0) \preceq {\it nf}(s^i(0)+s^j(0))$
and $s^i(0) \preceq {\it nf}(s(s^i(0)))$.
Hence, $E$ leads to regular congruence classes by Thm.~\ref{eqc 2}.
For example, there are just five ways to obtain a term of value $s^3(0)$
from normal-form terms using $s$ and $(+)$.
Accordingly,
$$N_3
\sortdef s(N_2) \mid N_3+N_0 \mid N_2+N_1 \mid N_1+N_2 \mid N_0+N_3$$
defines $\eqc{s^3(0)}{E}{}$.

If their respective preconditions are met,
Thms.~\ref{eqc 1} and~\ref{eqc 2} allow us to automatically
compute a grammar describing the congruence classes
wrt.\ a given theory $E$.
In practice,
with Thm.~\ref{eqc 1} we have the problem that $E$ is rarely 
given
by ground equations only,
while Thm.~\ref{eqc 2} requires properties that are sufficient but not
necessary.
For example, not even for the whole theory
from Fig.~\ref{$E$-generalization of $0$ and $\suc^4(0)$} (left)
is there an ordering such that
$s^i(0) \preceq 0 = {\it nf}(s^i(0)*0)$
and $0 \preceq s^i(0) = {\it nf}(0+s^i(0))$.

So far, it seems best to set up a \emph{grammar scheme}
for a given $E$ manually.
For example, for $E$ from
Fig.~\ref{$E$-generalization of $0$ and $\suc^4(0)$} (left),
it is easy to write a program that reads $n \in \N$ and
computes a grammar describing the congruence classes
$\eqc{0}{E}{}, \ldots, \eqc{s^n(0)}{E}{}$:
The grammar consists of the rules for $N_0$ and $N_t$ from
Fig.~\ref{intersection grammar} (top) and one rule
$$N_i \sortdef s(N_{i-1}) \mid \bigmid_{j=0}^i N_j+N_{i-j}
\mid \bigmid_{j \cdot k = i} N_j*N_k$$
for each $i=\,1n\i$.
Similarly, grammar schemes for the theory of
list operators like ${\it append}$,
${\it reverse}$, etc.\ and other theories
can be implemented.

The lack of a single algorithm that computes a grammar for each
$E$ from a sufficiently large class of equational theories
restricts the applicability of our $E$-generalization
method to problems where $E$ is not changed too often.
Using grammar schemes, this restriction can be relaxed somewhat.
The grammar-scheme approach is also applicable to theories given by
conditional equations or even other formulas, as long as they lead to
regular congruence classes and the schemes are computable.

A further problem is that
not all equational theories lead to regular congruence classes.
Consider, for example, subtraction $(-)$ on natural numbers.
We have $0 \eq \suc^i(0) - \suc^i(0)$
for all $i \in \N$.
Assume that $(\eq)$
is a refinement of some $(=_\G)$ with finitely many classes;
then $\suc^i(0) =_\G \suc^j(0)$ for some $i \neq j$.
If $(=_\G)$ is a congruence relation,
$0 =_G \suc^i(0) - \suc^i(0) =_\G \suc^i(0) - \suc^j(0)$,
although $0 \mathrel{\not=_E} \suc^i(0) - \suc^j(0)$.
Hence, $\eqc{0}{E}{} = \eqc{0}{\G}{}$ is impossible.
Thus, if an operator like
$(-)$ is defined in $E$, we cannot $E$-generalize using our
method.

\newcommand{\hg}[1]{{\it hg}(#1)}

However,
we can still compute an \emph{approximation} of the
$E$-generalization set in such cases
by artificially cutting off grammar rules
after a maximum number of alternatives.
This will result in a set that still contains only correct
$E$-generalizations but that is usually incomplete.
For example, starting from the grammar rules
$$\renewcommand{\arraystretch}{1}
\begin{array}{ccccccccccc}
N_0 & \sortdef & 0 &&
	& \mid & N_0 - N_0
	& \mid & N_1 - N_1
	& \mid & N_2 - N_2	\\
N_1 & \sortdef &&& \suc(N_0)
	& \mid & N_1 - N_0
	& \mid & N_2 - N_1 \\
N_2 & \sortdef &&& \suc(N_1)
	& \mid & N_2 - N_0 & ,	\\
\end{array}$$
we obtain only $E$-generalization terms $t$ whose evaluation
never exceeds the value $2$ on any subterm's instance.
Depending on the choice of the \emph{cut-off point},
the resulting $E$-generalization set may suffice for a given
application.

\subsection{More Expressive Grammar Formalisms}
\LABEL{More expressive grammar formalisms}

To overcome the limited expressiveness of pure regular tree grammars,
we tried to extend the results of Sect.~\ref{The core method}
to automata with equality tests.


For automata with equality tests between siblings
\cite{Bogaert.Tison.1992,Comon.Dauchet.Gilleron.1999}
or, equivalently, shallow systems of sort
constraints \cite{Uribe.1992}, we have the problem that this language
class is not closed under inverse substitution application
\cite[Sect.~5.2]{Bogaert.Tison.1992}.
For example, consider the grammar
$$\renewcommand{\arraystretch}{1}
\begin{array}{l@{\hspace{2cm}}l}
	\begin{array}[t]{l*{6}{@{\;}l}}
	N_0 & \sortdef & 0 &&& \mid & N_t-N_t [x_1=x_2]	\\
	N_t & \sortdef & 0 & \mid & \suc(N_t) & \mid & N_t-N_t	\\
	\end{array}
&
	\begin{array}[t]{l*{4}{@{\;}l}}
	N_{xy} & \sortdef & \multicolumn{3}{@{}l}{N_x - N_y}	\\
	N_x & \sortdef & x & \mid & \suc(N_x)	\\
	N_y & \sortdef & y & \mid & \suc(N_y)	\\
	\end{array}
\\
\end{array}$$
in which $N_0$ describes the
congruence class $\eqc{0}{E}{}$ wrt.\ the defining equations
of $(-)$, where the constraint $[x_1=x_2]$ requires the left and right
operand of $(-)$ to be equal.
For 
$\sigma = \subst{x \mapsto 0, y \mapsto \suc(0)}$,
the language
$\eqc{0}{E}{\sigma}$
is not recognizable because otherwise 
$\eqc{0}{E}{\sigma} \cap \L(N_{xy})
= \set{\suc^{i+1}(x) - \suc^i(y) \mid i \in \N}$
would have to be recognizable, too.
However, the latter set cannot be represented by a grammar with equality
tests between siblings.
For reduction automata
\cite{Caron.Coquide.Dauchet.1995,%
Comon.Dauchet.Gilleron.1999}
and generalized reduction automata \cite{Caron.Comon.Coquide.1994},
it seems to be still unknown whether they are closed under inverse
substitution application.


The approach using universal
substitutions $\tau_1,\tau_2$
strongly
depends on the fact that
$\bigwedge_{j=1}^n t_j \in \L(N_{ij})
\Ra f(\,1n{t_\i}) \in \L(N) ,$
which is needed in the proof of Lem.~\ref{norm}.2.
In other words, the rule
$N \sortdef \ldots f(\,1n{N_{i\i}}) $ 
is \emph{not} allowed to have additional constraints.
The straightforward approach to incorporate equality constraints
into the proof of Lem.~\ref{hsg lemma nondet}
is to show additionally that
$\forall t',t'' \in \T{}{}{\V}: \allowbreak
t' \sigma_i = t'' \sigma_i \Ra t' \sigma \tau_i = t'' \sigma \tau_i$
for $i=1,2$.
Note that syntactic equality is required by the form of constraints.
However, this leads to a contradiction:

\LEMMA{}{
If $\Sigma$ contains at least one function symbol of
arity $\geq 1$ and one constant,
there cannot exist $\tau_1$ and $\tau_2$ such that
$$\forall \sigma_1,\sigma_2 \;
\exists \sigma \;
\forall t',t'' \in \T{}{}{\V} \;
\forall i = 1,2: \;\;
t' \sigma_i = t'' \sigma_i \Ra t' \sigma \tau_i = t'' \sigma \tau_i .$$
}
\PROOF{
If such $\tau_i$ existed,
we could construct a \emph{contradictory} sequence
$\tpl{\sigma_1^{(j)},\sigma_2^{(j)}}_{j \in \N}$ as follows.
Let $\sigma_1^{(j)} = \subst{x \mapsto f^j(c)}$ and
$\sigma_2^{(j)} = \subst{x \mapsto c}$,
where $f(\ldots), c \in \Sigma$;
for the sake of simplicity, we assume that $f$ is
only unary.
Since
$x \sigma_1^{(j)} = f^j(c) \sigma_1^{(j)}$,
we have
$x \sigma^{(j)} \tau_1 = f^j(c) \sigma^{(j)} \tau_1 = f^j(c)$.
Similarly,
$x \sigma_2^{(j)} = c \sigma_1^{(j)}$
implies
$x \sigma^{(j)} \tau_2 = c$.
%
Since $x \sigma^{(j)}$
is mapped by $\tau_1$ and $\tau_2$ to terms starting with
different symbols, it must be a variable,
%
and $\tau_1$ must include the mapping
$x \sigma^{(j)} \mapsto f^j(c)$ for each $j \in \N$.
This is impossible because $\dom \tau_1$ can only be finite.
\qed
}

For these reasons, our approach remains limited to
ordinary regular tree grammars, i.e.\ 
those without any equality constraints.
The following lemma shows that we cannot find a more sophisticated
grammar formalism that can handle \emph{all} equational theories,
anyway.

\LEMMA{General Uncomputability of $E$-Generalization}{ \LABEL{Post}
There are equational theories $E$ such that it is undecidable
whether the set of constrained $E$-generalizations for certain
$t_1,t_2,\sigma_1,\sigma_2$ is empty.
Such a set cannot be represented by any grammar formalism that
is closed wrt.\ language intersection and allows
testing for emptiness.
}
\PROOF{
We encode a version of Post's \emph{correspondence problem}
into an $E$-generalization problem for a certain $E$.
Let a set $\set{ \,1n{\tpl{a_\i,b_\i}} }$ of pairs of nonempty
strings over a finite alphabet $A$ be given.
It is known to be undecidable whether there exists a sequence
$1,\,2m{i_\i}$ such that $m \geq 1$ and
$a_1 \cdot \:{\cdot}2m{a_{i_\i}} = b_1 \cdot \:{\cdot}2m{b_{i_\i}}$
\cite[Sect.~2.7]{Schoning.1997}, where ``$\cdot$'' denotes string
concatenation.
Let
$\Sigma
= A
\cup \set{ \,1n\i }
\cup \set{(\cdot), f, a, b}$,
and let $E$ consist of the following equations:
$$\renewcommand{\arraystretch}{1}
\begin{array}{rcll}
(x \cdot y) \cdot z & = & x \cdot (y \cdot z)	\\
f(a,x \cdot i,y \cdot a_i) & = & f(a,x,y) & \mbox{ for } i=\,1n\i \\
f(b,x \cdot i,y \cdot b_i) & = & f(b,x,y) & \mbox{ for } i=\,1n\i , \\
\end{array}$$
where $x$, $y$, and $z$ are variables.
Then, the congruence class
$\eqc{f(a,1,a_1)}{E}{}$ and
$\eqc{f(b,1,b_1)}{E}{}$
equals the set of all admitted Post sequences of $a_i$ and $b_i$,
respectively.
Let $\sigma_1 = \subst{x \mapsto a}$ and
$\sigma_2 = \subst{x \mapsto b}$,
then
the set of constrained $E$-generalizations
$\eqc{f(a,1,a_1)}{E}{\sigma_1} \cap \eqc{f(b,1,b_1)}{E}{\sigma_2}$
is nonempty iff the given correspondence problem has a solution.
\qed
}

\section{Learning Predicate Definitions}
\LABEL{Learning predicate definitions}

In this section, we relate $E$-generalization to Inductive Logic
Programming (ILP),
which seems to be the closest approach to machine learning.
We argue in favor of
an \emph{outsourcing} of equational tasks
from Horn program induction algorithms, similar to what
has long been common practice in the area of deduction.
From a theoretical point of view,
a general-purpose theorem-proving algorithm is
complete wrt.\ equational formulas, too, if all necessary instances of
the \emph{congruence axioms} $s \eq t \Ra f(s) \eq f(t)$
and $s \eq t \land p(s) \Ra p(t)$ are supplied.
However, in practice it proved to be much more efficient to handle
the equality predicate $(\eq)$ separately
using specially tailored methods, like
$E$-unification for fixed, or paramodulation for varying $E$.

Similarly, we show that integrating $E$-anti-unification into an
ILP algorithm helps to restrict the hypotheses-language bias and thus
the search space.
In particular, learning of determinate clauses can be reduced to
learning of atoms using generalization wrt.\
an $E$ defining a function for each determinate predicate.

\begin{figure}
\begin{center}
\begin{tabular}[t]{@{}lr@{$\;$}c@{$\;$}l@{}}
Necessity: & $B$ & $\not\impl$ & $F^+$	\\
Sufficiency: & $B \land h$ & $\impl$ & $F^+$	\\
Weak Consistency: & $B \land h$ & $\not\impl$ & ${\it false}$	\\
Strong Consistency: & $B \land h \land F^-$ & $\not\impl$
	& ${\it false}$ \\
\end{tabular}
\caption{Requirements for Hypothesis Generation 
	According to \cite{Muggleton.1999}}
\LABEL{Requirements to hypothesis generation}
\end{center}
\end{figure}

We investigate the \emph{learning} of a
definition of a new predicate symbol $p$ in four
different settings.
In all cases, we are given a conjunction $F \Lra F^+ \land F^-$
of positive and negative ground \emph{facts},
and a \emph{background theory} $B$ describing an equational theory $E$.
In this section,
we always assume that $E$ leads to regular congruence classes.
We generate a \emph{hypothesis} $h$ that must \emph{explain} $F$
using $B$.
From Inductive Logic Programming, up to four requirements for such a
hypothesis are known
\cite[Sect.~2.1]{Muggleton.1999}.
They are listed in
Fig.~\ref{Requirements to hypothesis generation}.
More precisely, the \emph{Necessity} requirement does not impose a
restriction on $h$, but forbids any generation of a (positive)
hypothesis, provided
the positive facts are explainable without it.
Muggleton remarks that this requirement can be checked by a conventional
theorem prover before calling hypothesis generation.
The \emph{Weak} and \emph{Strong Consistency} requirements coincide if
there are no negative facts; 
otherwise, the former is entailed by the latter.
In \cite[Sect.~5.2.4]{Dzeroski.1996},
only \emph{Sufficiency} (called \emph{Completeness} there) and
\emph{Strong Consistency} are required.

We show under which circumstances
$E$-generalization can generate a hypothesis
satisfying the \emph{Sufficiency} and both \emph{Consistency}
requirements.
The latter are meaningful in an equational setting only if we require
that $(\eq)$ is \emph{nontrivial},
i.e.\ $\exists x,y: \lnot \; x \eq y$.
Without this formula, ${\it false}$
could not even be derived from an
equational theory, however nonsensical.

Given $E$, we require our logical background theory $B$ to entail
the reflexivity, symmetry and transitivity axiom for $(\eq)$,
a congruence axiom for $(\eq)$  wrt.\ each function 
$f \in \Sigma$ and each predicate $p$ occurring in $B$,
the universal closure of each equation in $E$, and
the nontriviality axiom for $(\eq)$.
As a more stringent alternative
to the Necessity requirement,
we assume
that $B$ is not contradictory and
that the predicate symbol $p$ for which a definition has
to be learned does not occur in $B$, 
except in $p$'s congruence axiom.

\subsection{Atomic Definitions}
\LABEL{Atomic definitions}

To begin with, we investigate the learning of
a definition of a 
unary predicate $p$ by an atom $h \Lra p(t)$.
Let $F^+ \Lra \bigwedge_{i=1}^n p(t_i)$
and $F^- \Lra \bigwedge_{i=n+1}^m \lnot p(t_i)$.
For sets $T^+,T^-$ of ground terms and an arbitrary term $t$,
define
$$\renewcommand{\arraystretch}{1}
\begin{array}{llll}
\hypp{t}{T^+}
	& \Lra 
	& \forall t' \in T^+ \; \exists \hsigma: \; t \hsigma \eq t'
	& \mbox{ and}	\\
\hypn{t}{T^-}
	& \Lra 
	& \forall t' \in T^- \; \forall \hsigma: \; t \hsigma \not\eq t'
	& \mbox{ .}	\\
\end{array}$$
We name the substitutions
$\hsigma$ instead of $\sigma$ in order to identify them as given by
$\hypp{}{}$ and $\hypn{}{}$ in the proofs below.

\LEMMA{Requirements}{ \LABEL{Herbrand}
Let $t_i, t'_i$ be ground terms for $i=\,1m\i$ and
let $t$ be an arbitrary term.
Let $F^+ \Lra \bigwedge_{i=1}^n p(t_i)$
and $F^- \Lra \bigwedge_{i=n+1}^m \lnot p(t_i)$.
Let $T^+ = \set{ \,1n{t_\i} }$
and $T^- = \set{ \,{n+1}m{t_\i} }$.
Then:
\begin{enumerate}
\item	$B \land p(t) \impl \:{\lor}1{n'}{p(t'_\i)}$
	~ iff ~
	$t \sigma \.\eq t'_i$
	for some $i$, $\sigma$.
\item $B \land p(t) \impl F^+$  ~ iff ~
	$\hypp{t}{T^+}$.
\item $B \land p(t) \land F^- \not\impl {\it false}$
	~ iff ~
	$\hypn{t}{T^-}$.
\end{enumerate}
}
\PROOF{
\begin{enumerate}
\item The \emph{if} direction is trivial.
	To prove the \emph{only if} direction, observe that
	$B$ has a Herbrand model containing no instances of $p$.
	If we add the set
	$\set{ p(t'')
	\mid \exists \sigma \mbox{ ground}: t \sigma \eq t''}$
	to that model,
	we get a Herbrand model of $B \land p(t)$.
	In this model, $\:{\lor}1{n'}{p(t'_\i)}$ holds only if
	some $p(t'_i)$ holds, since they are all ground.
	This implies in turn that
	$p(t'_i)$ is among the $p(t'')$, i.e.\
	$t \sigma \eq t'_i$ for some ground substitution $\sigma$.
\item Follows from 1.\ for $n'=1$.
\item Follows from 1.\ for $\lnot F^- \Lra \:{\lor}1{n'}{p(t'_\i)}$,
	paraphrasing Strong Consistency as
	$B \land p(t) \not\impl \lnot F^-$.
\qed
\end{enumerate}
}

The following Lem.~\ref{hyp lr}, and Lem.~\ref{hyp r} below for the
determinate case, are the workhorses of this section.
They show how to apply $E$-generalization to obtain a 
hypotheses term set from given positive and negative example term sets.
In the theorems based on these lemmas,
we only need to enclose the hypotheses
terms as arguments to appropriate predicates.

\LEMMA{Hypotheses}{ \LABEL{hyp lr}
For each finite set $T^+ \cup T^-$ of ground terms,
we can compute a regular set $H = \Hyp{\ref{hyp lr}}{T^+}{T^-}$
such that
for all $t \in \T{}{}{\V}$:
$$\renewcommand{\arraystretch}{1}
\begin{array}{rlll}
t \in H & \Ra
	& \hypp{t}{T^+} \land \hypn{t}{T^-}
	& \mbox{ , and}	\\
\exists \sigma: \; t \sigma \in H
	& \La & \hypp{t}{T^+} \land \hypn{t}{T^-} & \mbox{ .}	\\
\end{array}$$
}
\PROOF{
Let $T^+ = \set{\,1n{t_\i}}$,
let $\G$ be a grammar defining $\,1n{\eqc{t_\i}{E}{}}$.
Obtain the universal substitutions $\,1n{\tau_\i}$ for $\G$ from
Lem.~\ref{hsg lemma nondet}.
All $\tau_i$ have the same domain.
Using the notations from Def.~\ref{norm def}, let
$$S = \set{\sigma
\mid \dom \sigma = \dom \tau_1
\land \forall x \in \dom \sigma \; \exists \NN \in \NM: \;
x \sigma = \tNN{\NN}} .$$
The set $S$ is finite, but large;
it has $\abs{\NM}^{\abs{\NM}^n}$ elements.
Define
$H^+ = \bigcap_{i=1}^n \eqc{t_i}{E}{\tau_i}$
and
$H^-
= \bigcup_{t' \in T^-} \bigcup_{\sigma \.\in S} \eqc{t'}{E}{\sigma}$.
Define $H = H^+ \setminus H^-$; all these sets are regular tree
languages and can be computed
using standard grammar algorithms.
\begin{itemize}
\item For $t \in H$,
	we trivially have $\hypp{t}{T^+}$.
	Assume that $\hypn{t}{T^-}$ does not hold,
	i.e.\ $t \sigma \eq t'$ for some $\sigma$ and $t' \in T^-$.
	Since $\var(t) \subseteq \dom \tau_1$ by construction,
	we may assume w.l.o.g.\ $\dom \sigma = \dom \tau_1$;
	hence $\tNNs{\sigma} \in S$.
	From Lem.~\ref{norm}.2, we get
	$t \sigma \in \eqc{t'}{E}{}
	\Ra t \tNNs{\sigma} \in \eqc{t'}{E}{}
	\Ra t \in \eqc{t'}{E}{\tNNs{\sigma}}$,
	contradicting $t \not\in H^-$.
\item If $\hypp{t}{T^+} \land \hypn{t}{T^-}$,
	then $t \hsigma_i \eq t_i$ for some $\hsigma_i$, ~ $i=\,1n\i$.
	Using Lem.~\ref{hsg lemma nondet},
	we get some $\sigma$ such that
	$t \sigma \in \eqc{t_i}{E}{\tau_i}$,
	hence $t \sigma \in H^+$.
	If we had $t \sigma \in \eqc{t'}{E}{\sigma'}$
	for some $t' \in T^-$ and $\sigma' \in S$,
	then $t \sigma \sigma' \eq t'$,
	contradicting $\hypn{t}{T^-}$.
\qed
\end{itemize}
}

\THEOREM{Atomic Definitions}{ \LABEL{appl lr}
Let $\,1m{t_\i}$ be ground terms.
Let $F^+ \Lra \bigwedge_{i=1}^n p(t_i)$ 
and $F^- \Lra \bigwedge_{i=n+1}^m \lnot p(t_i)$
be given.
We can compute a regular set
$H = \Hyp{\ref{appl lr}}{F^+}{F^-}$ such that
\begin{itemize}
\item each $p(t) \in H$ is a hypothesis
	satisfying the \emph{Sufficiency} and the
	\emph{Strong Consistency} requirement wrt.\ $F^+$, $F^-$; and
\item for each hypothesis satisfying these requirements and having the
	form $p(t)$,
	we have $p(t \sigma) \in H$ for some $\sigma$.
\end{itemize}
}
\PROOF{
Define $T^+ = \set{t_i \mid i=\,1n\i }$
and $T^- = \set{t_i \mid i=\,{n\.+1}m\i }$.
By Lem.~\ref{Herbrand}.2 and~3,
$p(t)$ is a hypothesis satisfying the Sufficiency and the Strong
Consistency requirement iff $\hypp{t}{T^+}$
and $\hypn{t}{T^-}$, respectively.
By Lem.~\ref{hyp lr},
we may thus choose
$H = \set{ p(t) \mid t \in \Hyp{\ref{hyp lr}}{T^+}{T^-} }$,
which is again a regular tree language.
\qed
}

The time requirement of the computation from Thm.~\ref{appl lr}
grows very quickly if negative examples are given.
Even for deterministic grammars,
up to $(m-n) \cdot \abs{\NT}^{\abs{\NT}^n}$ inverse substitution
applications are needed,
each requiring a renamed copy of the original grammar.
If only positive examples are given, the time complexity is
$\O(\abs{\G}^n+\abs{\G}^{n_o+1})$, 
which allows nontrivial practical applications.

By way of an example,
consider again the equational theory $E$ from
Fig.~\ref{$E$-generalization of $0$ and $\suc^4(0)$} (left).
Let $F^+ \Lra 0 \leq 0 \land 0 \leq s(0)$
and $F^- \Lra {\it true}$
be given.
In Ex.~\ref{exm 1}, we already computed a grammar
$\G$ describing
$\eqc{0}{E}{} = \L_\G(N_0)$ and $\eqc{s(0)}{E}{} = \L_\G(N_1)$,
see Fig.~\ref{intersection grammar} (top).
The congruence class of, say, $\tpl{0,0}$ could be defined by the
additional grammar rule $N_{\tpl{0,0}} \sortdef \tpl{N_0,N_0}$.
Instead of that rule, we add $N_{0 \leq	0} \sortdef (N_0 \leq N_0)$
to the grammar,
anticipating that any
$\tpl{t_1,t_2} \in \Hyp{\ref{hyp lr}}{}{}$ will be transformed
to $(t_1 \leq t_2) \in \Hyp{\ref{appl lr}}{}{}$
by Thm.~\ref{appl lr}, anyway.
Similarly, we add the rule $N_{0 \leq 1} \sortdef (N_0 \leq N_1)$.
The universal substitutions we obtain following the construction of
Lem.~\ref{hyp lr} are simply $\tau_1$ and $\tau_2$ from Ex.~\ref{exm 1}.
We do not extend them to also include variables like
$\vNN{\set{N_{0 \leq 0}},\set{N_{0 \leq 1}}} = v_{0 \leq 0, 0 \leq 1}$
in their domain because neither
$v_{0 \leq 0, 0 \leq 1} \in \Hyp{\ref{appl lr}}{}{}$
nor
$v_{\tpl{0,0}, \tpl{0,1}} \in \Hyp{\ref{hyp lr}}{}{}$
would make sense.
From a formal point of view, retaining the $\tau_i$ from
Ex.~\ref{exm 1} restricts
$\Hyp{\ref{appl lr}}{}{}$ and $\Hyp{\ref{hyp lr}}{}{}$
to predicates and terms of the form
$t_1 \leq t_2$ and $\tpl{t_1,t_2}$, respectively.

After lifting the extended $\G$ wrt.\ $\tau_1,\tau_2$,
we obtain the grammar $\G_{12}$ from Fig.~\ref{intersection grammar}
(bottom), extended by some rules like
$N_{0 \leq 0,0 \leq 1} \sortdef (N_{00} \leq N_{01})$.
By Thm.~\ref{appl lr},
each element of
$\Hyp{\ref{appl lr}}{F^+}{F^-} = \L_{\G_{12}}(N_{0 \leq 0,0 \leq 1})$
is a hypothesis satisfying the Sufficiency and the Strong Consistency
requirement.
%
Using the variable naming convention from Ex.~\ref{exm 1},
members of $\Hyp{\ref{appl lr}}{}{}$ are, e.g.:
{
\newcommand{\0}{l@{\hspace*{0.5cm}}rl}
\newcommand{\1}{@{\hspace*{1.4cm}}}
$$\renewcommand{\arraystretch}{1}
\begin{array}{@{}\0\1\0\1\0@{}}
1. &  v_{00} & \leq v_{01}
	& 3. & v_{00} * v_{00} & \leq v_{01}
	& 5. & 0 & \leq v_{01}	\\
2. & v_{00} & \leq v_{01} * v_{01}
	& 4. & v_{00} * v_{01} & \leq v_{11}*v_{01}
	& 6. & v_{00} & \leq v_{00} + v_{01}	\\
\end{array}$$%
}%
Hypothesis 1 intuitively means
that $(\leq)$ relates every possible pair of terms.
Hypotheses~2 and~3
indicate that $F^+$ was chosen too specifically, viz.\ all examples
had quadratic numbers as the right or left argument of $(\leq)$.
Similarly, hypothesis~5
reflects the fact that all left arguments are actually zero.
While $0 \leq x$ is a valid \emph{law},
$x \leq y \Lra x \eq 0$ is not a valid \emph{definition}.
Similarly, no variant of $x \leq x$ can be found in
$\Hyp{\ref{appl lr}}{}{}$
because it does not cover the second example from $F^+$.
Hypothesis 6 is an acceptable defintion of the $(\leq)$ relation
on natural numbers; it corresponds to
$x \leq y \Lra \exists z \in \N: \; y \eq x+z$.

If we take $F^+$ as above, but $F^- \Lra s(0) \not\leq 0$,
we get $S$ as the set of all $2^4$ substitutions with domain
$\set{v_{00},\ldots,v_{11}}$
and range
$\set{0,s(0)}$.
The resulting grammar for $\Hyp{\ref{appl lr}}{}{}$
is too large to be shown here.
%
$\Hyp{\ref{appl lr}}{}{}$
will no longer contain hypotheses~1 to~4
from above; they are subtracted as members of the union
$\bigcup_{\sigma \in S} \eqc{s(0) \leq 0}{E}{\sigma}$:
choose $\sigma = \subst{v_{00} \mapsto s(0), v_{01} \mapsto 0}$
for 1 to 3, and
$\sigma
= \subst{v_{00}~\mapsto~s(0), \allowbreak
v_{01}~\mapsto~s(0), \allowbreak
v_{11}~\mapsto~0}$
for 4.
$$\renewcommand{\arraystretch}{1.0}
\begin{array}[t]{@{}r@{\;}lll@{}}
(v_{00} \leq v_{01})
      & \subst{v_{00} \mapsto s(0), v_{01} \mapsto 0}
      & \eq
      & (s(0) \leq 0) \\
(v_{00} \leq v_{01} * v_{01})
      & \subst{v_{00} \mapsto s(0), v_{01} \mapsto 0}
      & \eq
      & (s(0) \leq 0) \\
(v_{00} * v_{00} \leq v_{01})
      & \subst{v_{00} \mapsto s(0), v_{01} \mapsto 0}
      & \eq
      & (s(0) \leq 0) \\
(v_{00} * v_{01} \leq v_{11} * v_{01})
      & \subst{v_{00} \mapsto s(0), v_{01} \mapsto s(0),
              v_{11} \mapsto 0}
      & \eq
      & (s(0) \leq 0) \\
\end{array}$$
None of the hypotheses 5 and 6 is eliminated, since
$$\renewcommand{\arraystretch}{1.0}
\begin{array}[t]{@{}r@{\;}l@{\;}l@{\;}l c
                      r@{\;}l@{\;}l@{\;}l c
                      r@{\;}l@{\;}l@{\;}l @{}}
(0 \leq v_{01}) & \sigma & \eq & (s(0) \leq 0)
        & \Ra
        & 0 & & \eq & s(0)      \\
(v_{00} \leq (v_{00}+v_{01})) & \sigma & \eq & (s(0) \leq 0)
        & \Ra
        & v_{00} & \sigma & \eq & s(0)
        & \Ra
        & (v_{00}+v_{01}) & \sigma & \neq_E & 0 .       \\
\end{array}$$

\subsection{Clausal Definitions}
\LABEL{Clausal definitions}

We now demonstrate how $E$-generalization can be incorporated into
an existing Inductive Logic Programming method to learn clauses.
To be concrete, we chose the method of
\emph{relative least general generalization (r-lgg)},
which originates from \cite{Plotkin.1970} and forms the basis of the
{\sc Golem} system \cite{Feng.Muggleton.1990}.
We show how to
extend it to deal with a given equational background theory $E$.

\THEOREM{Clausal Definitions}{ \LABEL{clauses eau}
Let two ground clauses $C_1$ and $C_2$ be given.
We can compute a regular set $H = {\it lgg}_E(C_1,C_2)$
such that:
\begin{itemize}
\item each $C \in H$ is a clause that
	$E$-subsumes $C_1$ and $C_2$; and
\item each clause $E$-subsuming both $C_1$ and $C_2$
	also subsumes an element of $H$.
\end{itemize}
}
\PROOF{
Let
$M =
\set{ \tpl{L_1,L_2}
\mid L_1 \in C_1 \land L_2 \in C_2 \land L_1 \mbox{ fits } L_2}$.
Assuming
$M = \set{ \tpl{p_i(t_{1i}),p_i(t_{2i})} \mid i=\,1m\i }$,
let
$T^+ = \set{ \tpl{\,1m{t_{1\i}}} , \tpl{\,1m{t_{2\i}}} }$
and $T^- = \set{}$.
Let
$H = \set{ \set{p_i(t_i) \mid i=\,1m{\i}}
\mid \tpl{\,1m{t_\i}} \in~\Hyp{\ref{hyp lr}}{T^+}{T^-} } $.
$H$ is again a regular tree language,
because the regular $\Hyp{\ref{hyp lr}}{T^+}{T^-}$ is
the image of $H$ under the tree homomorphism that maps
$\set{ \,1m{p_\i(x_\i)} }$ to $\tpl{ \,1m{x_\i} }$,
cf.\ \cite[Thm.~7 in Sect.~1.4]{Comon.Dauchet.Gilleron.1999}.
%
%

If $\set{ \,1m{p_\i(t_\i)} } \in H$,
i.e.\ $\hypp{\tpl{ \,1m{t_\i}} }{T^+,T^-}$,
then
$\set{
p_1(t_1 \hsigma_i), \allowbreak \ldots, \allowbreak p_m(t_m \hsigma_i)
}
\subseteq_E C_i$
for $i=1,2$,
where $\hsigma_i$ are the substitutions from the definition of
$\hypp{}{}{}$.
	Conversely,
	let some clause $C$
	$E$-subsume both $C_1$ and $C_2$.
	We assume w.l.o.g.\
	$C = \set{ \,1n{p_\i(t_\i)} }$
	and $t_j \sigma_i \eq t_{ij}$
	for some $\sigma_i$
	for $i=1,2$ and $j=\,1n\i$.
	By Lem.~\ref{hsg lemma nondet},
	some $\sigma$ exists such that
	$t_j \sigma \tau_i \eq t_{ij}$.
	Choosing
	$t'_j = t_j \sigma$ for $j=\,1n\i$ and
	$t'_j = \vNN{\NNt{t_{1j}},\NNt{t_{2j}}}$
	for $j=\,{n+1}m\i$,
	we obtain
	$t'_j \tau_i \eq t_{ij}$ for $j=\,1m\i$.
	Hence, $\hypp{\tpl{ \,1m{t'_\i}} }{T^+,T^-}$ holds,
	i.e.\ $C \sigma \subseteq \set{ \,1m{p_\i(t'_\i)} } \in H$.

To compute $H$,
the grammar defining all $\eqc{t_{ij}}{E}{}$
must be extended to also define $\eqc{\tpl{\,1m{t_{i\i}}}}{E}{}$.
Since only nonterminals for congruence classes are added,
no additional language intersections are necessary to compute the
extended~$\tau_i$.
\qed
}

For an empty $E$,
we have ${\it lgg}_E(C_1,C_2) = \set{ {\it lgg}(C_1,C_2) }$,
and Thm.~\ref{clauses eau} implies Plotkin's ${\it lgg}$ theorem
\cite[Thm.~3]{Plotkin.1970} as a special case.
In the terminology of 
Fig.~\ref{Requirements to hypothesis generation},
we have $F^+ \Lra C_1 \land C_2$ and $F^- \Lra {\it true}$.
The Consistency requirement is
satisfied if some predicate symbol $p$
different from $(\eq)$ occurs in
both $C_1$ and $C_2$ but not in $B$, 
except for $p$'s congruence axiom.
In this case, each hypothesis $h$ will have the form
$p(\ldots) \lor \ldots$,
hence $B \land h$ cannot be contradictory.
The set ${\it lgg}_E(C_1,C_2)$ is a subset of, but is not equal to,
the set of all 
hypothesis clauses satisfying Sufficiency.
Usually, there are other
clauses that imply both $C_1$ and $C_2$ but do not
$E$-subsume both.
The same limitation applies to Plotkin's syntactical ${\it lgg}$.

Theorems~\ref{clauses eau} and~\ref{appl lr} share a special case:
${\it lgg}_E(\set{p(t_1)},\set{p(t_2)})$
from Thm.~\ref{clauses eau} equals
$\Hyp{\ref{appl lr}}{p(t_1) \land p(t_2)}{{\it true}}$
from Thm.~\ref{appl lr}.
In this case, Thm.~\ref{appl lr} is stronger because it ensures
that the result set contains \emph{all} sufficient hypotheses.
On the other hand, Thm.~\ref{clauses eau} allows a more general form of
both hypotheses and examples.


However, even
Thm.~\ref{clauses eau} cannot generate all sufficient and strongly
consistent hypotheses that can be expressed in first order predicate
logic.
For example, $F = \set{p(a,b), p(a,c)}$
can be explained wrt.\ an empty $E$
either by:
\begin{enumerate} 
\item $q \land (q \ra p(a,b) \land p(a,c))$,
\item $p(a,b) \land p(a,c)$,
\item $p(a,b) \land b \eq c$, or
\item $p(a,x) \lor x \eq d$.
\end{enumerate}
None of these hypotheses
has a form that can be generated by Thm.~\ref{appl lr}.
Only hypothesis 4 has a form admitted by Thm.~\ref{clauses eau};
however, it does not $E$--subsume any member of $F$,
although it implies both of them.


\newcommand{\fm}[1]{{\bf #1}}
\newcommand{\ml}[1]{{\bf #1}}

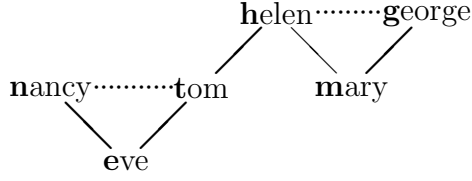
\begin{figure}
\begin{center}
\begin{picture}(5.5,2.5)
\put(3.600,2.250){\makebox(0.000,0.000){\fm{h}elen}}
      \put(3.800,2.050){\line(1,-1){0.600}}
\thicklines
      \put(3.400,2.050){\line(-1,-1){0.600}}
      \multiput(4.1,2.15)(0.1,0){9}{$\cdot$}
\put(5.600,2.200){\makebox(0.000,0.000){\ml{g}eorge}}
      \put(5.400,2.050){\line(-1,-1){0.600}}
\put(0.600,1.200){\makebox(0.000,0.000){\fm{n}ancy}}
      \put(0.800,1.050){\line(1,-1){0.600}}
      \multiput(1.15,1.15)(0.1,0){11}{$\cdot$}
\put(2.600,1.250){\makebox(0.000,0.000){\ml{t}om}}
      \put(2.400,1.050){\line(-1,-1){0.600}}
\put(4.600,1.200){\makebox(0.000,0.000){\fm{m}ary}}
\put(1.600,0.250){\makebox(0.000,0.000){\fm{e}ve}}
\end{picture}
\end{center}
\caption{Background knowledge in family relations example}
\LABEL{Background knowledge in family relations example}
\end{figure} 

To illustrate Thm.~\ref{clauses eau},
consider a well-known example about learning family relations.
We use the abbreviations
{\it d--daughter},
{\it p--parent},
{\it f--female},
{\it e--eve},
{\it g--george},
{\it h--helen},
{\it m--mary},
{\it n--nancy}, and
{\it t--tom}.
Let the background knowledge
$K \Lra p(h,m) \land p(h,t) \land p(g,m) \land p(t,e) \land P(n,e)
\land f(h) \land f(m) \land f(n) \land f(e)$
and the positive examples
$F_1 \Lra d(m,h)$ and $F_2 \Lra d(e,t)$
be given (cf.\ Fig.~\ref{Background knowledge in family relations
example}).

By generalizing relative to $K$,
i.e.\ by computing
${\it lgg}((F_1 \la K),(F_2 \la K))$,
and by eliminating all body literals containing a
variable not occurring in the head literal,
the clausal definition of the {\it daughter} relation
$d(v_{me},v_{ht}) \la p(v_{ht},v_{me}) \land f(v_{me}) \land K$
results.

In addition, using the abbreviation {\it s--spouse},
let the equations
$E = \set{
s(g) \.= h, \allowbreak
s(h) \.= g, \allowbreak
s(n) \.= t, \allowbreak
s(t) \.= n }$
be given.
The congruence classes of, e.g., $g$ and $h$ can be described by
$N_g \sortdef g \mid s(N_h)$ and $N_h \sortdef h \mid s(N_g)$.
We obtain by Thm.~\ref{clauses eau} all clauses of the form
$d(v_{me},t_{ht})
\la p(t'_{ht},v_{me}) \land p(t_{gn},v_{me}) \land f(v_{me})$
for any
$t_{ht}, t'_{ht} \in \eqc{h}{E}{\tau_1} \cap \eqc{t}{E}{\tau_2}$
and
$t_{gn} \in \eqc{g}{E}{\tau_1} \cap \eqc{n}{E}{\tau_2}$.

In order to obtain a constrained clause as before,
we first choose some $t_{ht}$ for the head literal,
and then choose $t'_{ht}$ and $t_{gn}$ from the filtered sets
$\eqc{h}{E}{\tau_1} \cap \eqc{t}{E}{\tau_2} \cap \T{}{}{\var(t_{ht})}$
and
$\eqc{g}{E}{\tau_1} \cap \eqc{n}{E}{\tau_2} \cap \T{}{}{\var(t_{ht})}$,
respectively.
We use the standard intersection algorithm for tree grammars mentioned
in Sect.~\ref{The core method} for filtering,
which in this case requires linear time in the grammar size for 
$\eqc{h}{E}{\tau_1} \cap \eqc{t}{E}{\tau_2}$
and $\eqc{g}{E}{\tau_1} \cap \eqc{n}{E}{\tau_2}$.
Choosing the smallest solutions for $t_{ht}$, $t'_{ht}$, and $t_{gn}$,
we obtain
$d(v_{me},v_{ht})
\la p(v_{ht},v_{me}) \land p(s(v_{ht}),v_{me}) \land f(v_{me})$,
which reflects the fact
that our background knowledge did not describe any
concubinages.


If
$p(h,m)$ is removed from $K$,
we still get
$d(v_{me},v_{ht}) \la p(s(v_{ht}),v_{me}) \land f(v_{me})$
in a similar way,
while the \emph{classical} relative ${\it lgg}$, not considering $E$,
does not yield a meaningful clause.

Although the knowledge about spouses could be encoded by additional
clauses $s(g,h) \land s(h,g) \land s(n,t) \land s(t,n)$ in $K$, rather
than in $E$,
it would require a weaker literal selection strategy to get a
clause like
$d(v_{me},v_{ht})
\la s(v_{ht},v_{gn}) \land p(v_{gn},v_{me}) \land f(v_{me})$.
The latter clause is determinate, but not constrained.

\subsection{Atomic Determinate Definitions}
\LABEL{Atomic determinate definitions}

Below, we prove a learning
theorem similar to Thm.~\ref{appl lr}, 
but that yields only those atomic hypotheses
$p(s,t)$ that define a determinate predicate $p$.
Formally, we are looking for those $p(s,t)$ that satisfy
$$\forall s'_1,s'_2,t'_1,t'_2:
(B \land p(s,t) \land s'_1 \eq s'_2
\impl p(s'_1,t'_1) \land p(s'_1,t'_2))
\Ra (B \impl t'_1 \eq t'_2) .$$
In such cases,
we say that the hypothesis $p(s,t)$ is determinate.
Determinacy of a hypothesis is essentially a \emph{semantic} property
\cite[Sect.~5.6.1]{Dzeroski.Lavrac.1994};
it is even undecidable for certain background theories.
Let 
$$\renewcommand{\arraystretch}{1}
\begin{array}{r c l@{\;}c@{\;}c@{\;}c@{\;}l l}
f(a,\tpl{1, \,2m{i_\i}})
	& \eq 
	& \tpl{ & a_1  & \cdot  
	& \:{\cdot}2m{a_{i_\i}},
	& 1, \,2m{i_\i}}
	& \mbox{ and}	\\
f(b,\tpl{1, \,2m{i_\i}})
	& \eq 
	& \tpl{ & b_1  & \cdot 
	& \:{\cdot}2m{b_{i_\i}}, 
	& 1, \,2m{i_\i}}
	& \mbox{ ,}	\\
\end{array}$$
then $p(x,\tpl{y,z}) \Lra f(y,z) \eq x$
defines a determinate $p$ if Post's correspondence problem from
Lem.~\ref{Post} has no solution, 
although $E$ admits regular congruence classes.
In order to compute the set of all determinate hypotheses, 
we have to make a little detour 
by defining a notion of \emph{weak determinacy},
which is equivalent to a simple \emph{syntactic} criterion 
(Lem.~\ref{rg uniq}).

Since $B$ does not imply anything about $p$ except its
congruence property,
we assume
$B \Lra B'' \land 
{(\forall x,y,x',y': 
x \eq x' \land y \eq y' \land p(x,y) \ra p(x',y'))}$,
where $p$ does not occur in $B''$.
We replace the full congruence axiom about $p$ by a partial one:
$B' \Lra B'' \land 
{(\forall x,y,y':
y \eq y' \land p(x,y) \ra p(x,y'))}$.
We call a hypothesis $p(s,t)$ weakly determinate if
$$\forall s',t'_1,t'_2:
{(B' \land p(s,t) \impl p(s',t'_1) \land p(s',t'_2))}
\allowbreak \Ra \allowbreak {(B' \impl t'_1 \eq t'_2)} .$$

For example, 
using $E$ from Fig.~\ref{$E$-generalization of $0$ and $\suc^4(0)$}
(left),
$p(x*y,x+y)$ is a weakly, but not ordinarily, determinate hypothesis.
      Abbreviating $s^n(0)$ by $n$, we have
      $B_1 \land p(x*y,x+y) \impl p(4*3,7) \land p(4*3,8)$,
      since $4*3 \eq 2*6$
      and $8 \eq 2+6 \not\eq 7$.
      However,
      $B_3 \land p(x*y,x+y) \impl p(4*3,t')$
      implies $x \sigma = 4$, $y \sigma = 3$, and $x \sigma + y \sigma
      \eq t'$
      by Lem.~\ref{Herbrand 3}.2.
      %
      Hence, only $t' \eq 3+4 \eq 7$ is possible.

We define
for sets $T^+,T^-$ of ground-term pairs and arbitrary terms $s,t$:
$$\renewcommand{\arraystretch}{1}
\begin{array}{llll}
\hypp{s,t}{T^+}
	& \Lra
	& \forall \tpl{s',t'} \in T^+ \; \exists \sigma: \;
	s \sigma = s' \land t \sigma \eq t'
	& \mbox{ and}	\\
\hypn{s,t}{T^-} 
	& \Lra
	& \forall \tpl{s',t'} \in T^- \; \forall \sigma: \;
	s \sigma \neq s' \lor t \sigma \not\eq t'
	& \mbox{ .}	\\
\end{array}$$
We have a lemma similar to Lem.~\ref{Herbrand},
with a simlar proof, which is omitted here.

\LEMMA{Weak Requirements}{ \LABEL{Herbrand 3}
Let $s_i,t_i, s'_i,t'_i$ be ground terms
and
$s,t$ arbitrary terms.
Let $F^+ \Lra \bigwedge_{i=1}^n p(s_i,t_i)$
and $F^- \Lra \bigwedge_{i=n+1}^m \lnot p(s_i,t_i)$.
Let $T^+ = \set{ \,1n{\tpl{s_\i,t_\i}} }$
and $T^- = \set{ \,{n+1}m{\tpl{s_\i,t_\i}} }$.
Then:
\begin{enumerate}
\item	$B' \land p(s,t) \impl \:{\.\lor}1{n'}{p(s'_\i,t'_\i)}$
	iff
	$s \sigma \.= s'_i \land t \sigma \.\eq t'_i$
	for some $i$, $\sigma$.
\item $B' \land p(s,t) \impl F^+$  ~ iff ~
	$\hypp{s,t}{T^+}$.
\item $B' \land p(s,t) \land F^- \not\impl {\it false}$
	~ iff ~
	$\hypn{s,t}{T^-}$.
\qed
\end{enumerate}
}

\LEMMA{Syntactic Criterion}{ \LABEL{rg uniq}
\begin{enumerate}
\item If $\var(t) \subseteq \var(s)$, then the hypothesis
	$p(s,t)$ is weakly determinate.
\item Each weakly determinate hypothesis $p(s,t)$
	has a weakly determinate instance $p(s \sigma,t \sigma)$
	with $\var(t \sigma) \subseteq \var(s \sigma)$
	and $B' \impl p(s,t) \lra p(s \sigma,t \sigma)$.
\end{enumerate}
}
\PROOF{
\begin{enumerate}
\item If $B' \land p(s,t) \impl p(s',t'_1) \land p(s',t'_2)$,
	we have
	$s \sigma_1 = s' = s \sigma_2
	\land t \sigma_1 \eq~t'_1
	\land t \sigma_2 \eq t'_2$
	by Lem.~\ref{Herbrand 3}.2.
	Hence, $\sigma_1$ and $\sigma_2$
	coincide on $\var(s) \supseteq \var(t)$,
	and we get $t'_1 \eq t \sigma_1 = t \sigma_2 \eq t'_2$.
\item Define
	$\sigma
	= \subst{ x \mapsto a \mid x \in \var(t) \setminus \var(s)}$,
	where $a$ is an arbitrary constant.
	Then $\var(t \sigma) \subseteq \var(s \sigma)$ by construction.
	Since
	$B' \impl p(s,t) \ra p(s \sigma,t \sigma)$,
	we have the situation that
	$p(s \sigma,t \sigma)$ is again weakly determinate.

	Since $s \sigma = s$,
	we have
	$B' \land p(s,t) \impl p(s,t) \land p(s,t \sigma)$.
	Since $p(s,t)$ is weakly determinate,
	this implies $t \eq t \sigma$.
	Since $B'$ ensures that $p$ is $E$-compatible in its right
	argument,
	we have $B' \land p(s \sigma, t \sigma) \impl p(s,t)$.
\qed
\end{enumerate}
}

\LEMMA{Equivalence for Constructor Terms}{ \LABEL{det, weak det}
Let $s$ be a constructor term.
\begin{enumerate}
\item $p(s,t)$ is a weakly determinate hypothesis
	iff it is a determinate one.
\item 
	$B \land p(s,t) \impl p(s',t')$
	iff $B' \land p(s,t) \impl p(s',t')$,
	if $s'$ is an instance of $s$.
\item 
	$B \land p(s,t) \impl p(s',t')$
	iff $B' \land p(s,t) \impl p(s',t')$,
	if $s'$ is a constructor term.
\end{enumerate}
}
\PROOF {
All \emph{if} directions follow from $B \impl B'$.
In particular, weak determinacy always implies determinacy.
\begin{enumerate}
\item Obtain some $\sigma$ from Lem.~\ref{rg uniq}.2 such that
	$p(s \sigma,t \sigma)$ is again weakly determinate,
	$\var(s \sigma) \supseteq \var(t \sigma)$,
	and $B' \impl p(s,t) \lra p(s \sigma,t \sigma)$.
	From the latter, we get
	$B \impl p(s,t) \lra p(s \sigma,t \sigma)$.

	Hence, if
	$B \land p(s,t) \land s'_1 \eq s'_2
	\impl p(s'_1,t'_1) \land p(s'_2,t'_2)$ holds,
	we have
	$s \sigma \sigma_1 \eq s'_1 \eq s'_2 \eq s \sigma \sigma_2$
	and
	$t \sigma \sigma_1 \eq t'_1 \land t \sigma \sigma_2 \eq t'_2$
	for some $\sigma_1, \sigma_2$
	by Lem.~\ref{Herbrand}.2.
	Since $s$ is a constructor term,
	we have
	$x \sigma \sigma_1 \eq x \sigma \sigma_2$
	for each $x \in \var(s \sigma) \supseteq \var(t \sigma)$.
	Hence,
	$t'_1 \eq t \sigma \sigma_1 \eq t \sigma \sigma_2 \eq t'_2$.
\item Obtain
	$s \hsigma \eq s'$ and $t \hsigma \eq t'$ for some $\hsigma$
	from Lem.~\ref{Herbrand}.2 and the Def.\ of $\hypp{}{}$.
	Since $s \sigma' = s'$ for some $\sigma'$,
	we have
	$s \sigma' \eq s \hsigma$.
	Since $s$ is a constructor term,
	we have $x \sigma' \eq x \hsigma$ 
	for all $x \in \var(s) \supseteq \var(t)$.
	Hence, $t \sigma' \eq t \hsigma \eq t'$.
	By Lem.~\ref{Herbrand 3}.2,
	this implies $B' \land p(s,t) \impl p(s',t')$.
\item Follows from~2,
	since
	$s \hsigma \eq s'$ implies that $s'$ is an instance of $s$.
\qed
\end{enumerate}
}

\newsavebox{\greybox}
\newlength{\greygrain}
\setlength{\greygrain}{0.002cm}
%
\newcommand{\grey}[2]{
\savebox{\greybox}{\multiput(0,0)(0.1,0){#1}
                                        {\rule{\greygrain}{\greygrain}}}
\multiput(0.0,0)(0,0.1){#2}{\usebox{\greybox}}
\multiput(0.05,0.05)(0,0.1){#2}{\usebox{\greybox}}
}
\newcommand{\redundant}{%
        $\scriptstyle\blacktriangleleft$%
        \hspace{-.05cm}%
        \rule[0.04cm]{0.2cm}{0.05cm}%
}
\begin{figure}
\begin{center}
\begin{picture}(13.7,6.1)
        \setlength{\greygrain}{0.020cm}
\put(1.250,1.050){\framebox(4.000,2.000)[tl]
                        { \sf Lem.\ref{det, weak det}.1}}
        \put(1.250,1.050){\grey{40}{20}}
\put(5.250,3.050){\framebox(7.000,1.000)[tr]
                        {\sf Lem.\ref{det, weak det}.1 }}
        \put(5.250,3.050){\grey{70}{10}}
\put(3.350,1.150){\framebox(2.000,4.000)[tl]{ \sf Lem.\ref{rg uniq}.1}}
        \put(3.350,1.150){\grey{20}{40}}
\put(8.350,1.150){\framebox(4.000,4.000)[tr]{\sf Lem.\ref{rg uniq}.2 }}
        \setlength{\greygrain}{0.008cm}
        \put(8.350,1.150){\grey{40}{40}}
\put(8.300,4.100){\makebox(4.000,1.000){$p(x\.*y,x\.+y\.+z\.*0)$}}
\put(8.300,2.100){\makebox(4.000,1.000){$p(x,x\.*x\.+z\.*0)$}}
\put(8.300,1.100){\makebox(4.000,1.000){$p(x\.*x,x\.+z\.*0)$}}
\put(8.300,4.100){\makebox(0.000,1.000){\redundant}}
\put(8.300,2.100){\makebox(0.000,1.000){\redundant}}
\put(8.300,1.100){\makebox(0.000,1.000){\redundant}}
\thicklines
\put(3.250,2.050){\framebox(5.150,2.100)[tl]{~ \sf Lem.\ref{hyp r}}}
\put(1.200,1.000){\framebox(11.200,4.200)[tl]{ \sf Lem.\ref{hyp lr}}}
\put(1.300,4.100){\makebox(2.000,1.000){$p(x\.*y,z)$}}
\put(1.300,3.100){\makebox(2.000,1.000){$p(x,z)$}}
\put(5.300,4.100){\makebox(3.000,1.000){$p(x\.*y,x\.+y)$}}
\put(5.300,2.100){\makebox(3.000,1.000){$p(x,x\.*x)$}}
\put(5.300,1.100){\makebox(3.000,1.000){$p(x\.*x,x)$}}
%
\put(5.400,5.400){\makebox(7.000,0.000)[b]
                        {$\overbrace{\rule{7cm}{0cm}}^{}$}}
\put(5.400,5.700){\makebox(7.000,0.000)[b]{weak det.}}
\put(3.300,0.800){\makebox(5.000,0.000)[t]
                        {$\underbrace{\rule{5cm}{0cm}}_{}$}}
\put(3.300,0.500){\makebox(5.000,0.000)[t]{$\var(s) \supseteq \var(t)$}}
\put(1.100,1.050){\makebox(0.000,2.000)[r]
                        {det.$\left\{\rule{0cm}{1.2cm}\right.$}}
\put(12.500,2.050){\makebox(0.000,2.000)[l]
                        {$\left.\rule{0cm}{1.2cm}\right\}$cstr.}}
\end{picture}
\end{center}
\caption{Possible cases wrt.\ weak and ordinary determinacy}
\LABEL{Karnaugh}
\end{figure}
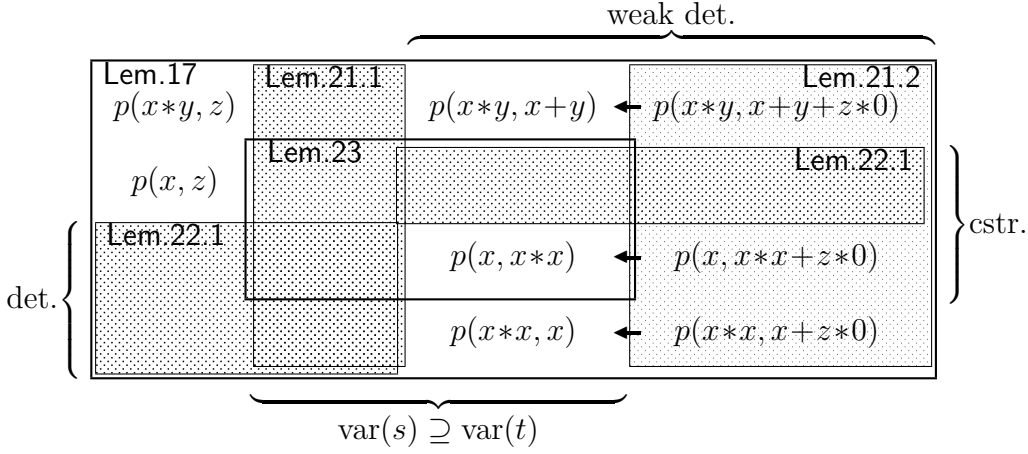

The Karnaugh 
diagram in Fig.~\ref{Karnaugh}
summarizes the relations between the criteria from Lem.~\ref{rg uniq}
and~\ref{det, weak det}, weak and ordinary determinacy.
It gives an example hypothesis for each possible case.
%
Lemma~\ref{det, weak det} ends our little detour.
It ensures that weak and ordinary determinacy
coincide if we supply only constructor terms to the \emph{input}
argument of a hypothesis $p$.
On the one hand, 
this is a restriction because we cannot learn a hypothesis
like $p(x*x,x)$, which defines a partial function realizing the integer
square root.
On the other hand, it is often desirable
that a hypothesis correspond to an \emph{explicit} definition,
i.e.\ that it can be applied like a rewrite rule
to a term $s$ by purely syntactical pattern matching.
Tuples built using the operator $\tpl{\ldots}$
are the most frequently occurring special cases
of constructor terms.
For example,
a hypothesis $p(\tpl{x,y},x + 2 * y)$
may be preferred to
$p(\tpl{2 * x, y},2 * (x+y))$
because the former is explicit and implies the latter wrt.\
$E$ from Fig.~\ref{$E$-generalization of $0$ and $\suc^4(0)$} (left).
Lemma~\ref{det, weak det}.2 allows us to instantiate
$\tpl{x,y}$ from the former hypothesis arbitrarily, even 
with non-constructor terms like
$\tpl{2*1,z_1+z_2}$.

The following lemma corresponds to Lem.~\ref{hyp lr},
but leads to reduced algorithmic time complexity.
It does not need to compute universal substitutions
because it uses constrained $E$-generalization from Def.~\ref{eau v}.
It still permits negative examples,
handling them more efficiently than Lem.~\ref{hyp lr}.
They may make sense even if
only determinate predicates are to be learned
because they allow us to
exclude certain undesirable hypotheses without
committing to a fixed 
function
behavior.

\LEMMA{Weakly Determinate Hypotheses}{ \LABEL{hyp r}
For each finite set of ground term pairs
$T^+ \cup T^-$,
we can compute a regular set $H = \Hyp{\ref{hyp r}}{T^+}{T^-}$
such that for all $s,t \in \T{}{}{\V}$:
$$\renewcommand{\arraystretch}{1}
\begin{array}{rlll}
\tpl{s,t} \in H
	& \Ra
	& \hypp{s,t}{T^+} \land \hypn{s,t}{T^-}
		\land \var(s) \supseteq \var(t)
	& \mbox{ , and} \\
\exists \sigma: \; \tpl{s \sigma,t \sigma} \in H
        & \La
	& \hypp{s,t}{T^+} \land \hypn{s,t}{T^-}
		\land \var(s) \supseteq \var(t)
	& \mbox{ .}   \\
\end{array}$$
}
\PROOF{
Assume $T^+ = \set{ \tpl{s_i,t_i} \mid i=\,1n\i }$
and $T^- = \set{ \tpl{s_i,t_i} \mid {i=\,{n\.+1}m\i} }$.
For $\set{\,1n\i} \subseteq I \subseteq \set{\,1m\i}$,
let $s_I$ be the most specific
syntactical generalization of $\set{s_i \mid i \in I}$,
with $s_I \sigma_{I,i} = s_i$ for each $i \in I$.
Such an $I$ is called \emph{maximal}
if
$\forall \set{\,1n\i} \subseteq I' \subseteq \set{\,1m\i}: \;
s_I = s_{I'} \Ra I' \subseteq I$,
where $(=)$ denotes term equality up to renaming.

For example, if $T^+ = \set{ \tpl{a\.+a, t_a} }$
and $T^- = \set{ \tpl{b\.+b, t_b}, \tpl{b\.+c, t_c} }$,
then $\set{1,2}$ and $\set{1,2,3}$ are maximal, but $\set{1,3}$ is not.
Since $s_{\set{1,2}} = x+x$ can be instantiated to $a+a$ and $b+b$,
we must merely ensure that
$t_{\set{1,2}} \subst{x \mapsto a} \not\eq t_a$
and $t_{\set{1,2}} \subst{x \mapsto b} \not\eq t_b$
in order to obtain $\hypn{s_{\set{1,2}},t_{\set{1,2}}}{T^-}$.
However, for $s_{\set{1,3}} = x+y$, it is not sufficient to ensure
$t_{\set{1,3}} \subst{x \mapsto a, y \mapsto a} \not\eq t_a$
and $t_{\set{1,3}} \subst{x \mapsto b, y \mapsto c} \not\eq t_c$.
Since
$s_{\set{1,3}}$ happens to be instantiable to $b\.+b$ as well,
$\hypn{s_{\set{1,3}},t_{\set{1,3}}}{T^-}$ could be violated
if $t_{\set{1,3}} \subst{x \mapsto b, y \mapsto b} \eq t_b$.
Therefore, only 
generalizations $s_I$ of maximal $I$ should be considered.

Let $\I$ be the set of all maximal $I$.
For $I \in \I$, let
$T_I
= \bigcap_{i=1}^n \eqc{t_i}{E}{\sigma_{I,i}}
\setminus \bigcup_{i \in I, i > n} \eqc{t_i}{E}{\sigma_{I,i}}$.
Each such set $T_I$ can be computed
from $\,1m{ \eqc{t_\i}{E}{} }$
by standard tree grammar algorithms.
Given the grammar for each $T_I$, it is easy to compute a grammar for
their \emph{tagged union}
$H = \set{ \tpl{s_I,t_I} \mid I \in \I \land t_I \in T_I}$.
To prove the properties of $H$,
first observe the following:
\begin{enumerate}
\item We always have
	$\var(t_I) \subseteq \dom \sigma_{I,1} \subseteq \var(s_I)$.
	The first inclusion follows from
	$t_I \in T_I \subseteq \eqc{t_1}{E}{\sigma_{I,1}}$,
	the second from the definition of $\sigma_{I,1}$.
\item If $I$ is maximal and $s_I \sigma = s_i$
	for some $i \in \set{\,1m\i}$ and $\sigma$, then $i \in I$:
	\\
	Since $s_I \sigma_{I,j} = s_j$ for $j \in I$
	and $s_I \sigma = s_i$,
	the term $s_I$ is a common generalization of the set
	$\set{ s_j \mid j \in I} \cup \set{s_i}$.
	Hence, its most special generalization,
	viz.\ $s_{I \cup \set{i}}$, is an instance of $s_I$.
	Conversely, $s_{I \cup \set{i}}$ is trivially
	a common generalization of $\set{ s_j \mid j \in I}$;
	hence $s_I$ is an instance of $s_{I \cup \set{i}}$.
	Therefore, $s_{I \cup \set{i}} = s_I$,
	which implies $i \in I$ because $I$ is maximal.
\end{enumerate}
\begin{itemize}
\item If $I \in \I$ and $t_I \in T_I$,
	then trivially
	$s_I \sigma_{I,i} = s_i$
	and $t_I \sigma_{I,i} \eq t_i$ for each $i \leq n$.
	Assume
	$s_I \sigma = s_i$ and $t_I \sigma \eq t_i$
	for some $\sigma$ and some $i>n$.
	By (2), we have $i \in I$,
	and therefore $s_I \sigma_{I,i} = s_i$.
	Hence, $\sigma_{I,i}$ and $\sigma$ coincide on
	$\var(s_I) \supseteq \var(t_I)$, using (1).
	We get $t_I \sigma_{I,i} = t_I \sigma \eq t_i$,
	which contradicts $t_I \not\in \eqc{t_i}{E}{\sigma_{I,i}}$.
\item If $s,t$ are given such that
	$\hypp{s,t}{T^+}$ and $\hypn{s,t}{T^-}$ hold,
	let
	$I = \set{\,1n\i}
	\cup \set{ i \mid n < i \leq m
	\land \exists \sigma'_i: \; s \sigma'_i = s_i}$.
	Then, $s$ is a common generalization of
	$\set{s_i \mid i \in I}$,
	and we have $s \sigma = s_I$ for some $\sigma$.

	We show $I \in \I$:
	Let $I'$ be such that $s_{I'} = s_I$ and let $i \in I'$,
	then
	$s \sigma \sigma_{I',i}
	= s_I \sigma_{I',i}
	= s_{I'} \sigma_{I',i}
	= s_i$,
	hence $i \in I$.
	Since $i$ was arbitrary, we have $I' \subseteq I$,
	i.e.\ $I$ is maximal.

	For $i \leq n$, we have
	$s \sigma \sigma_{I,i}
	= s_I \sigma_{I,i}
	= s_i
	= s \sigma_i$.
	In other words, $\sigma \sigma_{I,i}$ and the $\sigma_i$
	obtained from $\hypp{s,t}{T^+}$
	coincide on $\var(s) \supseteq \var(t)$.
	Hence,
	$t \sigma \sigma_{I,i}
	= t \sigma_i
	\eq t_i$,
	i.e.\ $t \sigma \in \eqc{t_i}{E}{\sigma_{I,i}}$.
	For $i > n$ and $i \in I$,
	we still have $s \sigma \sigma_{I,i} = s_i$,
	as above.
	Hence $t \sigma$ cannot be a member of
	$\eqc{t_i}{E}{\sigma_{I,i}}$.
	Therefore, $\tpl{s \sigma,t \sigma} \in H$.
	\qed
\end{itemize}
}

\THEOREM{Atomic Determinate Definitions}{ \LABEL{appl r3}
Let $F^+ \Lra \bigwedge_{i=1}^n p(s_i,t_i)$
and $F^- \Lra \bigwedge_{i=n+1}^m \lnot p(s_i,t_i)$
be given
such that each $t_i$ is ground and each $s_i$ is a ground
constructor term.
Then, we can compute a regular set $H = \Hyp{\ref{appl r3}}{F^+}{F^-}$
such that
\begin{itemize}
\item each $p(s,t) \in H$ is a determinate hypothesis
        satisfying the \emph{Sufficiency} and the
        \emph{Strong Consistency} requirement wrt.\
	$F^+$, $F^-$; and
\item for each determinate hypothesis satisfying these requirements
	and having the form $p(s,t)$
	with $s$ constructor term,
        we have $p(s \sigma,t \sigma) \in H$ for some $\sigma$.
\end{itemize}
}
\PROOF{
Let $T^+ = \set{ \tpl{s_i,t_i} \mid i=\,1n\i }$
and $T^- = \set{ \tpl{s_i,t_i} \mid i=\,{n+1}{m}{\i} }$.
Define
$H = \set{ p(s,t) \mid \tpl{s,t} \in \Hyp{\ref{hyp r}}{T^+}{T^-} }$,
which is again a regular tree language.
%
%
\begin{itemize}
\item If $p(s,t) \in H$,
	then $\tpl{s,t} \in \Hyp{\ref{hyp r}}{T^+}{T^-}$,
	i.e.\ $\hypp{s,t}{T^+}$, $\hypn{s,t}{T^-}$ and
	$\var(s) \supseteq \var(t)$ hold.
	By Lem.~\ref{rg uniq}.1,
	$p(s,t)$ is weakly determinate;
	by Lem.~\ref{Herbrand 3}.2 and~3,
	it satisfies the requirements wrt.\ $B'$.
	By construction of Lem.~\ref{hyp r},
	$s$ is a constructor term.
	Hence, by Lem.~\ref{det, weak det}.1 and~3,
	$p(s,t)$ is determinate
	and satisfies the requirements wrt.\ $B$,
	respectively.
\item Let $p(s,t)$ be a
	determinate hypothesis satisfying the requirements wrt.\ $B$,
	where $s$ is a constructor term.
	By Lem.~\ref{det, weak det}.1 and~3,
	it is also weakly determinate and satisfies the requirements
	wrt.\ $B'$, respectively.
	Obtain $\sigma$ from Lem.~\ref{rg uniq}.2
	such that $p(s \sigma,t \sigma)$ additionally satisfies
	$\var(t \sigma) \subseteq \var(s \sigma)$.
	%
	By Lem.~\ref{Herbrand 3}.2 and~3,
	we then have $\hypp{s \sigma,t \sigma}{T^+}$ 
	and $\hypn{s \sigma,t \sigma}{T^-}$,
	respectively.
	By Lem.~\ref{hyp r}, we have
	$\tpl{s \sigma \sigma',t \sigma \sigma'} 
	\in \Hyp{\ref{hyp r}}{T^+}{T^-}$
	for some $\sigma'$,
	i.e.\ $p(s \sigma \sigma',t \sigma \sigma') \in H$.
\end{itemize}
To compute $H$,
the union of up to $(m-n) \cdot 2^{m-n}$,
the intersection of $n$
and the difference between two grammars are needed.
No additional grammar intersection is necessary to compute any
universal substitution.
\qed
}


From a theoretical point of view,
learning relations by classical ILP can be \emph{simulated} by learning
functions into a set ${\it bool}$ by $E$--generalization.
This is similar to the simulation of theorem proving by
a rewrite system defining appropriate rules for each junctor,
and
faced with similar efficiency problems in practice.
Nevertheless, we can simulate a small example to illustrate the use of
Thm.~\ref{appl r3} here.

\begin{figure}
\begin{tabular}{@{}|l|l@{$\;$}l@{$\;$}l@{$\;$}l@{$\;$}l@{$\;$}l|@{}}
\hline
smiling   & y  & y  & y  & y  & n  & n        \\
holding   & ba & fl & sw & sw & sw & fl       \\
hasTie    & y  & y  & y  & n  & n  & n        \\
headShape & sq & oc & rd & sq & oc & rd       \\
bodyShape & sq & oc & oc & oc & rd & oc       \\
\hline
friendly  & y  & y  & n  & n  & n  & n        \\
\hline
\end{tabular}
\hfill
%
\begin{tabular}{@{}|r@{$\;$}l@{}l@{}l@{}l@{}l@{}l@{}l@{$\;$}l|@{}}
\hline
$F^+ = \{$
& $p(\langle$ & $y,$ & $ba,$ & $y,$ & $sq,$ & $sq$ & $\rangle,y)$, &\\
& $p(\langle$ & $y,$ & $fl,$ & $y,$ & $oc,$ & $oc$ & $\rangle,y)$, &\\
& $p(\langle$ & $y,$ & $sw,$ & $y,$ & $rd,$ & $oc$ & $\rangle,n)$, &\\
& $p(\langle$ & $y,$ & $sw,$ & $n,$ & $sq,$ & $oc$ & $\rangle,n)$, &\\
& $p(\langle$ & $y,$ & $sw,$ & $n,$ & $oc,$ & $rd$ & $\rangle,n)$, &\\
& $p(\langle$ & $n,$ & $fl,$ & $n,$ & $rd,$ & $oc$ & $\rangle,n)$
      & $\}$  \\
\hline
\end{tabular}
\caption{Training data in robot example}
\LABEL{rob fig}
\end{figure}

In \cite[Sect.~5.2.1]{Dzeroski.Lavrac.1994},
a description of friendly and unfriendly robots is learned from a set of
examples.
For each robot, attributes like
${\it smiling}, {\it hasTie} \in \set{{\it yes}, {\it no}}$,
${\it holding} \in \set{{\it balloon}, {\it flag}, {\it sword}}$,
and
${\it headShape}, {\it bodyShape}
\in \set{{\it square}, \allowbreak
{\it octagon}, \allowbreak {\it round}}$
are given
(see Fig.~\ref{rob fig} (left), which uses obvious abbreviations).
Using attribute--value learners, a boolean expression
like
${\it fr}
\Lra {\it sm} = {\it y} \allowbreak
\land ({\it ho} = {\it ba} \lor {\it ho} = {\it fl})$
is learned as a description of a friendly robot.
As an improvement,
in \cite[Sect.~5.3.1]{Dzeroski.Lavrac.1994}
a derived attribute
${\it sameShape}$, defined by
${\it sameShape} \Lra ({\it headShape} = {\it bodyShape})$,
is added manually for each robot from the training set.
This results in a simpler description of a friendly robot,
viz.\
${\it friendly} \Lra {\it sameShape}$.

In order to duplicate both examples by $E$--generalization,
it is sufficient to let $E$ define the $(=)$ relation on attribute
values and the junctors $(\land)$, $(\lor)$, and $(\lnot)$.
A grammar defining the congruence classes $\eqc{{\it true}}{E}{}$
and $\eqc{{\it false}}{E}{}$ can be automatically obtained using
Thm.~\ref{eqc 1} or Lem.~\ref{eqc 2}.
One of its rules looks
$N_{\it false} \sortdef
(N_{\it balloon} \.= N_{\it flag}) \ldots \mid
(N_{\it true} \land N_{\it false}) \ldots \mid
(N_{\it false} \lor N_{\it false}) \ldots \mid
(\lnot N_{\it true})$.
The given attribute values are converted to $F^+$ as shown in
Fig.~\ref{rob fig} (right).
Using Thm.~\ref{appl r3},
both above descriptions appear in the hypotheses set
$\Hyp{\ref{appl r3}}{}{}$.
In contrast to the attribute--value learner approach, it is not
necessary to provide ${\it sameShape}$
explicitely
as an additional
attribute.
Since $E$ defines $(=)$ anyway, the second description from above
appears in the form
$p(\tpl{\ldots,v_{hd},v_{bd}},
(v_{\it hd}~=~v_{\it bd}))$.
%

Further examples of the application of Thm.~\ref{appl r3} are given in
Sect.~\ref{Applications}.

\subsection{Clausal Determinate Definitions}
\LABEL{Clausal determinate definitions}

We now show that learning a semi-determinate clause by ${\it lgg}$
can be simulated by learning an equivalent constrained clause
using $E$-generalization.
By analogy to the above, obtain the background theory $B'$ from $B$
by replacing the full congruence axiom for $p_0$ with a partial one.
Lemma~\ref{det lit rem 2} shows
how a semi-determinate clause $C$ can be transformed into an
equivalent constrained clause $\dlr{C}$.
Theorem~\ref{det lit rem 5} simulates ${\it lgg}$-learning of $C$ by
${\it lgg}^c_E$-learning of $\dlr{C}$.

\LEMMA{Determinate Literal Removal}{ \LABEL{det lit rem 2}
Let a semi-determinate clause
$C \Lra 
(p_0(s_0,t_0)
\la \bigOp{i}{\bigwedge}1n{p_\i(s_\i,x_\i)}
\land~\bigOp{i}{\bigwedge}1m{q_\i(t_\i)})$
be given such that $p_i(s_i,x_i) \Lra g_i(s_i) \eq x_i$.
Let
$\sigma = \:{}n1{ \subst{x_\i \mapsto g_\i(s_\i)} }$.
Then,
$\dlr{C} \Lra 
(p_0(s_0 \sigma,t_0 \sigma) 
\la \bigOp{i}{\bigwedge}1m{q_\i(t_\i \sigma)})$
is a constrained clause
that defines the same relation for~$p_0$ wrt.\ $B'$,
and hence also wrt.\ $B$.
}
\PROOF{
From the properties of semi-determinacy,
we have $s_0 \sigma = s_0$.
Since $p_0$ does not occur in $B'$ outside its partial
congruence axiom,
we can use the following property of SLD resolution 
\cite{Clark.1979}:
$B' \land (p_0(s_0,t_0) \la C) \impl p_0(s,t)$
iff
$s_0 \sigma' = s \land t_0 \sigma' \eq t$
and $B' \impl C \sigma'$ for some $\sigma'$.
A similar property holds for $\dlr{C}$.

The proofs of both directions are based on establishing
$x_i \sigma \sigma' \eq x_i \sigma'$.
This property follows from $p_i(s_i \sigma',x_i \sigma')$ and the
definitions of $g_i$ and $\sigma$,
when 
$(B' \land C \impl p_0(s,t)) \Ra (B' \land \dlr{C} \impl p_0(s,t))$
is proved.
When the converse direction is shown,
it is established by extending $\sigma'$ 
to $\var(\dlr{C}) \cup \set{ \,1n{x_\i} }$
defining $x_i \sigma' = x_i \sigma \sigma'$.
\qed
}

\THEOREM{Clausal Determinate Definitions}{ \LABEL{det lit rem 5}
We use the abbreviation
$D 
= \set{ \lnot p_i(s,t) \allowbreak
\mid \allowbreak
s,t \in \T{}{}{\set{}} 
\allowbreak \land p_i \mbox{ determinate}
\allowbreak \land B' \impl p_i(s,t) }$.
Let two ground
Horn clauses $C_1$ and $C_2$ be given,
such that 
each body literal of each $C_i$ is entailed by $B'$ and is not an
element of $D$.
We can compute a regular tree language
$H = {\it lgg}^c_E(C_1,C_2)$
such that:
\begin{itemize}
\item each member
	$C \in H$
	is a 
	constrained clause that $E$-subsumes
	$C_1$ and $C_2$;
\item and for each semi-determinate
	clause
	$C \subseteq {\it lgg}(C_1 \cup C'_1, C_2 \cup C'_2)$
	with $C'_1,C'_2 \subseteq D$,
	~
	$\dlr{C}$ subsumes some member of $H$.
\end{itemize}
}
\PROOF{
For $i=1,2$,
let $p_0(s_{0i},t_{0i})$ be the head literal of $C_i$.
Let
$M$ be the set of all pairs
$\tpl{L_1,L_2}$
of body literals $L_1$ from $C_1$ and $L_2$ from $C_2$ such that
$L_1$ fits $L_2$.
Assuming
$M = \set{
\tpl{q_j(t_{j1}),q_j(t_{j2})} \mid j=\,1k\i }$,
define
$T^+ = \set{
\tpl{ s_{01} , \tpl{ t_{01}, \,1k{t_{\i1}} } } ,
\tpl{ s_{02} , \tpl{ t_{02}, \,1k{t_{\i2}} } } }$
and
$T^- = \set{}$.
Define
$H = \set{ p_0(s_0,t_0) \la \:{\land}1k{q_\i(t_\i)}
\mid \tpl{ s_0 , \tpl{ t_0, \,1k{t_\i} } }
\in \Hyp{\ref{hyp r}}{T^+}{T^-} } $.

$H$ is again a regular tree language
because
$\Hyp{\ref{hyp r}}{T^+}{T^-}$ is
the image of $H$ under the tree homomorphism that maps
the term
$p_0(x_0,y_0) \la \:{\land}1k{q_\i(y_\i)}$
to $\tpl{ x_0 ,\tpl{ y_0, \,1k{y_\i} } }$.
Since $T^- = \set{}$,
the set $\Hyp{\ref{hyp r}}{T^+}{T^-}$ 
contains at least one element
$\tpl{s_0,\tpl{t'_0, \,1k{t'_\i}}}$,
and the left component of an element of $\Hyp{\ref{hyp r}}{T^+}{T^-}$
is always $s_0$.
\begin{itemize}
\item For each clause
	$p_0(s_0,t_0) \la \:{\land}1k{q_\i(t_\i)}$ in $H$,
	we have $\var(s_0) \supseteq \var(t_0, \,1k{t_\i})$
	and $\hypp{ s_0 , \tpl{ t_0, \,1k{t_\i}} }{T^+,T^-}$
	by Thm.~\ref{hyp r}.
	Hence, 
	$\set{ p_0(s_0,t_0),
	\,1k{\lnot q_\i(t_\i)} } \; \hsigma_i
	\subseteq_E C_i$.
\item Assume 
	$(p_0(s_0,t_0)
	\la \:{\land}1n{p_\i(s_\i,x_\i)}
	\land \:{\land}1m{q_\i(t_\i)})
	\subseteq {\it lgg}(C_1 \cup C'_1,C_2 \cup C'_2)$
	is a semi-determinate clause.
	Then,
	$(\lnot q_j(t_j \sigma_i)) \in C_i$ for some $\sigma_i$
	---
	we assume w.l.o.g.\
	$t_j \sigma_i \eq t_{ji}$.
	Moreover,
	$\lnot p_j(s_j \sigma_i,x_j \sigma_i)$ is a member of 
	$C'_i \subseteq D$,
	implying that
	$x_j \sigma_i \eq g_j(s_j \sigma_i) = x_j \sigma \sigma_i$
	is entailed by $B'$,
	where $\sigma$ denotes the substitution from 
	$\dlr{C}$ computation by
	Lem.~\ref{det lit rem 2}.
	Since $\dom \sigma = \set{ \,1n{x_\i} }$,
	we have
	$x \sigma \sigma_i \eq x \sigma_i$
	for all variables $x$.
	Therefore,
	$t_j \sigma \sigma_i \eq t_j \sigma_i \eq t_{ji}$,
	and $s_0 \sigma \sigma_i = s_0 \sigma_i = s_{0i}$
	because $\var(s_0)$ is disjoint from
	the domain of $\sigma$.
	%
	Hence,
	we can extend the clause
	$\dlr{C} 
	= \set{p_0(s_0 \sigma,t_0 \sigma),
	\,1m{\lnot q_\i(t_\i \sigma)}}$
	to some superset
	$\set{p_0(s_0 \sigma,t_0 \sigma),
	\,1m{\lnot q_\i(t_\i \sigma)} , \,{m+1}k{\lnot q_\i(t'_\i)} }$
	that satisfies
	$\hypp{}{}$
	and 
	$\var(s_0 \sigma) \supseteq \var(t_j \sigma) \cup \var(t'_{j'})$
	and is therefore a member of~$H$.
\end{itemize}
To compute ${\it lgg}^c_E$,
the grammar defining $\eqc{t_{ji}}{E}{}$ must be extended by two rules
to define $\eqc{\tpl{s_{0i},\tpl{t_{0i}, \,1k{t_{\i i}}}}}{E}{}$
as well.
One intersection of the two extended grammars is needed;
no universal substitution needs to be computed.
\qed
}

The form of Thm.~\ref{det lit rem 5} differs from that of
Thm.~\ref{clauses eau} because neither $C$ nor $\dlr{C}$ need
$E$-subsume the other.
To establish some similarity between the second assertion of the two
theorems,
note that a subsumed clause defining a predicate leads to a more
specific definition that its subsuming clause.
Let $C'$ subsume $C_1$ and $C_2$ and contain a nontrivial head
literal $p_0(\ldots)$.
Then $C'$ also subsumes $C = {\it lgg}(C_1,C_2)$.
By Thm.~\ref{det lit rem 5}, $\dlr{C}$ subsumes
some member of ${\it lgg}^c_E(C_1,C_2)$.
That member thus leads to a more specific definition of
$p_0$ than $C'$.

In order to duplicate a most flexible ${\it lgg}$ approach,
Thm.~\ref{det lit rem 5} allows
a literal pre- and postselection strategy,
to be applied before and after ${\it lgg}$ computation,
respectively.
Both may serve to eliminate undesirable body literals from the 
${\it lgg}$ result clause.
Preselection can be modeled using the $C_i$ and $C'_i$,
while postselection is enabled by choosing
$C \subsetneq {\it lgg}(C_1 \cup C'_1,C_2 \cup C'_2)$.
In all cases,
Thm.~\ref{det lit rem 5} 
provides a corresponding constrained clause
from ${\it lgg}^c_E(C_1,C_2)$,
which is equivalent to, or more specific than, $C$.

Similar to Thm.~\ref{clauses eau},
each $C \in {\it lgg}^c_E(C_1,C_2)$ is sufficient wrt.\
$F^+ \Lra C_1 \land C_2$ and $F^- \Lra {\it true}$.
Each such
$C$ is consistent if some predicate symbol $q$ occurs in both $C_1$
and $C_2$, but not in $B$, except for its congruence axiom.

Again similar to the nondeterminate case, 
$\Hyp{\ref{appl r3}}
{p_0(s_{01},t_{01}) \land p_0(s_{02},t_{02})}{\it true}$
equals
${\it lgg}^c_E(\set{p_0(s_{01},t_{01})},\set{p_0(s_{02},t_{02})})$
from Thm.~\ref{det lit rem 5}.
In this common special case,
Thm.~\ref{appl r3}, but not Thm.~\ref{det lit rem 5},
ensures that the result set contains \emph{all} sufficient hypotheses.

On the other hand,
Thm.~\ref{det lit rem 5} ensures that for
each purely determinate clause,
i.e.\ a clause without any nondeterminate $q_i$ in its body,
${\it lgg}^c_E(C_1,C_2)$ contains a clause leading to an equivalent, or
more specific, definition of $p_0$.
In other words,
${\it lgg}$-learning of purely determinate clauses can be
simulated by ${\it lgg}^c_E$-learning of atoms.

\begin{figure}
\begin{center}
$$\begin{array}[t]{@{}l|l||l|l@{}}
\cline{2-3}
F & \multicolumn{2}{|c|}{p_0(b,bbb) \;\land\; p_0(\eps,b)}	\\
\cline{2-3}
P
	& a(\eps,y,y)
	& a(\eps,y) = y \hspace*{1cm} a(x,\eps) = x
	& E	\\
& a([v \mid x],y,[v \mid z]) \la a(x,y,z)
	& a(a(x,y),z) = a(x,a(y,z))	\\
\cline{2-3}
K_q
	& \multicolumn{2}{|c|}{%
		q(\eps,d) \;\land\;
		q(b,d) \;\land\;
		q(c,e) \;\land\;
		q(bb,d) \;\land\;
		q(bc,e) \;\land\;
		q(bcb,e)}	\\
\cline{2-3}
	K_a
&
	\begin{array}[t]{@{}l@{}l@{}l@{}l@{}l@{}}
	a(\eps,\eps,\eps)
		& \land	\\
	a(\eps,b,b)
		& \land
		& a(b,\eps,b)
			& \land	\\
	a(\eps,bb,bb)
		& \land
		& a(b,b,bb)
		& \land
		& \ldots	\\
	a(\eps,bbb,bbb)
		& \land
		& a(b,bb,bbb)
		& \land
		& \ldots	\\
	\end{array}
&
	\begin{array}[t]{@{}rlll@{}l@{}}
	N_\eps & \sortdef \eps \mid & a(N_\eps,N_\eps)	\\
	N_b & \sortdef b \mid & a(N_\eps,N_b) & \mid a(N_b,N_\eps) \\
	N_{bb} & \sortdef & a(N_\eps,N_{bb}) & \mid a(N_b,N_b)
		& \ldots	\\
	N_{bbb} & \sortdef & a(N_\eps,N_{bbb})
		& \mid a(N_b,N_{bb})
		& \ldots \\
	\end{array}
&
	\G
\\
\cline{2-3}
	{\it lgg}
&
	\begin{array}[t]{@{}lrl@{}}
	\multicolumn{2}{l}{p_0(v_{b,\eps},v_{bbb,b})}	\\
	& \la & a(v_{b,\eps},v_{b,\eps},v_{bb,\eps})	\\
	& \land & a(v_{bb,\eps},b,v_{bbb,b})	\\
	& \land & q(v_{bb,\eps},d)	\\
	\end{array}
&
	\begin{array}[t]{@{}l@{}}
	\multicolumn{1}{@{}l@{}}{%
		p_0(v_{b,\eps},a(a(v_{b,\eps},v_{b,\eps}),b))} 	\\
	\\
	\\
	\multicolumn{1}{@{}r@{}}{\la q(a(v_{b,\eps},v_{b,\eps}),d)} \\
	\end{array}
&
	{\it lgg}_E^c
\\
\cline{2-3}
\end{array}$$
\caption{Comparison of ILP Using ${\it lgg}$ and ${\it lgg}^c_E$}
\LABEL{append fig}
\end{center}
\end{figure}

Let us compare the ILP methods using ${\it lgg}$ and
${\it lgg}^c_E$ in an example.
Assume part of the background knowledge describes lists
with an associative
append operator and a neutral element $\eps$ (nil).
The topmost two lines of Fig.~\ref{append fig}
show a Horn program $P$ and an equational theory $E$,
each of which formalizes that knowledge,
where $v,x,y,z$ denote variables, $a$ denotes \emph{append} and
$b,c,d,e$ below will denote some constants.

Moreover, let a conjunction $K_q$
of 
facts about a predicate $q$ be given,
as shown in the third line of Fig.~\ref{append fig}.
We abbreviated, e.g., $q([b,c,b],[e])$ to $q(bcb,e)$.
Let $F \Lra p_0(b,bbb) \land p_0(\eps,b)$ be given.
Let us assume for now that a preselection strategy chose
$K'_q \Lra q(\eps,d) \land q(bb,d)$.

Neither ${\it lgg}$ nor ${\it lgg}^c_E$ can use the first part of
background knowledge directly.
Most ILP systems, including {\sc Golem},
restrict background theories to sets of ground literals.
Hence, they cannot directly use equality as background knowledge
because
this would require formulas like 
$p_0(x,y) \la eq(y,y') \land p_0(x,y')$ in the
background theory.
While Plotkin's ${\it lgg}$ is also defined for nonground clauses,
it has not been defined for clause sets like $P$.
Moreover,
since $F$ contains only ground literals, all relevant arguments of body
literals must be ground to obtain the necessary variable bindings in the
generalized clause.
For example, a clause like
$$\renewcommand{\arraystretch}{1}
\begin{array}{l l@{}c@{}@{}c l@{}c@{}c@{}c@{} l@{}c@{}c@{}c@{} lll}
& (p_0( & v_{b,\eps}, & v_{bcb,c}
        & ) \la a( & v_{x,\eps}, & c, & v_{xc,c}
        & ) \land a( & v_{xy,x}, & v_{x,\eps}, & v_{xyx,x}
        & ) & \land \ldots ) \\
=
{\it lgg}( & (p_0( & b, & bcb
        & ) \la a( & x, & c, & xc
        & ) \land a( & xy, & x, & xyx
        & ) & \land \ldots ),   \\
& (p_0( & \eps, & c
        & ) \la a( & \eps, & c, & c
        & ) \land a( & x, & \eps, & x
        & ) & \land \ldots ) & ) \\
\end{array}$$
would lack bindings
like $v_{b,\eps} = v_{x,\eps}$ and $v_{bcb,c} = v_{xyx,x}$.
Thus, we have to derive the conjunction $K_a$
of all ground facts implied by
$P$ that could be relevant in any respect.

On the other hand, 
$E$ has to be transformed into a grammar $\G$
in order to compute ${\it lgg}^c_E$.
We can do this by
Thm.~\ref{eqc 2} with $(\prec)$ as the lexicographic path
ordering, which is commonly used to prove termination of the rewrite
system associated with $E$
\cite[Sect.~5.3]{Dershowitz.Jouannaud.1990}.
Alternatively, we
could instantiate a predefined grammar scheme like
$$N_{x_1\ldots x_n}
\sortdef \bigmid_{n=0} \; \eps \mid \bigmid_{n=1} \; x_1
\mid \bigmid_{i=0}^n \; a(N_{x_1 \ldots x_i},N_{x_{i+1} \ldots x_n}) .$$
At least
we do not have to rack our brains over the question 
of which terms might be
relevant --- it is sufficient to define the congruence classes of all
terms occurring in $F$ or $K'_q$.

Lines~4 to~8 of Fig.~\ref{append fig} show the preprocessed form
$K_a$ and $\G$ of $P$ and $E$, respectively.
Observe that a ground literal
$a(s,t,u)$ in the left column corresponds to a grammar alternative
$N_u \sortdef \ldots a(N_s,N_t)$ in the right one.
It is plausible to assume that there are
at least as many literals in
$K_a$ as there are alternatives in $\G$.
Next, we compute
$$\renewcommand{\arraystretch}{1}
\begin{array}{rlclllcllll}
{\it lgg}(
	& (p_0(b,bbb) \la
	& K_a
	& \land
	& K'_q),
	& (p_0(\eps,b) \la
	& K_a
	& \land
	& K'_q)
	& )
	& \mbox{ and}	\\
{\it lgg}^c_E(
	& (p_0(b,bbb) \la
	&
	&
	& K'_q),
	& (p_0(\eps,b) \la
	&
	&
	& K'_q)
	& )
	&	\\
\end{array}$$
and apply some literal postselection strategy.
A sample result is shown in the bottom part of Fig.~\ref{append fig}.
More precisely,
${\it lgg}^c_E$ results in the set of all terms
$p_0(v_{b,\eps},t_{bbb,b}) \la q(t_{bb,\eps},d)$
for any
$$\renewcommand{\arraystretch}{1}
\begin{array}{cccccl}
t_{bbb,b} 
	& \in 
	& \eqc{bbb}{E}{\subst{v_{b,\eps} \mapsto b}}
	& \cap 
	& \eqc{b}{E}{\subst{v_{b,\eps} \mapsto \eps}}
	& \mbox{ and}	\\
t_{bb,\eps} 
	& \in 
	& \eqc{bb}{E}{\subst{v_{b,\eps} \mapsto b}}
	& \cap 
	& \eqc{\eps}{E}{\subst{v_{b,\eps} \mapsto \eps}}
	& \mbox{ .}	\\
\end{array}$$
The choice of $t_{bbb,b}$ and $t_{bb,\eps}$
on the right-hand side in Fig.~\ref{append fig}
corresponds to the choice of body literals about $a$ on the left;
both sides are equivalent definitions of $p_0$.
If the post\-selection strategy chooses
$a(v_{b,\eps},b,v_{bb,b})
\land a(v_{bb,b},v_{b,\eps},v_{bbb,b})
\land a(v_{b,\eps},v_{b,\eps},v_{bb,\eps})
$
on the left,
we need only to choose
$t_{bbb,b} = a(a(v_{b,\eps},b),v_{b,\eps})$
on the right
to duplicate that result.
However,
if preselection chooses different literals about $q$,
e.g.\
$K'_q \Lra q(bc,e) \land q(c,e)$,
we have to recompute the grammar $\G$ to include definitions for
$\eqc{bc}{E}{}$ and $\eqc{c}{E}{}$.

The ${\it lgg}^c_E$ result
clause is always a constrained one, whereas ${\it lgg}$ yields a
determinate clause.
The reason for the latter
is that function calls have to be simulated by predicate
calls, requiring extra variables
for intermediate results.
The deeper a term in the ${\it lgg}^c_E$ clause
is nested, the longer the
extra variable chains are in the corresponding ${\it lgg}$ clause.
If $K'_q \Lra {\it true}$ is chosen,
${\it lgg}^c_E$ yields an atom rather than a proper clause.

When the ${\it lgg}^c_E$ approach is used,
the hypotheses search space is split.
Literal pre- and postselection strategies need to handle nondeterminate
predicates
only.
The preselected literals, i.e.\ $K'_q$, control the size and form of the
grammar $\G$.
Choices of, e.g., $t_{bbb,b} \in \L_\G(N_{bbb,b})$
can be made independently of pre- and postselection,
each choice leading to the condensed equivalent of a
semi-determinate clause.

Filtering of, e.g., $\L_\G(N_{bbb,b})$ allows us to ensure additional
properties of the result clause if they can be expressed
by regular tree languages.
For example, orienting each equation
from $E$ in Fig.~\ref{append fig}
left to right generates a canonical term-rewriting system $R$.
Since all terms in $E$ are linear, a grammar $\G_{\it NF}$
for the set of normal forms
wrt.\ $R$ can be obtained automatically from~$E$.
Choosing, e.g.,
$t_{bbb,b} \in \L_\G(N_{bbb,b}) \cap \L_{\G_{\it NF}}(N_{\it NF})$
ensures that no redundant clause like
$p_0(v_{b,\eps},a(v_{b,\eps},a(v_{b,\eps},b))
\la q(a(v_{b,\eps},v_{b,\eps}),d)$
can result.
In classical ILP, there is no corresponding filtering method
of similar simplicity.

\section{Applications}
\LABEL{Applications}

In this section, we apply $E$-generalization in three different
application areas.
In all cases, we use the paradigm of learning a determinate atomic
definition from positive examples only.
We intend to demonstrate that the notion of $E$-generalization can
help to solve even comparatively
ambitious tasks in Artificial Intelligence at the first attempt.
We make no claim to develop a single application to full maturity.
Instead, we cover a variety of different areas
in order to illustrate the flexibility of $E$-generalization.

\subsection{Candidate Lemmas in Inductive Proofs}
\LABEL{Candidate lemmas in inductive proofs}

Auxiliary lemmas play an important role
in automated theorem proving.
Even in pure first-order logic, where lemmas are not strictly
necessary
\cite{Gentzen.1932}, proofs may become exponentially longer
without them and are consequently harder to find.
In induction proofs, which exceed first-order logic owing to the
induction axiom(s), using lemmas may be unavoidable to 
demonstrate a certain theorem.

\begin{figure}
\begin{center}
\begin{tabular}[t]{@{}rl@{$\;$}ll@{}}
Claim: & \multicolumn{3}{l}{$(x*y)*z \equiv_E x*(y*z)$}	\\
Proof: & \multicolumn{3}{l}{Induction on $z$:}	\\
$z\.=0$: && $(x*y)*0$	\\
	& $\eq$ & $0$ & by 3.	\\
	& $\eq$ & $x*0$ & by 3.	\\
	& $\eq$ & $x*(y*0)$ & by 3.	\\
$z\.=\suc(z')$: && $(x*y)*\suc(z')$	\\
	& $\eq$ & $(x*y)*z'+x*y$ & by 4.	\\
\cline{3-3}
	& $\equiv_E$ & \multicolumn{1}{|l|}{$x*(y*z')+x*y$}
	& by I.H. \\
\cline{3-3}
	& $\equiv_E$ & $x*(y*z'+y)$ 
	& by {\large\bf ?\hspace{-0.44em}?\hspace{-0.44em}?}	\\
	& $\eq$ & $x*(y*\suc(z'))$ & by 4.	\\
\end{tabular}
\end{center}
\caption{Induction Proof Using 
	Fig.~\ref{$E$-generalization of $0$ and $\suc^4(0)$} (Left)}
\LABEL{Induction proof}
\end{figure}

By way of a simple example, consider the induction
proof in Fig.~\ref{Induction proof},
which uses the equational theory from Fig.~\ref{$E$-generalization of
$0$ and $\suc^4(0)$} (left).
At the position marked ``{\bf ?}'', the distributivity law is needed
as a lemma in order to continue the proof.
While this is obvious to a mathematically experienced reader, an
automated prover that does not yet know the law will get stuck
at this point and require a user interaction because the actual
term cannot be rewritten any further.
In this simple example, where only one lemma is required, the
\emph{cross-fertilization} technique of \cite{Boyer.Moore.1979}
would suffice to generate it automatically.
However, this technique generally fails if several lemmas are
needed.

In such cases, we try to simulate mathematical \emph{intuition}
by $E$-generalization in order to find a useful lemma
and allow the prover to continue;
i.e.\ to increase its level of automation.
We consider the last term $t_1$ obtained in the proof so
far (the surrounded term $x*(y*z')+x*y$ in the example)
and try to find a new lemma that could be applied next by the prover.
We are looking for a 
lemma of the form $t_1 \equiv_E t_2$ for some $t_2$
such that
$\forall \sigma \mbox{ ground}: \; t_1 \sigma \eq t_2 \sigma$
holds.
Using Thm.~\ref{appl r3}, we are able to compute the set of
all terms $t_2$ such that
$t_1 \sigma \eq t_2 \sigma$ holds
at least for finitely many given~$\sigma$.

We therefore choose some ground substitutions $\,1n{\sigma_\i}$
with $\var(t_1) = \set{x,y,z'}$ as their domain, and let
$F^+
\Lra \bigwedge_{i=1}^n
p(\tpl{x \sigma_i,y \sigma_i,z' \sigma_i},t_1 \sigma_i)$.
We then apply Thm.~\ref{appl r3} 
to this $F^+$ and $F^- \Lra {\it true}$.
(See Fig.~\ref{Lemma Generation}, 
where the partial congruence property of $p$
was used to simplify the examples in $F^+$.)
Using the notation from Lem.~\ref{hyp r},
we have just one $I$ in $\I$, viz.\ $I = \set{\,1n\i}$,
since we do not supply negative examples.
Therefore, we only have to compute
$T_I = \bigcap_{i=1}^n \eqc{t_1 \sigma_i}{E}{\sigma_i}$.

\begin{figure}
$$\begin{array}[t]{@{}lcc@{\;} *{18}{@{}c}@{}}
t_1 & = &&&&&&&&&& 
	x & *( & y & * & z' & ) + 
	& \multicolumn{2}{c}{x * y} 	\\
\hline
&&& p( \tpl{ & x \sigma_i & , & y \sigma_i & , & z' \sigma_i
	& }, & ( 
	& x & *( & y & * & z' & ) + 
	& \multicolumn{2}{c}{x * y}
	& ) \; \sigma_i & )	\\
\hline
F^+ & \Lra &
	& p(\tpl{  & 0 & , & \suc^3(0) & , & \suc^2(0)
	& }, & \multicolumn{9}{c}{0} && )	\\
&& \land & p(\tpl{  & \suc^2(0) & , & \suc(0) & , & 0
	& }, & \multicolumn{9}{c}{\suc^2(0)} && )	\\
&& \land & p(\tpl{  & \suc(0) & , & 0 & , & \suc(0)
	& }, & \multicolumn{9}{c}{0} && )	\\
\hline
H & \ni &
	& p(\tpl{  & v_{021} & , & v_{310} & , & v_{201}
	& },
	& 
	& v_{021} & *( & v_{310} & * & v_{201} & +& v_{310} & )
	&& ) \\
\hline
t_2 & = &&&&&&&&
	&& x & *( & y & * & z' & + & y & ) \\
\end{array}$$
\caption{Generation of Lemma Candidates by Thm.~\ref{appl r3}}
\LABEL{Lemma Generation}
\end{figure}

The most specific syntactical generalization $s_I$ of
$\set{ \,1n{\tpl{x,y,z'} \sigma_\i} }$
need not be $\tpl{x,y,z'}$ again.
However,
we always have $\tpl{x,y,z'} \sigma' = s_I$
for some~$\sigma'$.
If we choose $\sigma_i$
that are
\emph{sufficiently different},
we can ensure that $\sigma'$ has an inverse.
This is the case in Fig.~\ref{Lemma Generation},
where $\sigma'
= \subst{{x \mapsto v_{021},} \allowbreak \;
{y \mapsto v_{310},} \allowbreak \;
{z' \mapsto v_{201}}}$.
By Thm.~\ref{appl r3},
$B' \land p(\tpl{x,y,z'},t_I \sigma'^{-1})$
implies 
$p(\tpl{x,y,z'} \sigma_i,t_1 \sigma_i)$
for each $t_I \in T_I$ and $i=\,1n\i$.
Since it trivially also implies
$p(\tpl{x,y,z'} \sigma_i,t_I \sigma'^{-1} \sigma_i)$,
we obtain
$t_1 \sigma_i \eq t_I \sigma'^{-1} \sigma_i$
using determinacy.

Therefore, defining $t_2 = t_I \sigma'^{-1}$
ensures that $t_1$ and $t_2$ have the same value
under each sample substitution $\sigma_i$.
This is a necessary condition for $t_1 \equiv_E t_2$, but not
sufficient.
Before using a lemma suggestion $t_1 \equiv_E t_2$ to
continue the original proof, it must be checked
for validity by a recursive call to the induction prover itself.
Two simple restrictions can help to eliminate unsuccessful hypotheses:
\begin{itemize}
\item	Usually, only those equations $t_1 \equiv_E t_2$
	are desired that satisfy
	$\var(t_2) \subseteq \var(t_1)$.
	For example,
	it is obvious that
	$x*(y*z')+x*y \equiv_E x+v_{123}$
	is not universally valid.
	This restriction of the result set is already built into
	Thm.~\ref{appl r3}.
	Any lemma contradicting this restriction will not appear in the
	grammar language.
	However, all its instances that satisfy the restriction will
	appear.
\item	Moreover, if $E$ was given by a ground-convergent 
	term-rewriting system $R$
	\cite[Sect.~2.4]{Dershowitz.Jouannaud.1990},
	it makes sense to require $t_2$ to be in normal form
	wrt.\ $R$.
	For example,
	$x*(y*z')+x*y \equiv_E (x+0)*(y*z'+y*\suc(0))$
	is a valid lemma, but redundant, compared with
	$x*(y*z')+x*y \equiv_E x*(y*z'+y)$.
	The closed representation of the set $T_I$
	as a regular tree language
	allows us to easily eliminate such undesired terms $t_2$.

	For left-linear term-rewriting systems
	\cite[Sect.~2.3]{Dershowitz.Jouannaud.1990},
	the set of all normal-form terms is always representable as
	a regular tree language; hence terms in non-normal form can
	be eliminated by intersection.
	For rewriting systems that are not left-linear,
	we may still filter out a subset of all non-normal-form terms.

	If desired by some application, $T_I$ could also
	be restricted to those terms $t_I$ that satisfy
	$V' \subseteq \var(t_I) \subseteq V$
	for arbitrarily given variable sets $V',V$.
\end{itemize}

The more sample instances are used, the more
of the enumerated lemma candidates will be valid.
However, our method does not lead to \emph{learnability in the limit}
\cite{Gold.1967} because normally any result language will still
contain invalid equations --- regardless of the number of sample
substitutions.
It does not even lead to \emph{PAC-learnability} \cite{Valiant.1984},
there currently being no way to compute
the number of sample
substitutions depending on the required $\delta$ and $\epsilon$
accuracies.

\newcommand{\maxInfix}[2]{#1 \mathop{\uparrow} #2}
\newcommand{\minInfix}[2]{#1 \mathop{\downarrow} #2}

\begin{figure}
\begin{center}
\begin{tabular}{@{}lr@{$\;=\;$}llrr@{}}
Theory & \multicolumn{2}{c}{Lemma} & Rhs & No & Time	\\
\hline
$+$,$*$ & $(x+y)+z$ & $x+(y+z)$	& 1,1,3 & 6. & 21	\\
$+$,$*$ & $x*(y+z)$ & $x*y+x*z$ & 0,2,2 & 10. & 17	\\
$+$,$*$ & $x*y$ & $y*x$ & 0,0 & 3. & 0	\\
$+$,$*$ & $(x*y)*z$ & $x*(y*z)$	& 0,0,2 & 31. & 7	\\
$+$,$*$ & $(x*y)*z$ & $x*(y*z)$	& 0,0,2,4 & 3. & 22	\\
\hline
$+$,$-$,$*$,$/$,$\idiv$,$\mod$ & $x/z+y/z$ & $(x+y)/z$ 
	& 5,1,3 & 2. & 263324	\\
$+$,$-$,$*$,$/$,$\idiv$,$\mod$
	& $((x \mod z)\.+(y \mod z)) \mod z$ & $(x+y) \mod z$ 
	& 0,1,1 & 1. & 19206	\\
$+$,$-$,$*$,$/$,$\idiv$,$\mod$
	& $(x \idiv y) * y$ & $x - (x \mod y)$ 
	& 6,0,3 & 1. & 17304	\\
$+$,$-$,$*$,$/$,$\idiv$,$\mod$ & $x$ & $(x * y) \idiv y$
	& 2,1,3 & 4. & 17014 	\\
\hline
$+$,$*$,$<$
	& \multicolumn{1}{r@{$\;\Lra\;$}}{$x<y$}
	& $x*z < y*z$
	& 0,1,1 & 3. & 174958	\\
$+$,$*$,$<$
	& \multicolumn{1}{r@{$\;\Lra\;$}}{$x<y$}
	& $x+z < y+z$
	& 0,1,1 & 20. & 174958	\\
\hline
$+$,$*$,$<$,$\maxInfix{}{}$,$\minInfix{}{}$
	& \multicolumn{1}{r@{$\;\Lra\;$}}{$x < y$}
	& $x < \maxInfix{x}{y}$
	& 0,1,1 & 6. & 47128	\\
$+$,$*$,$<$,$\maxInfix{}{}$,$\minInfix{}{}$
	& $x$
	& $\minInfix{x}{(\maxInfix{x}{y})}$
	& 3,0,3 & 7. & 45678	\\
$+$,$*$,$<$,$\maxInfix{}{}$,$\minInfix{}{}$
	& $(\maxInfix{x}{y})+(\minInfix{x}{y})$
	& $x+y$
	& 5,1,5 & 2. & 42670	\\
\hline
$+$,$*$,${\it dp}$ & ${\it dp}(x)+{\it dp}(y)$ & ${\it dp}(x+y)$
	& 2,4 & 2. & 6	\\
$+$,$*$,${\it dp}$ & ${\it dp}(x)$ & $x+x$ & 0,4 & 4. & 1	\\
$+$,$*$,${\it dp}$ & $x*{\it dp}(y)$ & ${\it dp}(x*y)$
	& 0,0 & 13. & 1	\\
\hline
$\lnot$,$\land$,$\lor$,$\ra$
	& \multicolumn{1}{r@{$\;\Lra\;$}}{$\lnot (x \land y)$}
	& $y \ra \lnot x$
	& 1,1,1,0 & 1. & 308	\\
$\lnot$,$\land$,$\lor$,$\ra$
	& \multicolumn{1}{r@{$\;\Lra\;$}}{$\lnot (x \land y)$}
	& $\lnot x \lor \lnot y$
	& 1,1,1,0 & 6. & 308	\\
\hline
${\it ap}$,${\it rv}$ & ${\it ap}({\it rv}(x),{\it rv}(y))$
	& ${\it rv}({\it ap}(y,x))$ & 2,2,2 & 1. & 89 \\
${\it ap}$,${\it rv}$
	& ${\it ap}(x,{\it ap}(y,z))$
	& ${\it ap}({\it ap}(x,y),z)$
	& 3,3,2 & 1. & 296 \\
${\it ap}$,${\it rv}$ & $x$ & ${\it rv}({\it rv}(x))$
	& 0,2 & 4. & 1	\\
${\it ap}$,${\it rv}$ & ${\it rv}({\it rv}(x))$ & $x$
	& 0,2 & 1. & 1	\\
\hline
${\it ap}$,${\it rv}$,${\it ln}$
	& ${\it ln}({\it ap}(x,y))$ & ${\it ln}(x)+{\it ln}(y)$
	& 1,2 & 4. & 4	\\
${\it ap}$,${\it rv}$,${\it ln}$
	& ${\it ln}({\it cons}(x,{\it ap}(y,z)))$
	& $\suc({\it ln}(y)\.+{\it ln}(z))$
	& 2,3 & 10. & 21	\\
\hline
${\it cube}$ & ${\it lf}({\it cc}(x))$ & ${\it up}({\it lf}(x))$
	& 1,1 & 1. & 18	\\
${\it cube}$ & ${\it lf}({\it cc}(x))$ & ${\it cc}({\it up}(x))$
	& 1,1 & 2. & 	\\
\hline
\end{tabular}
\end{center}
\caption{Generated Lemma Candidates}
\LABEL{Generated lemma candidates}
\end{figure}

In the example from Fig.~\ref{Induction proof},
we get, among other equations, the lemma suggestion
$x*(y*z')+x*y \equiv_E x*(y*z'+y)$,
which allows the prover to continue successfully.
This suggestion appears among the first ten,
if $T_I$ is enumerated by increasing term size.
Most of the earlier terms are variants wrt.\ commutativity,
like $x*(y*z')+x*y \equiv_E (y+y*z')*x$.

Figure~\ref{Generated lemma candidates} shows some examples of
lemma candidates generated by our prototypical implementation
(see Sect.~\ref{Prototype implementation}).
The column \emph{Theory} shows the equational theory used.
We distinguish between the truncating integer division $(\idiv)$ and the
true division $(/)$.
For example, $7 \idiv 3 \eq 2$, while $7 / 3$ is not defined.
The grammar rules that realize these partial functions 
are something like
$N_2 \sortdef \ldots \mid N_6 \idiv N_3 \mid N_7 \idiv N_3
\ldots \mid N_6 / N_3$.
The integer remainder is denoted by $(\mod)$.

We embedded the two-element Boolean algebra $\set{0,1}$ into the
natural numbers, with $1$ corresponding to ${\it true}$.
This allows us to model relations like $(<)$ and logical junctors.
The functions $(\maxInfix{}{})$ and $(\minInfix{}{})$
compute the maximum and minimum of two numbers, respectively.
The function ${\it dp}$
doubles a natural
number in $0$-$\suc$ representation,
${\it ap}$
concatenates two lists in
${\it cons}$-${\it nil}$
representation,
${\it rv}$
reverses a list,
and ${\it ln}$
computes its length as a natural number.
The ${\it cube}$ theory formalizes the six possible three-dimensional
$90^\circ$-degree rotations of a cube, viz.\
\emph{left},
\emph{right},
\emph{up},
\emph{down},
\emph{clockwise} and
\emph{counter-clockwise},
as shown in
Fig.~\ref{Computation of a Construction Law} (right).

The 
column \emph{Lemma} shows the corresponding lemma, its right-hand side
having been supplied, its left-hand side generated by the above method.
Note, for example, the difference between the lemmas
$x = {\it rv}({\it rv}(x))$ and ${\it rv}({\it rv}(x)) = x$.
The 
column \emph{Rhs} indicates the size of $t_1 \sigma_i$ for $i=\,1n\i$,
which is a measure of the size of grammars to be intersected.
For arithmetic and list theories,
the value of each number $t_1 \sigma_i$ and the length of each list $t_1
\sigma_i$ is given, respectively.

The 
column \emph{No} shows the place in which the lemma's right-hand side
appeared in the enumeration sequence,
while the column \emph{Time} shows the required runtime in milliseconds
(compiled {\sc Prolog} on a 933 MHz PC).
Both depend strongly on the number and size of the example ground
instances.
The dependence of \emph{No} can be seen in lines~4 and~5.

The runtime also depends on the grammar connectivity.
In a grammar that includes, e.g., $(-)$ or $(<)$,
each nonterminal can be reached from any other,
while in a grammar considering, e.g., only $(+)$ and $(*)$,
only nonterminals for smaller values and $N_t$ can be reached.
If the grammar defines $N_t, N_0, \ldots, N_6$,
computing
$\eqc{0}{E}{\sigma_1}
\cap \eqc{1}{E}{\sigma_2} 
\cap \eqc{1}{E}{\sigma_3}$
leads to $8^3$
intersection nonterminals in the former case,
compared with $2 \cdot 3^2$ in the latter.
For this reason, runtimes are essentially independent of the \emph{Rhs}
sizes for the 2nd to 4th theory.
The exception in line 6 is due to a larger input grammar, which defined
nonterminals up to $N_{10}$ rather than $N_6$.

\begin{figure}
$\;$
\hfill
$\begin{array}[b]{@{}l@{}c*{5}{@{}c}*{2}{@{\;}c}@{}l@{}}
\multicolumn{10}{@{}l@{}}
	{\mbox{Given: series $0,1,4,9,\ldots$, and $k=3$}} \\
& \mbox{Lgth} && \multicolumn{3}{@{}c@{}}{\mbox{Suffix}}
	&&& \mbox{Next}	\\
\cline{2-2}
\cline{4-7}
\cline{9-9}
p( & \suc^3(0) & . & \suc^4(0) & . & \suc(0).0.nil
	&& , & \suc^9(0) & )	\\
p( & \suc^2(0) & . & \suc(0) & . & 0.nil
	&& , & \suc^4(0) & )	\\
p( & \suc(0) & . & 0 & . & nil
	&& , & \suc(0) & )	\\
\hline
p( & \suc(v_p) & . & v_1 & . & v_2
	&& , & \suc(v_p)\.*\suc(v_p) & )	\\
\end{array}$
\hfill
\hfill
\hfill
\hfill
\begin{picture}(3.3,3.3)
\put(0.900,1.100){\makebox(1.750,2.000){%
	\includegraphics{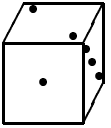}%
}}
\put(0.490,0.840){\line(1,0){0.360}}
	\put(0.490,0.840){\vector(0,1){0.360}}
	\put(0.490,0.840){\makebox(0.000,0.000)[tr]{$\it cl$}}
\put(2.400,0.840){\line(1,0){0.360}}
	\put(2.760,0.840){\vector(0,1){0.360}}
	\put(2.760,0.840){\makebox(0.000,0.000)[tl]{$\it cc$}}
\put(0.720,1.415){\line(-1,0){0.360}}
	\put(0.360,1.415){\vector(1,2){0.180}}
	\put(0.360,1.415){\makebox(0.000,0.000)[tr]{$\it lf$}}
\put(2.530,1.415){\line(1,0){0.360}}
	\put(2.890,1.415){\vector(1,2){0.180}}
	\put(2.890,1.415){\makebox(0.000,0.000)[tl]{$\it rg$}}
\put(0.900,0.740){\line(-1,-2){0.180}}
	\put(0.720,0.380){\vector(0,1){0.360}}
	\put(0.720,0.380){\makebox(0.000,0.000)[tl]{$\it up$}}
\put(0.900,2.990){\line(-1,-2){0.180}}
	\put(0.720,2.630){\vector(0,-1){0.360}}
	\put(0.720,2.630){\makebox(0.000,0.000)[br]{$\it dn$}}
\end{picture}
\hfill
$\;$
\caption{Law Computation by Thm.~\ref{appl r3}
	\hspace{1.5cm} Cube Rotations}
\LABEL{Computation of a Construction Law}
\end{figure}

\subsection{Construction Laws of Series}
\LABEL{Construction laws of series}

A second application of $E$-generalization consists in the
computation
of construction laws for term series, as in ordinary intelligence
tests.
The method is also based on Thm.~\ref{appl r3}
and is explained below.

For technical reasons, we write a series in reverse order as a
${\it cons}$-${\it nil}$ list, using an infix ``.'' for the reversed
${\it cons}$ to
enhance readability.
We consider suffixes of this list and append a number denoting its
length to each of them.
We use a binary predicate $p(l.s,n)$ to denote that the suffix $s$
of length $l$ leads to $n$ as the next series element.

We apply Thm.~\ref{appl r3}
to the $k$ last \emph{leads to}
relations obtained from the given series,
see Fig.~\ref{Computation of a Construction Law} (left),
where $k$ must be given by the user.
Each result has the form $p(l.s,n)$ and
corresponds to a rewrite rule $l.s \leadsto n$ that
computes the next term
from a series suffix and its length.
By construction, the rewrite rule is guaranteed to compute at least the
input terms correctly.
A notion of correctness is not formally defined for later terms, anyway.

\newcommand{\cube}[3]{%
	\raisebox{-0.2cm}{\includegraphics[scale=0.3]{#1#2#3}}%
}

\begin{figure}
\begin{center}
\begin{tabular}
	{@{}lll@{$\;$}rr@{}}
Theory & Series & Law & No & Time	\\
\hline
$+$,$*$ & $0;1,4,9$ & $v_p*v_p$ & 1. & 2797	\\
$+$,$*$ & $0;2,4,6$ & $\suc(\suc(v_1))$ & 1. & 3429	\\
$+$,$*$ & $0;2,4,6$ & $v_p+v_p$ & 3. & 3429  	\\
$+$,$*$ & $1,1;2,3,5$ & $v_1+v_2$ & 1. & 857 \\
\hline
$+$,$*$,${\it if}$,${\it ev}$
	& $0,1;2,1,4,1$ & ${\it if}({\it ev}(v_p),v_p,1)$
	& 13. & 13913 \\
$+$,$*$,${\it if}$,${\it ev}$
	& $0,0,1;1,0,0,1,1$ & ${\it ev}(v_2)$ & 1. & 61571	\\
$+$,$*$,${\it if}$,${\it ev}$
	& $0,0;1,0,0,1$ & ${\it ev}(v_1+v_2)$
	& 1. & 8573	\\
$+$,$*$,${\it if}$,${\it ev}$
	& $0;1,3,7$ & $\suc(v_1+v_1)$
	& 1. & 3714	\\
$+$,$*$,${\it if}$,${\it ev}$
	& $1,2,2,3,3,3,4;4,4,4$ & --- & & 8143	\\
\hline
${\it cube}$,${\it if}$,${\it ev}$ &
\cube 123, \cube 653; \cube 456, \cube 326, \cube 123
	& ${\it rg}({\it if}({\it ev}(v_p),v_1,\cube 624))$
	& 1. & 14713	\\
${\it cube}$,${\it if}$,${\it ev}$ &
\cube 123; \cube 246, \cube 312, \cube 154, \cube 231
	&${\it cl}({\it if}({\it ev}(v_p),{\it up}(v_1),{\it dn}(v_1)))$
	& 1. & 604234	\\
\hline
\end{tabular}
\end{center}
\caption{Computed Construction Laws}
\LABEL{Computed construction laws}
\end{figure}

Figure~\ref{Computed construction laws}
shows some computed construction laws.
Its first line exactly corresponds to the example in
Fig.~\ref{Computation of a Construction Law} (left).
The column \emph{Theory} indicates the equational theory used.
The ternary function ${\it if}$ realizes
${\it if} \cdot {\it then} \cdot {\it else}$,
with the defining equations
${\it if}(\suc(x),y,z) = y$ and ${\it if}(0,y,z) = z$,
and the unary function ${\it ev}$ 
returns $\suc(0)$ for even and $0$ for odd natural numbers.
Using ${\it if}$ and ${\it ev}$,
two series can be interleaved (cf.\ line~5).

The column \emph{Series} shows the given term series,
$\suc^n(0)$ being abbreviated to $n$.
The number $k$ of suffixes supplied to our procedure corresponds to the
number of series terms to the right of the semi-colon.
Any computed
hypothesis 
must explain all these series terms, 
but none of the earlier ones.
The column \emph{Law} shows the computed hypothesis.
The place within the series is denoted by $v_p$,
the first term having place 0, the second place 1,
and so on.
The previous 
series term and the last but one 
are denoted by $v_1$ and $v_2$, respectively.

The column \emph{No} shows the place in which the law appeared in the
enumeration sequence.
In line~5, some
formally smaller (wrt.\ height) terms are enumerated before
the term shown in Fig.~\ref{Computed construction laws},
but are nevertheless equivalent to it.
%
The column \emph{Time} shows the required runtime in
milliseconds on a 933 MHz
machine, again strongly depending on $k$ and the size of series terms.

The strength of our approach does not lie in its finding a plausible
continuation of a given series, but rather in building,
from a precisely limited set of operators, a nonrecursive
\emph{algorithm} for computing the next series terms.
Human superiority in the former area is demonstrated in
line~9, where no construction law was found.
The strength of the approach in the latter area became clear by the
series $0,0;1,0,0,1$, shown in line~7.
We had not expected any construction law to exist at all,
because the
series has a period relative prime to 2 and the trivial solution
$v_3$ had been eliminated by the choice of $k$ (a construction law
must compute the first 1 from the preceding 0s).

It is decidable whether the result language $\Hyp{\ref{appl r3}}{}{}$
is finite;
in such cases,
we can make precise propositions about all construction laws that can be
expressed using the given signature and equational theory.
For example, from line~9 we can conclude that no
construction law can be built from the given operators.

\subsection{Generalizing Screen Editor Commands}
\LABEL{Generalizing screen editor commands}

By way of
another application, we employed $E$-generalization for
learning complex cursor-movement
commands of a screen-oriented editor like {\sc Unix} {\tt vi}.
For each $i,j \in \N$, let $p_{i,j}$ be a distinct constant
denoting
the position of a given file
at column $i$ and line $j$;
let $P = \set{p_{i,j} \mid i,j \in \N}$.
For the sake of simplicity, we assume that
the screen is large enough to display
the entire contents of the file, so, we do not deal with scrolling
commands for the present.

Assuming the file contents to be given, cursor-movement commands can
be modeled as partial functions from $P$ to itself.
For example,
$d(p_{i,j}) = p_{i,j+1}$ if $j+1 \leq {\it li}$, undefined otherwise,
models the \emph{down} command, where ${\it li}$ denotes the number of
lines in the file.
The constant $H = p_{1,1}$ models the \emph{home} command.
Commands may depend on the file contents.
For example,
$$W(p_{i,j}) =
\min \set{ i'
\mid i < i' \leq {\it co}(j)
\land {\it ch}(p_{i'-1,j}) \in {\it SP}
\land {\it ch}(p_{i',j}) \not\in {\it SP} } ,$$
if the minimum is defined,
models the \emph{next word} command,
where ${\it co}(j)$, ${\it ch}(p)$, and ${\it SP}$
denote the number of columns of line $j$,
the character at position $p$, and
the set of space characters, respectively.

From a given file contents, it is easy to compute a regular tree grammar
$\G$ that describes the congruence classes of all its positions in time
linear to the file size and the number of movement commands.
Figure~\ref{exm screen, grammar} gives an example.
For the sake of brevity, columns are ``numbered'' by lower-case
letters, and, e.g., $\L({\rm b2}) = \eqc{p_{b,2}}{E}{}$.
Note that the file contents happen to explain some 
movement commands.

\begin{figure}
\newcommand{\0}[1]{\mbox{\rm #1}}
{\tt
\renewcommand{\arraystretch}{1.2}
\begin{tabular}
	{@{}l@{$\;$}|*{23}{@{\hspace*{0.01cm}}c@{\hspace*{0.01cm}}}|@{}}
\multicolumn{1}{l}{}
  &\0a&\0b&\0c&\0d&\0e&\0f&\0g&\0h&\0i&\0j
  &\0k&\0l&\0m&\0n&\0o&\0p&\0q&\0r&\0s&\0t
  &\0u&\0v
  &\multicolumn{1}{@{\hspace*{0.02cm}}c@{\hspace*{0.02cm}}}{\0w}\\
\cline{2-24}
\01&C&U&R&S&O&R& &M&O&T&I&O&N& &C&O&M&M&A&N&D&S&:\\
\02&l& &l&e&f&t& & & & &H& &h&o&m&e& & & & & & & \\
\03&r& &r&i&g&h&t& & & &m& &m&a&t&c&h&i&n&g& &(&)\\
\04&u& &u&p& & & & & & &W& &n&e&x&t& &w&o&r&d& & \\
\05&d& &d&o&w&n& & & & &B& &p&r&e&v& &w&o&r&d& & \\
\cline{2-24}
\multicolumn{1}{l}{}	\\
\end{tabular}%
}%
\hfill
$\begin{array}{@{}c@{\;}r*{8}{@{}c}@{}}
\0{b2} & \sortdef
	& \0{l(c2)}
	& \mid & \0{r(a2)}
	& \mid & \0{u(b3)}
	& \mid & \0{d(b1)}
	& .	\\
\0{c2} & \sortdef
	& \0{l(d2)}
	& \mid & \0{r(b2)}
	& \mid & \0{u(c3)}
	& \mid & \0{d(c1)}
	& \mid	\\
	&      & \0{W(a2)} 
	& \mid & \0{W(b2)}
	& \mid & \multicolumn{3}{l}{\0{B(d2)} ... \0{B(k2)}}
	& .	\\
\0{k2} & \sortdef 
	& \0{l(l2)}
	& \mid & \0{r(j2)}
	& \mid & \0{ u(k3) }
	& \mid & \0{ d(k1) }
	& \mid	\\
	&      & \0{B(l2)}
	& \mid & \0{B(m2)}
	& \mid & \multicolumn{3}{l}{\0{W(c2)} ... \0{W(j2)}}
	& .	\\
\end{array}$
\caption{Example File Contents
	\hspace{1cm} Corresponding Grammar Excerpt}
\LABEL{exm screen, grammar}
\end{figure}

Using $E$-generalization,
two or more cursor movements can easily be generalized to obtain
a common scheme.
Given the start and end positions, $\,1n{s_\i}$ and $\,1n{e_\i}$,
we apply Thm.~\ref{appl r3} to
$F^+ \Lra \:{\land}1n{p(s_\i,e_\i)}$
and get a rule of the form $p(x,t)$,
where $t \in \T{}{}{\set{x}}$ is a term describing a command sequence
that achieves each of these movements.

For $n=1$,
we can compute the \emph{simplest} term that transforms a
given starting position into a given end position.
This is useful to
advise a novice user about advanced cursor-movement commands.
Imagine, for example, that a user had typed the commands
$l,l,l$, $l,l,l$, $l,l,l$
to get from position $p_{k,2}$ to $p_{b,2}$.
The term of least height obtained from Thm.~\ref{appl r3},
viz.\ $p(x,l(B(x)))$,
indicates that the same
movement could have been achieved by simply
typing the commands $B$, $l$.

Each command could
also be assigned its own degree of simplicity,
reflecting, for example, the number of modifier keys (like \emph{shift})
involved,
or distinguishing between simple and advanced commands.
In the former case, the simplest term minimized the overall numbers
of keys to be pressed.

No grammar intersection is needed if $n=1$.
Moreover, the lifting of $\G$ can be done in constant time in this case.
In the example,
it is sufficient to include an alternative
$\ldots x \ldots$ into the right-hand side of the rule for $\rm k2$.
Therefore, a simplest term can be computed in an overall time of
$\O(\abs{\G} \cdot \log \abs{\G})$.
Changes in the file content require recomputation of the
grammar and the minimum term sizes.
In many cases, but not if, for example, a parenthesis is changed,
local content changes require local grammar
changes only.
It thus seems worthwhile to investigate an incremental approach,
which should also cover weight recomputation.

For $n>1$,
the smallest term(s) in the result language may be used to implement an
intelligent approach to repeat the last $n$
movement command sequences.
For example, 
the simplest scheme common to the movements
$p(p_{m,2},p_{o,2})$ and $p(p_{n,4},p_{v,4})$
wrt.\ the file content
of Fig.~\ref{exm screen, grammar}
is computed as
$p(x,d(W(u(x))))$.
Since the computation time grows exponentially with $n$,
it should be small.

In our prototypical implementation, we considered in all
the {\tt vi} commands
{\tt h}, {\tt j}, {\tt k}, {\tt l}, {\tt H}, {\tt M}, {\tt L},
{\tt +}, {\tt -}, {\tt w}, {\tt b}, {\tt e}, {\tt W}, {\tt B},
{\tt E}, {\tt 0}, {\tt \$}, {\tt f}, {\tt F}, {\tt \%},
{\tt $\{$}, and {\tt $\}$}
and renamed some of them to give them more suggestive identifiers.
We allow search for single characters only.
In order to consider nontrivial string search commands
as well, the above approach
should be combined with (string)
grammar inference \cite{Sakakibara.1997,Honavar.Parekh.1999}
to learn regular search expressions.
Moreover, commands that change the file content should be included
in the learning mechanism.
And last but not least,
a satisfactory user interface for these learning features
is desirable, e.g.\ allowing us to define command macros from examples.

\subsection{Prototypical Implementation}
\LABEL{Prototype implementation}

\begin{figure}
\begin{center}
\begin{picture}(13.5,7.5)
\put(3.000,6.833){\vector(-2,-1){1.933}}		
\put(3.000,6.833){\vector(-3,-4){1.733}}		
\put(3.000,6.833){\vector(0,-1){2.333}}			
\put(3.000,6.833){\vector(3,-2){3.467}}			
\put(3.000,6.833){\line(4,-3){3.000}}
\put(6.000,4.583){\vector(0,-1){3.417}}			
\put(3.000,6.833){\vector(3,-1){6.967}}			
\put(7.000,6.833){\line(-3,-1){3.000}}
\put(4.000,5.833){\vector(-2,-1){2.633}}		
\put(7.000,6.833){\vector(-3,-2){3.467}}		
\put(7.000,6.833){\vector(0,-1){2.333}}			
\put(7.000,6.833){\line(-1,-2){0.933}}
\put(6.067,4.967){\vector(0,-1){3.800}}			
\put(7.000,6.833){\line(3,-1){3.333}}
\put(10.333,5.722){\vector(0,-1){1.222}}		
\put(7.000,6.833){\vector(3,-2){1.467}}			
\put(10.800,6.800){\vector(1,-1){1.600}}		
\put(10.800,6.800){\line(-6,-1){6.800}}			
\put(10.800,6.800){\line(-5,-1){7.070}}			
\put(10.800,6.800){\line(-4,-1){3.700}}			
\put(4.000,5.667){\vector(-2,-1){2.300}}		
\put(3.720,5.380){\vector(-1,-2){0.430}}		
\put(7.100,5.866){\vector(0,-1){1.366}}			
\put(10.333,4.167){\vector(-2,-3){1.550}}		
\put(10.333,4.167){\vector(0,-1){3.000}}		
\put(3.000,4.167){\vector(-1,-3){1.000}}		
\put(3.000,4.167){\vector(-3,-2){1.467}}		
\put(5.000,4.167){\vector(-3,-1){2.933}}		
\put(5.000,4.167){\vector(-1,-2){1.483}}		
\put(5.000,4.167){\vector(0,-1){2.333}}			
\put(0.333,4.167){\vector(0,-1){2.333}}			
\put(5.000,1.500){\vector(-3,-1){0.933}}		
\put(5.000,1.500){\vector(0,-1){1.000}}			
\put(5.000,1.500){\vector(3,-1){0.933}}			
\put(5.933,1.167){\line(0,1){1.500}}
\put(5.933,2.667){\vector(-2,1){2.933}}			
\put(2.000,6.667){\dashbox{0.1}(10.000,0.667){}}	
\put(2.000,4.000){\dashbox{0.1}(6.000,0.667){}}		
\put(9.333,4.000){\dashbox{0.1}(4.000,1.333){}}		
\put(0.000,0.000){\dashbox{0.1}(11.333,2.000){}}	
\thicklines
\put(10.167,6.833){\framebox(1.667,0.333){\function{edt cmds}}}
\put(6.167,6.833){\framebox(1.667,0.333){\function{lemmas}}}	
\put(2.167,6.833){\framebox(1.667,0.333){\function{series}}}	
\put(11.500,4.834){\framebox(1.667,0.333){\function{edt grm}}}	
\put(11.500,4.167){\framebox(1.667,0.333){\function{var grm}}}	
\put(9.500,4.167){\framebox(1.667,0.333){\function{nf grm}}}	
\put(6.167,4.167){\framebox(1.667,0.333){\function{synt au}}}	
\put(4.167,4.167){\framebox(1.667,0.333){\function{uc e-au}}}	
\put(2.167,4.167){\framebox(1.667,0.333){\function{cs e-au}}}	
\put(7.833,5.500){\framebox(1.667,0.333){\function{exm inst}}}	
\put(0.167,5.500){\framebox(1.667,0.333){\function{term eval}}}	
\put(0.167,4.167){\framebox(1.667,0.333){\function{enum}}}	
\put(0.500,2.833){\framebox(1.667,0.333){\function{weight}}}	
\put(4.167,1.500){\framebox(1.667,0.333){\function{max nt s}}}	
\put(0.167,1.500){\framebox(1.667,0.333){\function{finite}}}	
\put(9.500,0.833){\framebox(1.667,0.333){\function{complem}}}	
\put(5.167,0.833){\framebox(1.667,0.333){\function{intersect}}}	
\put(3.167,0.833){\framebox(1.667,0.333){\function{simplify}}}	
\put(1.167,0.833){\framebox(1.667,0.333){\function{member}}}	
\put(4.167,0.167){\framebox(1.667,0.333){\function{empty}}}	
\put(7.833,1.500){\framebox(1.667,0.333){\function{top nt}}}
\end{picture}
\caption{Prototype Architecture}
\LABEL{Prototype architecture}
\end{center}
\end{figure}
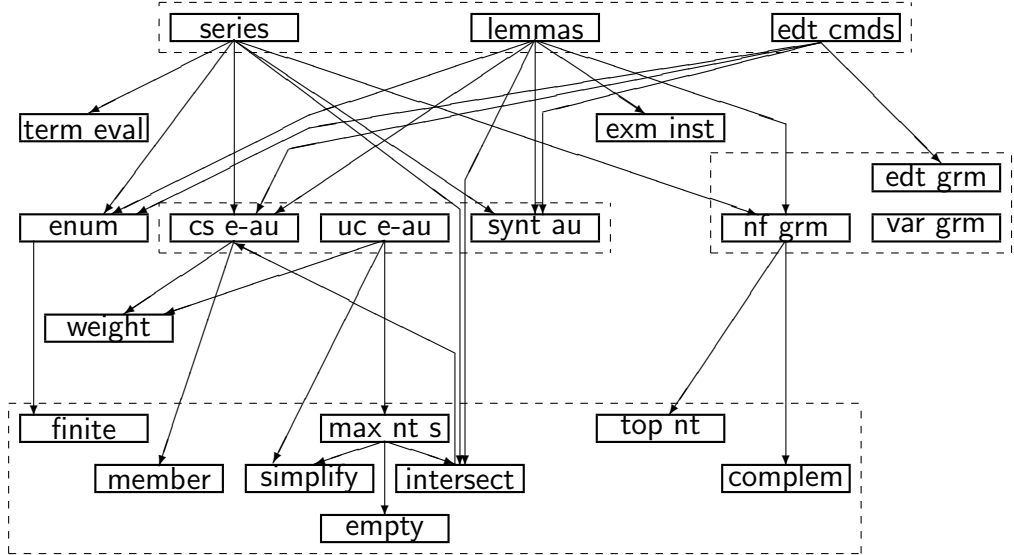

We built a prototypical implementation realizing the
$E$-generalization method from
Sect.~\ref{The core method} and the applications from
Sect.~\ref{Applications}.
It comprises about 4,000 lines of {\sc Prolog} code.
Figure~\ref{Prototype architecture} shows its architecture,
an arrow meaning that its source function uses its destination function.

The application module allows us to learn \function{series} laws,
candidate \function{lemmas},
and editor cursor commands (\function{edt cmds}).
The anti-unification module contains algorithms for
syntactic (\function{synt au}), constrained (\function{cs e-au})
and unconstrained (\function{uc e-au}) $E$-anti-unification.
The grammar-generation module can compute grammars for 
a given file content (\function{edt grm}),
for any set
$\set{ t \.\in \T{}{}{} \mid V \.\subseteq \var(t) \.\subseteq W}$
(\function{var grm}),
and for the set of normal forms wrt.\ $E$ (\function{nf grm}).
The grammar algorithms module allows us to test an $\L(N)$
for \function{finite}ness, \function{empt}iness,
and a given \function{member} $t$,
to compute \function{intersect}ion and
\function{complem}ent of two languages,
to \function{simplify} a grammar, and to generate
$\NM$ from Lem.~\ref{hsg lemma nondet} (\function{max nt s})
and a grammar for $\T{}{}{\set{}}$ (\function{top nt}).
For the sake of clarity, we omitted the dashed lines around
the pre- and postprocessing module.
The former merely contains code to choose
\function{exm inst}ances for lemma generation.
The latter does \function{term eval}uation to normal form,
and \function{enum}eration
and minimal \function{weight} computation for $\L(N)$.

The prototype still uses monolithic, specially tailored
algorithms for
$E$-anti-unification, as originally given in \cite{Heinz.1995},
rather than the combination of standard grammar algorithms described in
Sect.~\ref{The core method}.
For this reason, function \function{intersect} uses \function{cs e-au}
as a special case, viz.\ $\sigma_i = \set{}$, rather than vice versa.
However, all other \emph{uses} relations would remain unchanged in an
implementation strictly based on this paper.

Figure~\ref{runtimes} shows some measured
runtimes for $i$--fold simultaneous
$E$--anti--unification of arithmetic congruence classes.
The horizontal position indicates which classes were used as input,
ranging from $\eqc{0}{E}{}$ to $\eqc{20}{E}{}$.
A digit indicates the value of $i$; its index indicates $E$,
where $+$ means that $E$ just defines \emph{sum},
while $*$ means that $E$ defines \emph{sum} and \emph{product}.
The vertical position --- logarithmically scaled --- indicates the
required computation time in seconds on a 933 MHz PC under
compiled {\sc Prolog}.
For examples, the ``$5_*$'' near the upper left corner
means that it took $113$ seconds to generalize $3,3,3,3,3$ wrt.\
\emph{sum} and \emph{product}.

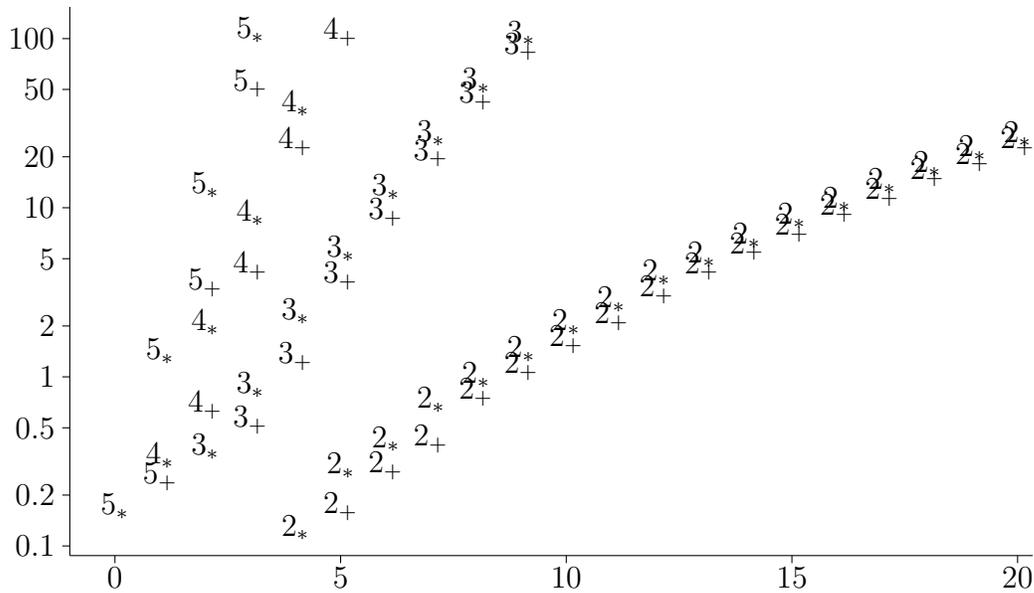
\begin{figure}
\begin{center}
\begin{picture}(13.9,7.8)
\put(1.000,0.500){\line(1,0){12.800}}
\put(1.600,0.500){\line(0,-1){0.100}}
\put(1.600,0.350){\makebox(0.000,0.000)[t]{$0$}}
\put(4.600,0.500){\line(0,-1){0.100}}
\put(4.600,0.350){\makebox(0.000,0.000)[t]{$5$}}
\put(7.600,0.500){\line(0,-1){0.100}}
\put(7.600,0.350){\makebox(0.000,0.000)[t]{$10$}}
\put(10.600,0.500){\line(0,-1){0.100}}
\put(10.600,0.350){\makebox(0.000,0.000)[t]{$15$}}
\put(13.600,0.500){\line(0,-1){0.100}}
\put(13.600,0.350){\makebox(0.000,0.000)[t]{$20$}}
\put(1.000,0.500){\line(0,1){7.300}}
\put(1.000,0.628){\line(-1,0){0.100}}
\put(0.800,0.628){\makebox(0.000,0.000)[r]{$0.1$}}
\put(1.000,1.305){\line(-1,0){0.100}}
\put(0.800,1.305){\makebox(0.000,0.000)[r]{$0.2$}}
\put(1.000,2.200){\line(-1,0){0.100}}
\put(0.800,2.200){\makebox(0.000,0.000)[r]{$0.5$}}
\put(1.000,2.878){\line(-1,0){0.100}}
\put(0.800,2.878){\makebox(0.000,0.000)[r]{$1$}}
\put(1.000,3.555){\line(-1,0){0.100}}
\put(0.800,3.555){\makebox(0.000,0.000)[r]{$2$}}
\put(1.000,4.450){\line(-1,0){0.100}}
\put(0.800,4.450){\makebox(0.000,0.000)[r]{$5$}}
\put(1.000,5.128){\line(-1,0){0.100}}
\put(0.800,5.128){\makebox(0.000,0.000)[r]{$10$}}
\put(1.000,5.806){\line(-1,0){0.100}}
\put(0.800,5.806){\makebox(0.000,0.000)[r]{$20$}}
\put(1.000,6.702){\line(-1,0){0.100}}
\put(0.800,6.702){\makebox(0.000,0.000)[r]{$50$}}
\put(1.000,7.378){\line(-1,0){0.100}}
\put(0.800,7.378){\makebox(0.000,0.000)[r]{$100$}}
\put(4.600,1.155){\makebox(0.000,0.000){$2_+$}}
\put(5.200,1.702){\makebox(0.000,0.000){$2_+$}}
\put(5.800,2.050){\makebox(0.000,0.000){$2_+$}}
\put(6.400,2.677){\makebox(0.000,0.000){$2_+$}}
\put(7.000,3.020){\makebox(0.000,0.000){$2_+$}}
\put(7.600,3.379){\makebox(0.000,0.000){$2_+$}}
\put(8.200,3.679){\makebox(0.000,0.000){$2_+$}}
\put(8.800,4.043){\makebox(0.000,0.000){$2_+$}}
\put(9.400,4.348){\makebox(0.000,0.000){$2_+$}}
\put(10.000,4.615){\makebox(0.000,0.000){$2_+$}}
\put(10.600,4.862){\makebox(0.000,0.000){$2_+$}}
\put(11.200,5.120){\makebox(0.000,0.000){$2_+$}}
\put(11.800,5.340){\makebox(0.000,0.000){$2_+$}}
\put(12.400,5.602){\makebox(0.000,0.000){$2_+$}}
\put(13.000,5.803){\makebox(0.000,0.000){$2_+$}}
\put(13.600,6.011){\makebox(0.000,0.000){$2_+$}}
\put(3.400,2.307){\makebox(0.000,0.000){$3_+$}}
\put(4.000,3.155){\makebox(0.000,0.000){$3_+$}}
\put(4.600,4.219){\makebox(0.000,0.000){$3_+$}}
\put(5.200,5.074){\makebox(0.000,0.000){$3_+$}}
\put(5.800,5.857){\makebox(0.000,0.000){$3_+$}}
\put(6.400,6.623){\makebox(0.000,0.000){$3_+$}}
\put(7.000,7.273){\makebox(0.000,0.000){$3_+$}}
\put(2.800,2.509){\makebox(0.000,0.000){$4_+$}}
\put(3.400,4.356){\makebox(0.000,0.000){$4_+$}}
\put(4.000,6.010){\makebox(0.000,0.000){$4_+$}}
\put(4.600,7.464){\makebox(0.000,0.000){$4_+$}}
\put(2.200,1.551){\makebox(0.000,0.000){$5_+$}}
\put(2.800,4.129){\makebox(0.000,0.000){$5_+$}}
\put(3.400,6.783){\makebox(0.000,0.000){$5_+$}}
\put(4.000,0.872){\makebox(0.000,0.000){$2_*$}}
\put(4.600,1.702){\makebox(0.000,0.000){$2_*$}}
\put(5.200,2.050){\makebox(0.000,0.000){$2_*$}}
\put(5.800,2.567){\makebox(0.000,0.000){$2_*$}}
\put(6.400,2.905){\makebox(0.000,0.000){$2_*$}}
\put(7.000,3.244){\makebox(0.000,0.000){$2_*$}}
\put(7.600,3.603){\makebox(0.000,0.000){$2_*$}}
\put(8.200,3.909){\makebox(0.000,0.000){$2_*$}}
\put(8.800,4.260){\makebox(0.000,0.000){$2_*$}}
\put(9.400,4.502){\makebox(0.000,0.000){$2_*$}}
\put(10.000,4.759){\makebox(0.000,0.000){$2_*$}}
\put(10.600,5.020){\makebox(0.000,0.000){$2_*$}}
\put(11.200,5.238){\makebox(0.000,0.000){$2_*$}}
\put(11.800,5.481){\makebox(0.000,0.000){$2_*$}}
\put(12.400,5.708){\makebox(0.000,0.000){$2_*$}}
\put(13.000,5.920){\makebox(0.000,0.000){$2_*$}}
\put(13.600,6.102){\makebox(0.000,0.000){$2_*$}}
\put(2.800,1.947){\makebox(0.000,0.000){$3_*$}}
\put(3.400,2.775){\makebox(0.000,0.000){$3_*$}}
\put(4.000,3.751){\makebox(0.000,0.000){$3_*$}}
\put(4.600,4.579){\makebox(0.000,0.000){$3_*$}}
\put(5.200,5.402){\makebox(0.000,0.000){$3_*$}}
\put(5.800,6.118){\makebox(0.000,0.000){$3_*$}}
\put(6.400,6.813){\makebox(0.000,0.000){$3_*$}}
\put(7.000,7.447){\makebox(0.000,0.000){$3_*$}}
\put(2.200,1.833){\makebox(0.000,0.000){$4_*$}}
\put(2.800,3.603){\makebox(0.000,0.000){$4_*$}}
\put(3.400,5.056){\makebox(0.000,0.000){$4_*$}}
\put(4.000,6.512){\makebox(0.000,0.000){$4_*$}}
\put(1.600,1.155){\makebox(0.000,0.000){$5_*$}}
\put(2.200,3.218){\makebox(0.000,0.000){$5_*$}}
\put(2.800,5.429){\makebox(0.000,0.000){$5_*$}}
\put(3.400,7.495){\makebox(0.000,0.000){$5_*$}}
\end{picture}
\caption{$E$--Anti--Unification runtime
      vs.\ size and number of grammars}
\LABEL{runtimes}
\end{center}
\end{figure}

All runtime figures given in this paper are taken from the prototype.
Currently, an efficiency-oriented re-implementation in C is planned.
%
%
%
%
We expect it to provide
a speed-up factor of between $10$ and $100$.
Moreover, it will use the available memory more efficiently,
thus
allowing us to run larger examples than when using {\sc Prolog}.
The {\sc Prolog} prototypical implementation and, in future,
the C implementation can be downloaded from the web page
\url{http://swt.cs.tu-berlin.de/~jochen/e-au}.

\section{Conclusions and Future Work}
\LABEL{Conclusions}

We presented a method
for computing a finite representation of the set of $E$-generalizations
of given terms and showed some applications.
$E$-generalization is able to cope
with
representation change in abstraction, making it a promising
approach to an old but not yet satisfactorily solved problem of
Artificial Intelligence.

Our approach is based on standard algorithms for
regular tree grammars.
It thus allows us to add
filtering components in a modular fashion, as needed by the
surrounding application software.
The closed form of an $E$-generalization set as a grammar and its
simple mathematical characterization make it easy to prove formal
quality properties if needed for an application.
Using a standard grammar language enumeration algorithm,
the closed form can be converted
to a succession form.

Our method cannot handle every equational theory $E$.
To use the analogy
with $E$-deduction,
our method corresponds to something between
$E$-unification (concerned
with a particular $E$ in each case)
and paramodulation (concerned with the large class of canonical $E$).
On the other hand,
neither partial functions nor conditional equations basically prevent
our method from being applicable.

In order to demonstrate
that $E$-generalization can be integrated into ILP
learning methods, we proved several ways of combining ${\it lgg}$-based
ILP and $E$-generalization.
Predicate definitions by atoms or clauses can be learned.
If desired,
the hypotheses space can be restricted to determinate hypotheses,
resulting in faster algorithms.

Learning of purely determinate clauses can be reduced to learning of
atoms by $E$-generalization.
An ${\it lgg}$-learner for constrained clauses
with built-in $E$-generalization
can learn a proper superclass, called semi-determinate predicates here.
%
We provide completeness properties for all our hypotheses sets.

Using $E$-generalization, the search space is split into 
two parts,
one concerned with selection of nondeterminate literals,
the other with selection of their argument terms.
While the first part is best handled by an elaborate strategy from
\emph{classical} ILP,
the second can be left to a grammar language enumeration strategy.
For example, the $\O(\abs{\G} \cdot \log \abs{\G})$ algorithm to find a
term of minimal complexity within a tree language
apparently has no corresponding selection algorithm for
determinate literals in classical ILP.
Separating both search space
parts allows us to modularize the strategy algorithms
and to use for each part one that best fits the needs of the surrounding
application.

Experiments with our prototypical implementation showed that
comparatively ambitious AI tasks are solvable at the first attempt
using $E$-generalization.
We focus on sketching applications in a number of different areas
rather than on perfectly elaborating a single application.
By doing so, we seek to demonstrate the flexibility of
$E$-generalization, which is a necessary feature for any approach to
be related to intelligence.

In \cite[Sect.~8]{Burghardt.Heinz.1996},
further applications were sketched, including
%
divergence handling in Knuth-Bendix completion,
guessing of Hoare invariants,
reengineering of functional programs, and
strengthening of induction hypotheses.
%
The method given in \cite{Burghardt.2002b}
to compute a finite representation
of the complete equational theory describing a given set of
finite algebras is essentially based on $E$-generalization, too.
It is shown there that the complete theory can be used to implement
fast special-purpose theorem provers for particular theories.

Based on this experience, we venture to suggest that $E$-generalization
is able to simulate an important aspect of human intelligence, and that
it is worth investigating further.
In particular, the restrictions
regular tree grammars impose on the background equational theory $E$
should be relaxed.
In this paper, we briefly looked at some well-known representation
formalisms that are more expressive than regular tree grammars
but with negative results.
It remains to be seen whether there are other more expressive
formalisms that can be used for $E$-generalization.

The attempt should also be made to combine it with higher-order
anti-unification \cite{Hasker.1995,Wagner.2002}.
Such a combination is expected to allow recursive functions to be
learned from examples.

As indicated above,
the applications of $E$-generalization could certainly be improved.
Lemma generation should be integrated into a real induction prover,
in particular
to test its behavior in combination with the \emph{rippling}
method \cite{Bundy.Harmelen.Ireland.1990}.
While rippling suggests checking homomorphic laws like
$f( \,1n{g(t_\i)} ) \eq g'(f( \,1n{t_\i} ))$ for validity,
$E$-generalization is able to suggest lemmas of arbitrary forms.
Empirical studies on series-based
intelligence tests, e.g.\ using geometrical theories about
\emph{mirror}, \emph{shift}, \emph{rotate}, etc.,
should look for a saturation effect:
Is there one single
reasonable equational
background theory that can solve a sufficiently large number of common
tests? And can a reasonable
\emph{intelligence quotient} be achieved by that theory?

Currently, we are investigating the use of $E$-generalization
in analogical reasoning \cite{Dastani.Indurkhya.Scha.1997},
a new application that does not fit into the schemas 
described in
Sect.~\ref{Learning predicate definitions}.
The aim is to allow problems
in intelligence tests 
to be stated in other ways
than mere linear series, e.g.\
to solve $(A:B) = (C:X)$, where $A,B,C$ are given terms and $X$ is 
a term which should
result from applying a rule to $C$ that at the same time transforms $A$
into $B$.

\ack{Ute Schmid, Holger Schlingloff and Ulrich Geske
provided valuable advice on presentation.}

\nocite{Carbonell.Michalski.Mitchell.1984}
\nocite{Dale.Moisl.Somers.1999}

\bibliographystyle{alpha}
\bibliography{lit,cog}

\newcommand{\etalchar}[1]{$^{#1}$}
\begin{thebibliography}{CDG{\etalchar{+}}99}

\bibitem[Baa91]{Baader.1991}
Franz Baader.
\newblock Unification, weak unification, upper bound, lower bound, and
  generalization problems.
\newblock In {\em Proc. 4th Conf. on Rewriting Techniques and Applications},
  volume 488 of {\em LNCS}, pages 86--91. Springer, 1991.

\bibitem[BH96]{Burghardt.Heinz.1996}
Jochen Burghardt and Birgit Heinz.
\newblock Implementing anti-unification modulo equational theory.
\newblock Arbeitspapier 1006, GMD, Jun 1996.

\bibitem[BM79]{Boyer.Moore.1979}
R.S. Boyer and J.S. Moore.
\newblock {\em A Computational Logic}.
\newblock Academic, New York, 1979.

\bibitem[BT92]{Bogaert.Tison.1992}
B.~Bogaert and Sophie Tison.
\newblock Equality and disequality constraints on direct subterms in tree
  automata.
\newblock In {\em Proc. {STACS} 9}, volume 577 of {\em LNCS}, pages 161--172.
  Springer, 1992.

\bibitem[Bur02a]{Burghardt.2002b}
Jochen Burghardt.
\newblock Axiomatization of finite algebras.
\newblock In {\em Proc.\ {KI} 2002}, number 2479 in LNAI, pages 222--234.
  Springer, 2002.

\bibitem[Bur02b]{Burghardt.2002c}
Jochen Burghardt.
\newblock Weight computation of regular tree languages.
\newblock {\em Journal of Applied Logic}, 2002.
\newblock submitted.

\bibitem[BvHSI90]{Bundy.Harmelen.Ireland.1990}
Alan Bundy, Frank van Harmelen, Alan Smaill, and Andrew Ireland.
\newblock Extensions to the rippling-out tactic for guiding inductive proofs.
\newblock In {\em Proc.\ 10th CADE}, volume 449 of {\em LNAI}, pages 132--146.
  Springer, 1990.

\bibitem[CCC{\etalchar{+}}94]{Caron.Comon.Coquide.1994}
Anne-C\'ecile Caron, Hubert Comon, Jean-Luc Coquid\'e, Max Dauchet, and Florent
  Jacquemard.
\newblock Pumping, cleaning and symbolic constraints solving.
\newblock In {\em Proc.\ ICALP}, volume 820 of {\em LNCS}, pages 436--449,
  1994.

\bibitem[CCD95]{Caron.Coquide.Dauchet.1995}
Anne-C\'ecile Caron, Jean-Luc Coquid\'e, and Max Dauchet.
\newblock Automata for reduction properties solving.
\newblock {\em Journal of Symbolic Computation}, 20(2):215--233, Aug 1995.

\bibitem[CDG{\etalchar{+}}99]{Comon.Dauchet.Gilleron.1999}
H.~Comon, M.~Dauchet, R.~Gilleron, F.~Jacquemard, D.~Lugiez, S.~Tison, and
  M.~Tommasi.
\newblock {\em Tree Automata Techniques and Applications}.
\newblock Available from \url{www.grappa.univ-lille3.fr/tata}, Oct 1999.

\bibitem[Cla79]{Clark.1979}
K.L. Clark.
\newblock Predicate logic as a computational formalism.
\newblock Research report, Imperial College, 1979.

\bibitem[DIS97]{Dastani.Indurkhya.Scha.1997}
M.~Dastani, B.~Indurkhya, and R.~Scha.
\newblock {\em An Algebraic Method for Solving Proportional Analogy Problems}.
\newblock Dublin City University, 1997.

\bibitem[DJ90]{Dershowitz.Jouannaud.1990}
N.~Dershowitz and J.-P. Jouannaud.
\newblock {\em Rewrite Systems}, volume~B of {\em Handbook of Theoretical
  Computer Science}, pages 243--320.
\newblock Elsevier, 1990.

\bibitem[DM84]{Dietterich.Michalski.1984}
T.~G. Dietterich and R.~S. Michalski.
\newblock {\em A Comparative Review of Selected Methods for Learning from
  Examples}, pages 41--82.
\newblock In Michalski et~al. \cite{Carbonell.Michalski.Mitchell.1984}, 1984.

\bibitem[DMS99]{Dale.Moisl.Somers.1999}
Dale, Moisl, and Somers, editors.
\newblock Marcel Dekker, New York, 1999.

\bibitem[DST80]{Downey.Sethi.Tarjan.1980}
Peter~J. Downey, Ravi Sethi, and Robert~E. Tarjan.
\newblock Variations on the common subexpression problem.
\newblock {\em JACM}, 27(4):758--771, Oct 1980.

\bibitem[D{\v{z}}e96]{Dzeroski.1996}
Sa{\v{s}}o D{\v{z}}eroski.
\newblock {\em Inductive Logic Programming and Knowledge Discovery in
  Databases}, pages 117--152.
\newblock MIT Press, 1996.

\bibitem[Fay79]{Fay.1979}
Fay.
\newblock First-order unification in an equational theory.
\newblock In {\em Proc.\ 4th Workshop on Automated Deduction}, 1979.

\bibitem[Gen32]{Gentzen.1932}
Gerhard Gentzen.
\newblock {\em {U}n\-ter\-su\-chun\-gen über das logische {S}chlie\-ßen}.
\newblock 1932.

\bibitem[Gol67]{Gold.1967}
E.~Mark Gold.
\newblock Language identification in the limit.
\newblock {\em Information and Control}, 10:447--474, 1967.

\bibitem[GS89]{Gallier.Snyder.1989}
Jean~H. Gallier and Wayne Snyder.
\newblock Complete sets of transformations for general {E}-unification.
\newblock {\em Theoretical Computer Science}, 67:203--260, 1989.

\bibitem[Has95]{Hasker.1995}
R.W. Hasker.
\newblock {\em The Replay of Program Derivations}.
\newblock PhD thesis, Univ. of Illinois at Urbana-Champaign, 1995.

\bibitem[Hei95]{Heinz.1995}
Birgit Heinz.
\newblock {\em {A}nti-{U}nifikation modulo {G}leichungstheorie und deren
  {A}nwendung zur {L}emmagenerierung}.
\newblock PhD thesis, TU Berlin, Dec 1995.

\bibitem[HP99]{Honavar.Parekh.1999}
Vasant Honavar and Rajesh Parekh.
\newblock {\em Grammar Inference, Automata Induction, and Language
  Acquisition}.
\newblock In Dale et~al. \cite{Dale.Moisl.Somers.1999}, 1999.

\bibitem[Kow73]{Kowalski.1973}
Robert Kowalski.
\newblock Predicate logic as programming language.
\newblock Memo~70, Dept.\ of Comp.\ Logic, School of Artif.\ Intell., Univ.\
  Edinburgh, 1973.

\bibitem[LD94]{Dzeroski.Lavrac.1994}
Nada Lavrac and Sa{\v{s}}o D{\v{z}}eroski.
\newblock {\em Inductive Logic Programming: Techniques and Applications}.
\newblock Ellis Horwood, New York, 1994.

\bibitem[McA92]{McAllester.1992}
David McAllester.
\newblock Grammar rewriting.
\newblock In {\em Proc. CADE--11}, volume 607 of {\em LNAI}. Springer, 1992.

\bibitem[MCM84]{Carbonell.Michalski.Mitchell.1984}
Ryszard~S. Michalski, Jaime~G. Carbonell, and Tom~M. Mitchell, editors.
\newblock {\em Machine Learning, An Artificial Intelligence Approach}.
\newblock Springer, 1984.

\bibitem[MF90]{Feng.Muggleton.1990}
S.~Muggleton and C.~Feng.
\newblock Efficient induction of logic programs.
\newblock In {\em Proc. 1st Conf. on Algorithmic Learning Theory, Tokyo}, pages
  368--381. Omsha, 1990.

\bibitem[Mug99]{Muggleton.1999}
Stephen Muggleton.
\newblock Inductive logic programming: Issues, results and the challenge of
  learning language in logic.
\newblock {\em Artificial Intelligence}, 114:283--296, 1999.

\bibitem[O'H92]{OHara.1992}
S.~O'Hara.
\newblock A model of the redescription process in the context of geometric
  proportional analogy problems.
\newblock In {\em Proc.\ AII '92, Dagstuhl, Germany}, volume 642 of {\em LNAI},
  pages 268--293. Springer, 1992.

\bibitem[Plo70]{Plotkin.1970}
Gordon~D. Plotkin.
\newblock A note on inductive generalization.
\newblock {\em Machine Intelligence}, 5:153--163, 1970.

\bibitem[Plo71]{Plotkin.1971}
Gordon~D. Plotkin.
\newblock A further note on inductive generalization.
\newblock {\em Machine Intelligence}, 6:101--124, 1971.

\bibitem[Pot89]{Pottier.1989b}
Loic Pottier.
\newblock Generalisation de termes en theorie equationelle. {C}as
  associatif-commutatif.
\newblock Report 1056, INRIA, 1989.

\bibitem[Rey70]{Reynolds.1970}
John~C. Reynolds.
\newblock Transformational systems and the algebraic structure of atomic
  formulas.
\newblock {\em Machine Intelligence}, 5:135--151, 1970.

\bibitem[Sak97]{Sakakibara.1997}
Yasubumi Sakakibara.
\newblock Recent advances of grammatical inference.
\newblock {\em Theoretical Computer Science}, 185:15--45, 1997.

\bibitem[Sch97]{Schoning.1997}
Uwe Schöning.
\newblock {\em Theoretische Informatik --- kurzgefaßt}.
\newblock Spektrum-Hochschultaschenbuch. Heidelberg, Berlin, 1997.

\bibitem[Sie85]{Siekmann.1985}
Jörg~H. Siekmann.
\newblock {\em Universal Unification}.
\newblock Univ. Kaiserslautern, 1985.

\bibitem[TW68]{Thatcher.Wright.1968}
J.W. Thatcher and J.B. Wright.
\newblock Generalized finite automata theory with an application to a decision
  problem of second-order logic.
\newblock {\em Mathematical Systems Theory}, 2(1), 1968.

\bibitem[Uri92]{Uribe.1992}
T.E. Uribe.
\newblock Sorted unification using set constraints.
\newblock In {\em Proc.\ CADE--11}, volume 607 of {\em LNCS}, pages 163--177,
  1992.

\bibitem[Val84]{Valiant.1984}
L.G. Valiant.
\newblock A theory of the learnable.
\newblock {\em Communications of the {ACM}}, 27:1134--1142, 1984.

\bibitem[Wag02]{Wagner.2002}
Ulrich Wagner.
\newblock Combinatorically restricted higher order anti-unification.
\newblock Master's thesis, TU Berlin, Apr 2002.

\end{thebibliography}






\end{document}